\documentclass[amsmath, amssymb,12pt, aps, onecolumn, notitlepage, eqsecnum, longbibliography, superscriptaddress]{revtex4-2}
   
\usepackage{graphicx}
\usepackage{calc}
\usepackage{braket}
\usepackage{slashed}
\usepackage{wasysym}
\usepackage{dsfont}
\usepackage{amsthm,amsmath,amsfonts,amssymb,verbatim,color}
\usepackage{lipsum}
\usepackage{subfigure}
\usepackage{bm}
\usepackage{epsfig,slashed}
\usepackage{hyperref}
\usepackage{flafter}
\usepackage{booktabs}
\usepackage{makecell} 
\usepackage{bbm} 
\usepackage[export]{adjustbox} 
\usepackage{cancel}
\usepackage{epstopdf}
\usepackage{tikz}
\usepackage{array,color}
\usepackage{float}

\newcommand{\hS}{{\hat S}}

\newcommand{\cH}{\hat{\cal H}}

\begin{document}

\begin{abstract}
    In geometrically frustrated (GF) magnets, conventional long-range order is suppressed due to the presence of primitive triangular structural units, and the nature of the ensuing ground state remains elusive. One class of candidate states, extensively sought in experiments and vigorously studied theoretically, is the quantum spin liquid (QSL), a magnetically-disordered state
    in which all spins participate in a quantum-coherent many-body state. Randomly located impurities, present in all materials, may prevent QSL formation and 
    instead lead to the formation of a spin-glass state.
    In this article, we review available data on the specific heat, magnetic susceptibility, and neutron scattering in GF materials. 
    Such data show that a pure GF magnet possesses a characteristic ``hidden energy scale'' significantly exceeded by the other microscopic energy scales in the material. When cooled
    down to a temperature below the hidden energy scale, a GF material develops significant short-range order that dominates its properties and, in particular, dictates
    the spin-glass transition temperature for
    experimentally accessible impurity densities. We review the manifestations of short-range
    order in the commonly observed thermodynamics quantities in GF materials,
    possible scenarios for the hidden energy scale, and related open questions.
\end{abstract}

\title{Short-range order and hidden energy scale in geometrically frustrated magnets}

	\author{A.P. Ramirez}
	\affiliation{Physics Department, University of California, Santa Cruz, California 95064, USA}

 	\author{S.V. Syzranov}
	\affiliation{Physics Department, University of California, Santa Cruz, California 95064, USA}

\maketitle

\tableofcontents

\section{Introduction}

In most magnetic materials at low temperatures, spins are arranged in a regular pattern, i.e. form long-range magnetic order. In some materials, however, 
it may be impossible to simultaneously minimize all magnetic interactions between spins, which may lead to competition between multiple magnetic states, and the material may possibly avoid long-range order.

In 1973, Anderson proposed~\cite{Anderson:firstQSL} that 
the ground state of
the Heisenberg model with nearest-neighbor antiferromagnetic (AF) interactions on a triangular lattice
may lack magnetic order and instead display liquid-like properties with strong quantum 
fluctuations.
Although this proposal turned out to be incorrect (the model develops the so-called $120^\circ$ order ~\cite{Bernu:triangularOrder,Capriotti:triangularOrder,White:traingularOrder}), it initiated
a series of theoretical studies~\cite{SavaryBalents:review,KnolleMoessner:review,Broholm:QSLreview} 
of ``quantum spin liquids'' (QSLs), magnetic 
states 
that lack long-range order (LRO) at $T=0$ and, broadly speaking, exhibit liquid-like 
properties due to strong quantum fluctuations.
The interest in QSLs has been driven by the possibility of numerous fundamental phenomena predicted in them, such as novel topological states, fractionalized excitations, and long-range entanglement, 
as well as the existence of anyon excitations in some QSLs, which can be used for robust quantum computation.

QSLs can be grouped into two theoretical categories that fit the broad description. The first includes models of spins on bipartite, e.g. honeycomb, lattices in which the interactions are tuned to prevent LRO. Spins on such lattices would
be ordered antiferromagnetically if the interactions were AF and short-ranged but avoid this order due to a more complicated structure of the 
interactions, exemplified by the Shastry-Sutherland~\cite{ShastrySutherland} and Kitaev~\cite{Kitaev:honeycob} models. Despite the existence of materials such as $RuCl_3$, considered as their approximate realizations~\cite{Iga:TmB4,Jackeli:SOMott},
these materials nevertheless undergo long-range order in zero magnetic field~\cite{Johnson:RuCl3ordering,Kubota:RuCl3ordering,Sears:RuCl3magneticOrder,Banerjee:RuCl3neutronScattering} instead of displaying clear signs of a QSL.
While interesting 
behavior suggestive of a QSL has recently been reported 
in $RuCl_3$ in sufficiently strong magnetic fields~\cite{Matsuda:RuCl3oscillations,CzajkaOng:RuCl3oscillations,ZhengYu:RuCl3QSL}, the
interpretation of the respective results as evidence for a
QSL is under dispute (see, e.g., Refs.~\cite{BruinTakagi:RuCl3magneticTransitions,BruinTakagi:RuCl3oscillations,LefrancoiceTaillefer:RuCl3noOscillations}).

A second class of systems, the so-called geometrically frustrated (GF) magnets, consists of materials in which the smallest structural unit in the magnetic sublattice is a triangle~\cite{Wannier:Ising,Anderson:OrderingFerrites,Ramirez:FrustratedThermodynamic}.
Such triangles make it impossible to minimize simultaneously all nearest-neighbor AF interactions between spins.
 This class includes large mineral families such as the spinels, pyrochlores, magnetoplumbites, jarosites, and delafossites~\cite{Lacroix:book}.
These materials show a suppression of LRO to temperatures far below the characteristic 
scales of AF interactions obtained from high-temperature measurements.
In the absence of LRO, strong quantum fluctuations in such materials are predicted to yield QSL states with fascinating fundamental
properties, such as many-body entanglement and topological
excitations~\cite{SavaryBalents:review,KnolleMoessner:review,Broholm:QSLreview,Chamorro:ChemReview,Zhou:QSLreview,Clark:QSLreview}.
Theoretical interest in such states has been mirrored by experimental searches that have produced a large number of magnets with triangular primitive units and greatly suppressed LRO transitions~\cite{Ramirez:FrustratedThermodynamic}.

\subsection*{What exactly is a quantum spin liquid?}

The term ``spin liquid'' commonly refers to a $T=0$ state of a magnetic
material that lacks LRO. The term ``{\it quantum} spin liquid''
is used to label a quantum-coherent state in which, in addition to the lack of LRO,
the quantum nature of spins gives rise to qualitatively new properties.
Despite some commonalities among the different QSL models, in our view, there is not a universal, commonly accepted
definition of a QSL.
Often, in theoretical work, a spin liquid is referred to as a QSL if it displays long-range
entanglement or topological properties, such as topological order
or fractionalized excitations, or can be described in terms of emergent gauge fields.

The ambiguity surrounding the theoretical description of QSLs is mirrored experimentally, where there is no consensus that QSLs have been conclusively identified in specific materials.
The experimental assignment of the QSL label to a particular material has relied on such features as the absence of a transition to LRO or the form of the inelastic structure factor measured by neutron scattering.
These features, however, are not unique to 
QSLs and allow for other explanations.
For example, LRO can be suppressed by
many other experimental factors, e.g., dimerization~\cite{Merchant:dimerization}, quenched
disorder~\cite{Kimchi:disorder}, low effective dimensionality~\cite{DeJongh:dimensionality},
or combinations of these.

Numerous excellent reviews dedicated to QSLs have been written
over the last decade (see, e.g., Refs.~~\cite{SavaryBalents:review,KnolleMoessner:review,Broholm:QSLreview,Chamorro:ChemReview,Zhou:QSLreview,Clark:QSLreview}). We refer the 
readers to these reviews for a detailed discussion of possible 
definitions of QSLs. In this article, we focus rather on the properties
of GF materials, the largest class of systems in which QSLs
are sought,
and especially those properties that are common across
classes of GF materials and thus reveal the
fundamental physics of magnetic states in GF systems
and their properties observable in experiment.

\subsection*{Goal of this review}

The aim of this review is to outline a minimal set of existing facts that
apply to all GF magnets in the hope of informing future theoretical work.
Thus, we will restrict discussion to the most common experimental quantities, because such data allow for intercomparison among all GF materials. 
While GF magnets have been experimentally probed in many different ways, the most common quantities are the magnetic 
susceptibility [$\chi(T)$], specific heat [$C(T)$], and the intensity of 
neutron scattering.


Other probes include NMR and $\mu$SR. We will not consider these here because, in our view, each has a significant level of interpretive uncertainty. Each one gives good information on the internal field at the probe site, but the identity of the site is either unknown in the case of $\mu$SR~\cite{Takeya:NiGaSnmr,Uemura:SCGOmuSR,Bono:KagomeMuSR} or not well defined, as shown by the two distinct signals in $NiGa_2S_4$, which were interpreted as evidence of stacking faults~\cite{Takeya:NiGaSnmr}.

It is useful to remember what $\chi(T)$, $C(T)$, and neutron scattering reveal about the state of a magnetic material.  The susceptibility, $\chi(T)$, is a long wavelength probe that is especially useful at high temperatures for determining the spin-spin interaction strength via Curie-Weiss analysis and at low temperatures for detecting transitions, e.g., to an LRO or a spin-glass state.
Most importantly for this work, $\chi(T)$ can detect non-magnetic (singlet) excitations only weakly. The specific heat, however, is sensitive to all excitations with a spin origin, including non-magnetic excitations, but also to those not spin-derived, such as phonons and charged excitations. 
A powerful aspect of the $C(T)$ probe is its ability to count degrees of freedom, provided the non-spin-derived excitations are accurately determined. Neutrons, like $\chi(T)$, can probe magnetic 
and structural excitations and can access them
over a large range of wavelengths, from mesoscopic distances down to interatomic length scales. The complementarity of these three probes is sufficient to demonstrate 
key universal features of GF magnets.
As we argue, among those universal features is the short-range magnetic order (SRO)
that forms below a certain characteristic temperature, which is significantly exceeded by the spin-spin interaction energy.


\subsection*{Materials to be discussed}

The strength of geometrical frustration in a material is characterized
by the ratio
\begin{align}
    f=\theta_W/T_c,
    \label{fRatio}
\end{align}
known as the ``$f$-ratio'' and introduced by one of us in Ref.~\cite{Ramirez:FrustratedThermodynamic},
where $\theta_W$ is the Weiss constant, the energy scale that characterizes the strength of spin interactions in a material,
and $T_c$ is the temperature below which the magnetic fluctuations are 
frozen.

Non-GF 3D materials undergo magnetic-ordering phase transitions
at temperatures $T_c\sim\theta_W$, and thus have $f$-ratios of order
unity. By contrast, strongly GF materials avoid ordering down to very
low temperatures, resulting in $T_c\ll \theta_W$ and large 
$f$-ratios $f\gg 1$. 

In systems that undergo spin-glass and magnetic ordering transitions,
the temperature $T_c$ corresponds to the critical temperature of 
the respective transition. The definition, however, can be extended
to the case of the freezing of the magnetic degrees of freedom that is
not accompanied by any phase transition, as discussed in
Ref.~\cite{PoppRamirezSyzranov}.

In this review, we will focus on materials with strong
geometrical frustration, corresponding to $f\gtrsim 10$, that have
also been subject to magnetic vacancy studies and a full complement of the simple experimental probes mentioned above. 
We show that a select group of such materials
develops SRO at intermediate ($T\sim |\theta_W|/10$) temperatures. We then argue that this SRO is a universal feature of all GF systems and is key to understanding ground states and their elementary excitations. 

Due to the above restrictions, many GF systems will not be extensively analyzed. Most evident among the omissions is the family of paratacamite compounds related to well-known Herbersmithite [$\gamma-ZnCu_3(OH)_6Cl_2$]~\cite{Helton:frustrationZnCuOHCl}, which consists of a kagome lattice of $Cu^{2+}$ spins. Herbertsmithite, and its relatives volborthite [$Cu_3V_2O_7(OH)_2\cdot 2H_2O$]~\cite{LaFontaine:DFTCu3V2O7(OH)2,HiroiTakigawa:Cu3V2O7(OH)2H2O,Hiroyuki:magnetizationStepsVolborthite}, vesignieite [$BaCu_3V_2O_8(OH)_2$]~\cite{MakotoTakigawa:susceptibilityVesigniete}, 
kapellasite [$Cu_3Zn(OH)_6Cl_2$]~\cite{MakotoTakigawa:susceptibilityVesigniete,Janson:kapellasiteSimulation}, haydeeite [$\alpha-Cu_3Mg(OH)_6Cl_2$]~\cite{Colman:haydeeiteSynthesis}, and barlowite [$Cu_4(OH)_6FBr$]~\cite{Han:Barlowite,Yue:Barlowite} all have $Cu^{2+}$ on a kagome lattice and exhibit a variety of ground states including FM and AF order. They all possess hydrogen, which is deleterious for neutron scattering, though some have been studied in their deuterated form, most importantly Herbertsmithite~\cite{Helton:frustrationZnCuOHCl}, and we will show these data below. Also, for these materials, we are unaware of systematic studies of spin vacancies, a technique that has yielded great insight into the ground state.

A second, large group of materials that eludes the present review is 4f electron systems, the frustration in which has been intensively studied especially in recent years~\cite{Chamorro:ChemReview}. Since 4f atoms in an ionically bonded solid have exchange coupling strengths at least an order of magnitude less than 3d systems, the $\theta_W$ values are usually in the range of a few Kelvins or less and, thus, the interaction can be strongly influenced by dipole-dipole (long range) forces which are rarely considered for d-row transition ions. Most importantly, however, 
geometrical frustration addresses the physics at energy scales two or more orders of magnitude less than the mean field temperature, $\theta_W$.  Thus, for $\theta_W$ values less than 1 K, measurements of $C(T)$ and $\chi(T)$ are needed down to $10 mK$, which presents a barrier for characterization. Firstly, obtaining $C(T)$ even down to $50 mK$ often requires great care to maximize thermal coupling to the sample. Secondly, commercial magnetometers only reach $300 mK$. Thus, obtaining susceptibility data down to the required temperatures is considerably more difficult than for systems with larger $\theta_W$ values. Finally, the splitting of the total angular momentum multiplet often leads to high-$m_J$ doublets and, while this poses no fundamental barrier to collective behavior, higher-lying multiplets often obscure the determination of effective moment from data taken above $2K$. 
Thus, in our view, the existing body of experimental work on 4f materials precludes the systematic intercomparison of both high and low energy scales as can be done for 3d systems.


\section{Characterization of GF magnetic materials}

\subsection{Types of quenched disorder}

In order to describe GF materials in the pure limit, one should understand their response to quenched disorder. One may distinguish here between two generic types of disorder: intrinsic and extrinsic.

Intrinsic disorder occurs without introducing quenched randomness in 
the synthesis process. For example, in $Y_2Mo_2O_7$, Jahn-Teller distortions of the $Mo$ octahedra result in randomness in the spin-spin interaction and a spin-glass (SG) transition at $T_g = 22K$~\cite{Greedan:YMoOglass,Thygesen:YMOdisplacements}.
In some materials, different ions have to randomly occupy the same crystallographic site
to ensure charge balance between cations and anions. For example, 
in $Fe_2TiO_5$, the $Fe^{3+}$ and $Ti^{4+}$ ions are randomly located on
both $A$ and $B$ sites of the pseudo-Brookite structure~\cite{LaBarre:surfboards}. In 
$NaCdCo_2F_7$, the $A$ site can be occupied by both $Na^{1+}$ and $Cd^{2+}$~\cite{KanckoColman:mixingNaCdCoF}.
Another example of such disorder comes from the inverse spinels, such as $B(AB)O_4$, 
in which site mixing occurs between the $A^{2+}$ and $B^{3+}$ ions~\cite{Goodenough:book}.

Extrinsic disorder refers to foreign atoms introduced
in the clean material.
Often, such disorder is represented by vacancies,
i.e. 
foreign non-magnetic atoms substituting for the 
magnetic atoms of the clean material.
Of course, non-magnetic vacancies will be at least subtly different
in atomic size and, therefore, will not bond with their neighboring cations in a manner identical to the magnetic ion, hence one expects
small steric distortion around the vacancies.
Extrinsic disorder may also be introduced by substitution on either a non-magnetic cation site or the anion site, which can lead to charge disorder or sterically induced randomness in the exchange interactions between magnetic ions~\cite{Martinho:ZnCrOsubstitution}.


Techniques for measuring disorder at the required precision are few. In most cases, purity control consists of identifying the presence or absence of unexpected diffraction peaks in X-ray diffraction (XRD). Such a technique, however, cannot detect impurities or disorder
other than by the observations of foreign crystallographic phases or broadening of the diffraction peaks.
In the case of $Y_2Mo_2O_7$, pair distribution function analysis of neutron and X-ray scattering was needed to determine the distribution of bond lengths responsible for interaction disorder and the observed spin glass behavior~\cite{Thygesen:YMOdisplacements,BoothSubramanian:YMOdisorder}. For deviations from the stoichiometry  larger than a few percent, energy-dispersive X-ray absorption can reveal impurity atoms, but its precision is limited.

\subsection{Importance of spatial dimensionality}

The role of spatial dimensionality is well-understood for non-frustrated systems exhibiting LRO and is manifested both in the critical behavior as well as the ground state and its elementary excitations~\cite{DeJongh:dimensionality}. Effective dimensionality less than three can arise either from a crystal structure that is lamellar (2D) or fibrous (1D), or may result from a collective structural distortion such as a cooperative Jahn-Teller effect that modulates the spin-spin interaction in a uniform manner~\cite{BlancRamirez:SaltEffectiveD}.

The role of spatial dimensionality is less well-understood for SGs.
Great effort has been devoted to the question of whether or not a 2D system can transition to an SG state, and the present consensus, based on computational work, is that spin glass freezing anomalies are only observable in 3D systems~\cite{FernandezParisi:Ising2Dtransition}. 
While certain layered materials such as $Rb_2(Cu,Co)F_4$~\cite{DekkerMydosh:SGrelaxation},
have been shown to exhibit SG behavior, such materials can be classified only as ``quasi-2D'', i.e.
while the intra-plane interactions are small compared to the inter-plane ones, the
3D nature of the intra-plane correlations is essential for the emergence of the SG phase.

While the consensus view is that purely 2D systems do not evince an SG transition, it is possible that the SG-transition temperature is actually strongly suppressed in quasi-2D
materials, owing to their effective 2D nature at short scales. 
Numerous authors seeking to establish the existence of 2D QSLs in layered materials
argued that the absence of an SG cusp in $\chi(T)$ for experimentally accessed temperatures 
rules out an SG state, which is seen as inimical to a QSL
(see, for example, Refs.~\cite{Shimizu:QSLclaim,Helton:frustrationZnCuOHCl,Ma:QSLclaim,Kitagawa:QSLclaim}).
The SG transition in such layered systems, however, may take place at temperatures below 
the observed range.


\subsection{Heisenberg vs. Ising spins}

This review focuses on strongly GF systems with Heisenberg spins. Ising spins on frustrating lattices present a different and informative complement to Heisenberg spins. In early theoretical work, Wannier considered 2D Ising spins on a triangular lattice and showed that, whereas ferromagnetic spin-spin interactions lead to LRO, antiferromagnetic interactions cause a total suppression of the transition and lead to a nonzero ground-state entropy (per spin)~\cite{Wannier:Ising}. The existence of such non-vanishing ground-state entropy for theoretical Ising models makes them "infinitely frustrated", a situation reflected experimentally by the lack of Ising-based magnets for which the \textit{f}-ratio can be defined by a measurable low-energy scale. Among the materials that have Ising spins but do not exhibit an ice-like state are $Co_3Mg(OH)_6Cl_2$~\cite{Fujihala:SpinIce} and $A_2Co(SeO_3)_2$, where $A = K$ or $Rb$~\cite{Zhong:LayeredTriangular}.

A possible reason behind the lack of experimental Ising GF magnets is related to the stringent conditions required to create local degeneracy. Ising spins occur due to, e.g., a distortion in the crystal field environment or due to spin-orbit coupling, i.e. a local structural anisotropy. Such anisotropy can lead to anisotropy in the spin-spin interaction and the subsequent selection of certain favored LRO states, thus destroying frustration. While the paucity of GF Ising systems is unfortunate from the perspective of comparing theory and experiment, their role in the development of quantum materials would be limited anyway due to the Ising single-ion potential well constraint at very low temperatures. (Transverse-field Ising materials are possible but present the additional requirement of a unique transverse field direction, which is more difficult to realize with a triangular motif.)

With the above caveat in hand, a special class of 3D Ising systems was pioneered by Harris, Bramwell et al., who described a formal correspondence between water ice and a class of Ising magnets which they called ``spin ice''~\cite{Harris:HoTiOice}.
The correspondence is, specifically, between two possible hydrogen atom positions around each oxygen in water ice and the two states of an Ising spin. The spin-ice materials class consists of Ising ions, such as $Ho^{3+}$ or $Dy^{3+}$,
occupying the A-site of the pyrochlore lattice, a sublattice consisting of corner-shared tetrahedra. Ramirez et al. 
measured~\cite{Ramirez:iceEntropy} the specific heat of the spin ice candidate,  $Dy_2Ti_2O_7$, and showed that
the ground-state entropy of the $Dy^{3+}$ ions 
is close to Pauling's estimate $\left(R/2\right)\ln(3/2)\approx 0.29 R\ln(2)$.

While it may seem as though the spin-ice state results purely from the frustration of interactions, several factors set spin ice apart from the Heisenberg GF systems, which are the focus of this review. First, $Dy_2Ti_2O_7$ is not strongly GF because $\theta_W=0.5K$, which is the same temperature region where the spin ice SRO sets in, thus $f\approx 1$. Second, when disordered, $Dy_2Ti_2O_7$ fails to exhibit an SG state~\cite{SnyderSchiffer:ices}. 
Third, because of the small spin-spin exchange interactions, long-range dipole-dipole interactions play an important role in determining the ground state energy~\cite{DenHertogGingras:spinIce}.
It is interesting to consider why $Dy_2Ti_2O_7$ doesn’t undergo LRO similar to other Ising systems on triangle-based lattices. This is most likely due to the non-collinear nature of the crystal field distortion of the rare-earth environment. The Ising axis for each ligand environment surrounding the $Dy^{3+}$ ions connects the vertex of the tetrahedron to its center. Thus, with four distinct crystalline directions, the distortion itself is sterically frustrated, which may help to explain the relatively small number of spin-ice materials.

Unlike Ising systems, no evidence has been found for missing entropy, i.e. extensive entropy of the ground states,
in the Heisenberg spin systems.  
While missing entropy, suggestive of spin-liquid behavior,
has been reported for $CuAl_2O_4$~\cite{Nirmala:SGinCuAlO} and $CuGa_2O_4$~\cite{Fenner:EntropyCuGlasses}, where $Cu^{2+}$ occupies the A-site of the spinel lattice, the conclusion about the missing entropy was reached based on the specific-heat 
extending only up to $20 K$ or 
$30 K$, respectively. At the same time, 
the inverse susceptibility $\chi^{-1}(T)$ 
in these materials deviates from 
linearity at a much higher temperature, implying that the remaining entropy will be found at higher temperatures.


\section{Magnetic susceptibility}

\subsection{The Curie-Weiss law}

\begin{figure}[t!]
    \centering
    \includegraphics[width=\linewidth]{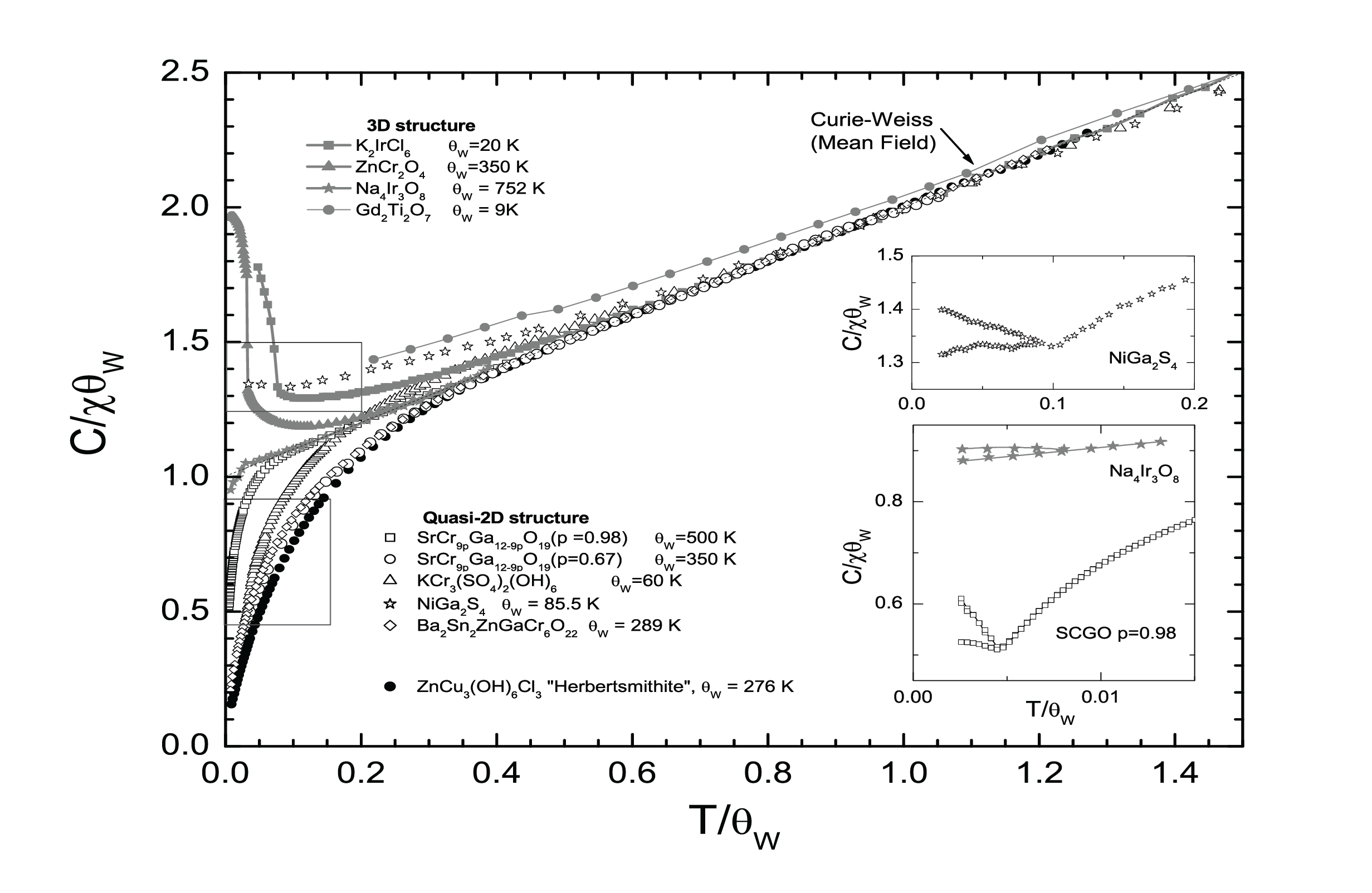}
    \caption{The dependence of the inverse susceptibility $\chi^{-1}(T)$
    on the temperature $T$ for several GF materials~\cite{Cooke:KIrCl,LaForge:quasispin,Okamoto:NaIrOQSL,Ramirez:GdTiO,SCGO:unpublished,Ramirez:KCrOHSO,NakatshujiMaeno:NaGaSneutron,HagemannCava:BaSnGaZnCrO,Helton:frustrationZnCuOHCl}. The susceptibility and the temperature
    are shown, respectively, 
    in units of $C/\theta_W$ and $\theta_W$, where $C$ and $\theta_W$
    are the Curie and Weiss constants. The upper and lower insets, represented by the boxes in the main frame, show, respectively,
    the spin-glass transitions in $NiGa_2S_4$ and $Na_4Ir_3O_8$ and $SrCr_{8.8}Ga_{3.2}O_{19}$.
    The dashed line corresponds to the Curie-Weiss law.
    }
    \label{fig:ChiVsT}
\end{figure}

The dependence of the inverse magnetic susceptibility $\chi(T)$ on temperature $T$ for several GF materials~\cite{Cooke:KIrCl,LaForge:quasispin,Okamoto:NaIrOQSL,Ramirez:GdTiO,SCGO:unpublished,Ramirez:KCrOHSO,NakatshujiMaeno:NaGaSneutron,HagemannCava:BaSnGaZnCrO,Helton:frustrationZnCuOHCl} 
with large $f$-ratios $f>10$ is shown in Fig.~\ref{fig:ChiVsT}.
At sufficiently high temperatures, GF materials exhibit the Curie-Weiss behaviour of the susceptibility
\begin{align}
    1/\chi(T) \propto T+\theta_W,
    \label{CWlaw}
\end{align}
where $\theta_W>0$ is the Weiss constant. This behavior is identical to the high-temperature
behavior of susceptibility in non-GF antiferromagnetic materials.
There is, however, a key difference between the Curie-Weiss laws in non-GF and GF systems.
In a non-GF system, it persists only down to the critical temperature $T_c\sim \theta_W$, at which 
the spins order in a regular pattern, and the susceptibility exhibits a singularity. By constrast, in GF systems,
the dependence~\ref{CWlaw} of the magnetic susceptibility on temperature
extends down to significantly lower temperatures $T\ll \theta_W$.

The value of $\theta_W$ obtained from the temperature dependence~\eqref{CWlaw} can be used to obtain the $f$-ratio [cf. Eq.~\ref{fRatio}], which
characterizes the degree of frustration.
Because the dependence~\eqref{CWlaw} implies 
that a material behaves effectively as free spins in the limit 
of high temperatures $T\gg \theta_W$,
the respective value of $\theta_W$ is faithful
only if the measured slope of $1/\chi(T)$ matches
the expected Hund's rule moment.


\subsection{Susceptibility and glass transitions}

While the linear, high-temperature behavior of the inverse susceptibility is a universal
feature of GF systems, there is a dichotomy in the low-temperature behavior between 
GF systems that eventually undergo LRO to an AF state and those that resist ordering.

At temperatures above magnetic and glass transitions, this dichotomy usually manifests
in the sign of the deviation of the magnetic susceptibility from the Curie-Weiss law.
When the system undergoes an AF transition, the high-temperature
$\chi(T)^{-1}$ typically bends upward on cooling (see Fig.~\ref{fig:ChiVsT}),
reflecting the appearance of AF correlations. 
However, if the system undergoes an SG glass transition instead of developing magnetic order,
$\chi(T)^{-1}$ typically bends downwards on cooling above the transition, a behavior that in conventional magnetic systems is referred to as a \textit{Curie tail} and similar to that coming from free spins. 

Such Curie tails can be explained by the presence of dilute vacancy defects.
As discussed in detail in Sec.~\ref{Sec:quasispins} below, a vacancy defect 
in a GF material is equivalent to an 
effective spin, a ``quasispin'', in terms of its magnetic susceptibility.
At sufficiently high temperatures, magnetic correlations in the system are short-ranged,
and the quasispins of vacancy defects can be considered non-interacting and independent of each other. Such quasispins act as effective free spins
and give rise to the Curie tail. At lower temperatures,
the quasispin-quasispin interactions become essential (see Sec.~\ref{Sec:quasispins} for a detailed discussion).

At the lowest temperatures, SG transitions are observed (see the insets in Fig.~\ref{fig:ChiVsT})
even in rather pure materials, such as the triangular lattice compound $NiGa_2S_4$~\cite{Nambu:NiGaS} and the spinel $ZnCr_2O_4$~\cite{LaForge:quasispin}. 
In $ZnCr_2O_4$, SG freezing is even observed deep in the AF temperature regime, demonstrating that an SG state can coexist with LRO. 

The novelty of SGs occurring in nearly pure GF materials was realized early in the development of GF magnetism~\cite{Ramirez:FrustratedThermodynamic,Greedan:YMoOglass}
and suggests an analogy with conventional SGs in metallic materials in which spins have densities less than 1\%, far below the nearest-neighbor percolation threshold. Such 
materials undergo SG freezing due to the long-range nature of the RKKY interaction~\cite{BinderYoung:review}. The analogy falters, however, in the following manner. While the RKKY interaction in conventional SGs exists above the glass temperature, $T_g$, there is no evidence for such an interaction for GF SGs above $T_g$. In addition, the medium in which the GF SG state forms is comprised of a dense population of spins, whereas conventional metallic SGs have no magnetic sources in the spatial region between the spins. Thus, the physical difference between the two classes of SGs is profound~\cite{Ramirez:FrustratedThermodynamic}, despite similarities in their collective, bulk behavior~\cite{Ramirez:T2,Gingras:YMoOsusceptiblityGlass}.     


\subsection{Quasispins}

\label{Sec:quasispins}

Vacancy defects, the most common type of quenched disorder in GF materials, have a fundamental effect on magnetic susceptibility. It has been found that, in terms of response to a magnetic field, vacancies act as effective free spins,
``quasispins'', with the Curie-like susceptibility [in units of $(g\mu_B)^2/k_B$]
\begin{align}
    \chi_\text{vac}=\frac{N_\text{vac}}{T}\langle\hS_z^2\rangle,
    \label{QuasispinDefinition}
\end{align}
where $N_\text{vac}$ is the density of the vacancy defects
and $\hS_z$ is the projection of the quasispin to the direction along which the susceptibility is measured.

The quasispin behavior~\eqref{QuasispinDefinition} was first identified~\cite{Schiffer:TwoPopulationModel} in $SrCr_{9p}Ga_{12-9p}O_{19}$ (where $N_\text{vac}\propto 1-p$) and attributed to ``orphan spins'', i.e. the spins of rare $Cr^{3+}$ ions that are disconnected from the bulk of the spins by the vacancies. The density of such ``orphan spins'' is rather low and insufficient to account for the value of the Curie constant in the contribution~\eqref{QuasispinDefinition}
to the susceptibility~\cite{LaForge:quasispin}, yet the ``orphan spin'' contribution would scale non-linearly with the defect density $N_\text{vac}$, in contradiction with the observed linear dependence~\cite{Schiffer:TwoPopulationModel}.

A more realistic picture of the vacancy susceptibility is the quasispin picture, in which each isolated vacancy 
is screened by the surrounding magnetic medium and 
behaves as a free spin.
Such behavior
should be expected in a generic 
material based on the fluctuation-dissipation relation
$\chi_{zz}={\langle\hat{M}_z^2\rangle}/{T}$ if the magnetization $\hat M_z$ of a correlated 
region around a vacancy commutes with the system's Hamiltonian.

The behavior and values of quasispins of vacancy defects have been a subject of intense research
in a variety of systems, such as 
2D Heisenberg antiferromagnets~\cite{SandvikDagottoScalapino,SachdevBuragonhainVojta,HoglundSandvik03,MaryasinZhitomirsky:VacancyPhaseDiagram,Maryasin_2015}, one-dimensional spin chains and ladders~\cite{SandvikDagottoScalapino,SunRamirezSyzranov:1Dquasispin,KatsuraTsujiyama,Wortis1974,Bogani,GoupalovMattis,ValkovShustin,Sedik:QuasispinInteraction}
and arrays of classical vector spins~\cite{WollnyVojta:vacancies,WollnyFritzVojta}. 
It has been argued in Ref.~\cite{Sedik:QuasispinInteraction} that
the variance $\langle \hS_z^2 \rangle$
of the quasispin of a vacancy defect 
and quasispin-quasispin interactions match the 
variance 
of the spins of the magnetic atom and spin-spin interactions in
the vacancy-free material, as long as the presence of the vacancy
does not change the short-range magnetic order around it and
the magnetization along the $z$ axis is a good quantum number.
This has been confirmed~\cite{SunRamirezSyzranov:1Dquasispin} by exact calculations for a 1D chain with nearest-neighbor and next-to-nearest-neighbor interactions, the minimal model in which vacancies exhibit quasispin behavior.


Quasispins exhibit further intriguing phenomena close to quantum phase transitions.
It has been demonstrated in Ref.~\cite{SachdevBuragonhainVojta} that in a material close to such a transition, the soft modes of the fluctuating order parameter renormalize the value of 
a vacancy quasispin and lead to an effective fractional quasispin. 
A fractional quasispin has been reported in numerical simulations
in Ref.~\cite{HoglundSandvik:FractionalSpin} for a bilayer antiferromagnetic system
in which the quantum phase transition is driven by the value of the interlayer
coupling.

The generic behavior of the magnetic susceptibility
in a material with dilute vacancies can be summarized by the equation
\begin{align}
    \chi_{zz}(T)=\frac{\langle\hS_z^2\rangle N_\text{vac}}{T}
    +\frac{N-b(T)N_\text{vac}}{N}\chi_\text{bulk}(T),
    \label{VacancySusceptibilityGeneric}
\end{align}
where $N$ is the density of magnetic atoms in the vacancy-free material;
$\chi_\text{bulk}(T)$ is the susceptibility of the vacancy-free material; $N_\text{vac}$ is the vacancy density,
and $b(T)$ is the effective ``vacancy size''~\cite{SunRamirezSyzranov:1Dquasispin}.
The first term in Eq.~\eqref{VacancySusceptibilityGeneric}
describes the contribution of the quasispins of vacancy defects, while the second term comes from the bulk spins diluted by the vacancies; each vacancy 
reduces the effective number of bulk spins by the ``vacancy size'' $b(T)$.

Equation~\eqref{VacancySusceptibilityGeneric} describes the susceptibility
of a GF magnet in the limit of dilute vacancies, in which the quasispins of 
isolated vacancies can be considered independently. With increasing the vacancy density $N_\text{vac}/N$, the 
susceptibility~\eqref{VacancySusceptibilityGeneric}
receives non-linear corrections in $N_\text{vac}$~\cite{Sedik:QuasispinInteraction}.


\section{Hidden energy scale and short-range order}

In this section, we review available data for the properties of the glass transitions, neutron-scattering results, and thermodynamic properties of GF magnetic materials.
All such data demonstrate the existence of the ``hidden energy scale'' in such materials, significantly exceeded by other microscopic energies, e.g. $\theta_W$, in them. When a material is cooled down below such a scale, it develops significant short-range magnetic order. This scale, albeit a property of a disorder-free material, also determines the temperature of the glass transition for realistic impurity concentrations.


\subsection{Glass transition temperature vs. amount of disorder}

\label{Sec:GlassTransitionHiddenScale}

\begin{figure}[ht!]
    \centering
    \includegraphics[width=0.65\linewidth]{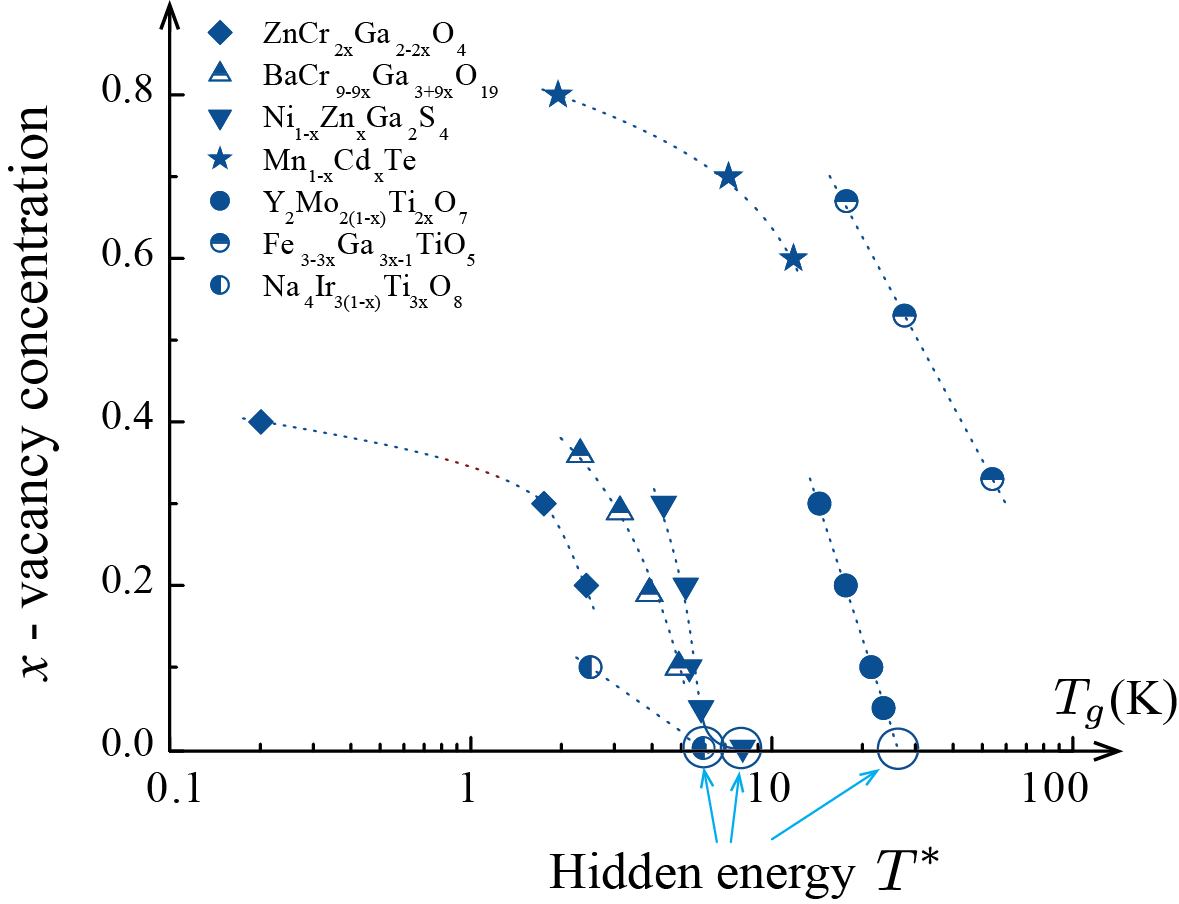}
    \caption{Adapted with modification from Ref.~\cite{Syzranov:HiddenEnergy}.
    The glass-transition temperature as a function of vacancy density 
    for GF magnetic materials. Extrapolation to the limit of vanishing vacancy densities reveals the hidden energy scale $T^*$.
    \label{fig:HiddenEnergy}
    }
\end{figure}

In Ref.~\cite{Syzranov:HiddenEnergy}, the spin-glass transition temperature $T_g$ has been analyzed as a function of the density of vacancy defects in all frustrated materials for which the data on such transitions are available. Figure~\ref{fig:HiddenEnergy} summarizes the discovered trends.

Although the glass transition is disorder-driven, increasing the density of vacancies, which contribute to the total disorder strength, lowers the glass transition temperature. In other words, contrary to common intuition and existing models (see, e.g. Refs.~\cite{BinderYoung:review,Dotsenko:GlassReview,Mezard:GlassReviewBook}),
purifying the material by reducing the number
of vacancies makes random spin freezing more favorable.

In the 
limit of vanishing density concentration, the glass transition temperature extrapolates to a scale of order $T^*\sim 10K$, which we dubbed the ``hidden energy scale'', in all analyzed frustrated compounds~\footnote{with the exception of the compounds $Fe_{3-3x}Ga_{3x-1}Ti_5$, in which a significantly higher glass-transition temperature may be attributable to the formation of antiferromagnetic clusters~\cite{LaBarre:surfboards,Phelan:surfboards}}. This scale is significantly exceeded by the 
Weiss constant~$\theta_W$, which characterizes the strength of spin-spin interactions and is typically a few hundred
Kelvin~\cite{Syzranov:HiddenEnergy}.

\begin{figure}[ht!]
    \centering
    \includegraphics[width=0.75\linewidth]{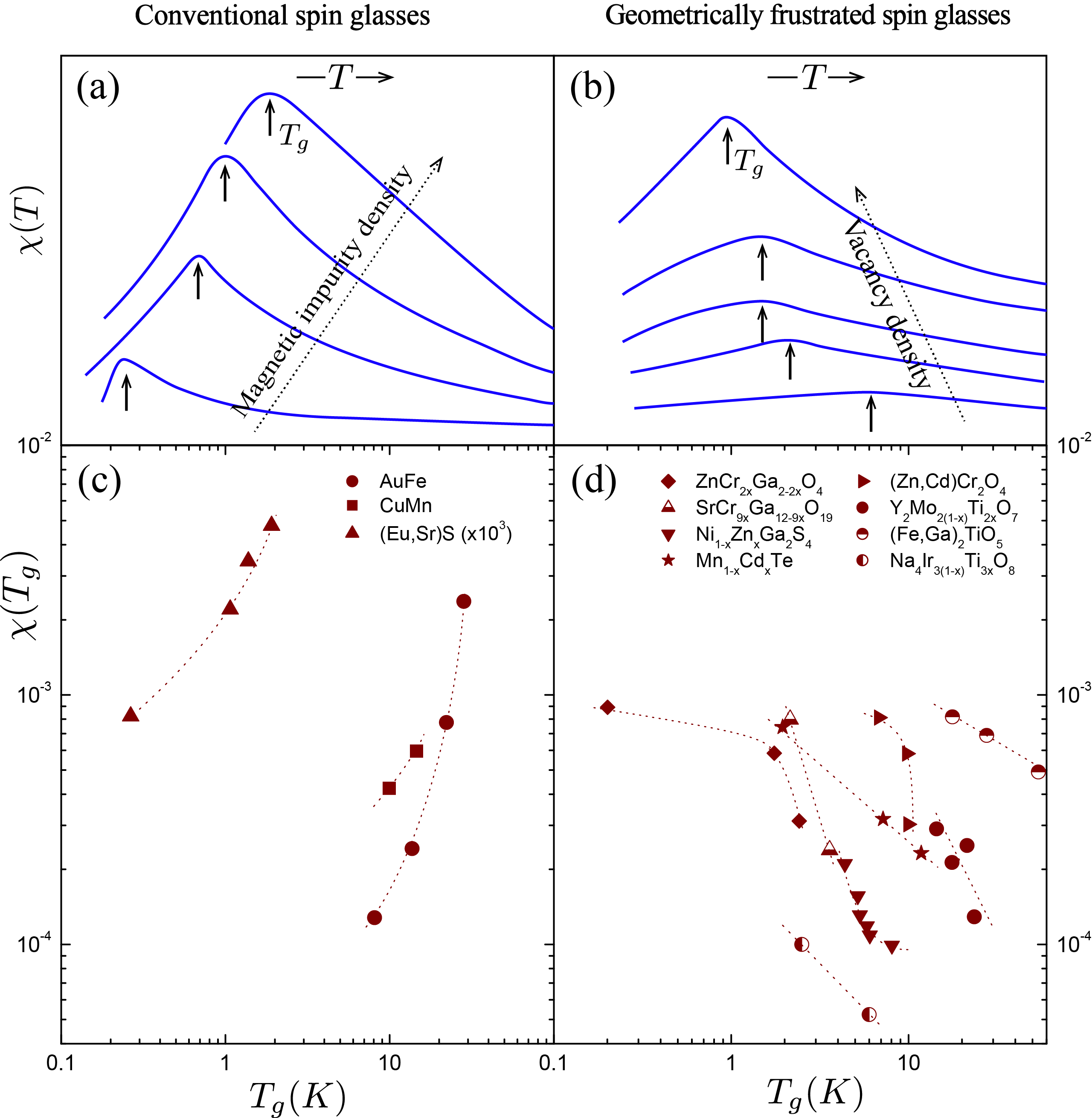}
    \caption{Adapted with modification from Ref.~\cite{Syzranov:HiddenEnergy}.
    The behavior of magnetic susceptibility $\chi(T)$ and
    glass transition temperatures as a function of impurity density
    in conventional glasses [panels (a) and (c)] and in geometrically frustrated materials [panels (b) and (d)].
    }
    \label{fig:ChiTplots}
\end{figure}

Another universal trend displayed by all GF magnets is the increase of magnetic susceptibility 
$\chi(T)$ with increasing the vacancy concentration, as shown in Figs.~\ref{fig:ChiTplots}(b) and (d). 
This trend, as well as the decrease of the SG transition temperature with adding impurities to a material, is opposite to the trends observed in conventional SGs (i.e. SGs in non-GF materials)\cite{BinderYoung:review}.

The dichotomy between between conventional SGs and those 
in GF materials comes from the different nature of quenched randomness, degrees of freedom that undergo SG freezing, and the clean medium.
Indeed, in conventional spin glasses, magnetic impurities are randomly located in a non-magnetic 
medium, and the SG transition consists in the freezing of the magnetic moments 
of those impurities.
With increasing the density of magnetic impurities, the strength of interactions
between them grows, and both the SG transition temperature and the susceptibility grow,
as shown in Fig.~\ref{fig:ChiTplots}a for the compound
$Eu_{x}Sr_{1-x}S$~\cite{Maletta:EuSrS}.

By contrast, in GF materials, the medium consists of magnetic atoms, while quenched disorder is 
dominated by randomly located vacancies,
i.e. non-magnetic atoms. As discussed in Sec.~\ref{Sec:quasispins},
each vacancy can be associated with a quasispin. 
Because the quasispin contribution to the susceptibility $\chi(T)$
is more singular at low temperatures
than the bulk susceptibility
[cf. Eq.~\eqref{VacancySusceptibilityGeneric}], it dominates the effect of changing the impurity concentration, resulting in the increase of the susceptibility with vacancy density~\cite{Syzranov:VacancyPhenomenological}, $d\chi(T)/dN_\text{vac}>0$.

While quenched disorder is necessary for SG transitions, we argue~\cite{Syzranov:HiddenEnergy,PoppRamirezSyzranov}
that the temperature of SG transitions in the analyzed GF compounds 
is determined by the properties of the clean GF medium rather than the 
amount of quenched disorder for experimentally achievable amounts.
As we clarify in what follows, the lowest-energy magnetic degrees of freedom
have a characteristic energy of the order of the hidden energy scale $T^*$
described above and are suppressed at lower temperatures.
In the presence of quenched disorder, this usually leads to spin-glass freezing 
at a temperature of order $T^*$~\cite{Syzranov:HiddenEnergy}.

The hidden energy $T^*$ being the property of the clean GF medium is consistent
with its value being of the same order of magnitude ($10K$) in a
broad class of materials despite different microscopic details and types of disorder.
Below,
we present further evidence for a crossover in the properties
of the clean GF medium at temperatures $T\sim T^*$ in the neutron scattering intensity
and discuss a microscopic scenario for the hidden energy scale.


\subsection{Neutron data}

\label{Sec:NeutronHiddenScale}

As we describe in this subsection,
the intensity of neutron scattering in GF materials confirms the 
existence of the hidden energy scale and can provide further insights 
into its nature.

Neutrons are an essential probe for understanding conventional antiferromagnets since they scatter elastically from atomic spins and can detect staggered order, even though this order produces no net magnetic moment. In the ordered state of a magnet possessing LRO, inelastic neutron scattering can also measure the energy-wavevector relationship of spin waves and, by appropriate modeling, extract the spin-spin interaction strengths. 

Because LRO is usually not seen in GF systems, information gleaned from elastic scattering is confined to broad intensity versus wavevector peaks. Such a diffuse scattering signal lies underneath sharp Bragg peaks caused by scattering of neutrons from the ionic nuclei. Most generally, magnetic scattering can be related to the generalized susceptibility, as discussed by Bramwell in Ref.~\cite{Lacroix:book}. In particular, the anisotropy in this susceptibility can reveal the nature of SRO, the prime example being the so-called ``pinch points'' in the diffuse scattering in spin ice~\cite{Henley:CoulombReview,Tabata:KagomeIce,Fennell:CoulombHoTiO}.  Pinch-point patterns of scattering result from the spin ice-rule expressing the lowest energy state of the four spin directions on a tetrahedron, namely two-in-two-out. We believe that such an analysis of scattering intensities will be critically important for understanding the ground state of GF systems, but the present discussion will be concerned simply with the presence or absence of the SRO revealed by such data.

\begin{figure}[H]
    \centering
    \includegraphics[width=0.8\linewidth]{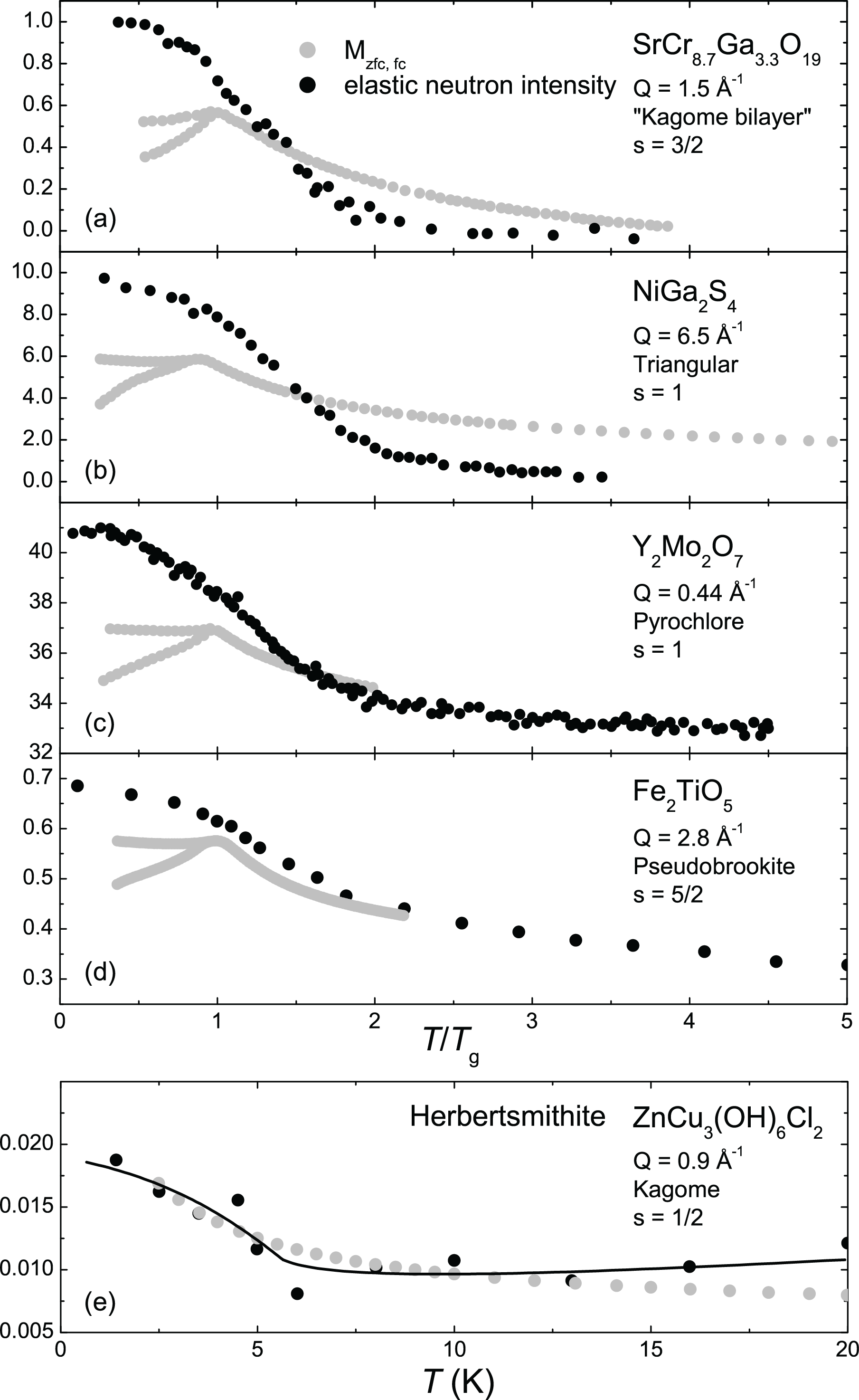}
    \caption{Neutron scattering intensity (black circles)~\cite{YangLee:SCGOjam,Stock:neutron,Gardner:YMoOneutron,LaBarre:surfboards,Helton:frustrationZnCuOHCl} and 
    magnetic susceptibility (grey circles)~\cite{YangLee:SCGOjam,Nambu:NiGaS,Greedan:YMoOglass,LaBarre:surfboards,Helton:frustrationZnCuOHCl} for various GF compounds as a function of temperature.
    }
    \label{fig:NeutronData}
\end{figure}

Broholm et al. were first to show that a GF system, 
$SrCr_{9p}Ga_{12-9p}O_{19}$ $(p = 0.79)$, exhibits a broad SRO scattering signal. They extracted a maximum spin-spin correlation length of $\xi=7 \pm 2\AA$ from the width of the broad peak, and showed that the scattering intensity at the corresponding wavelength increases rapidly on cooling through the hidden energy scale~\cite{Broholm:SCGOneutronScattering}. Similar data were obtained for a sample of $SrCr_{8.9}Ga_{3.1}O_{19}$ ($p = 0.97$, $T_g = 3.68 K$) and are shown in Figure~\ref{fig:NeutronData}a, along with the susceptibility on the same compound~\cite{YangLee:SCGOjam}.
Thus, we see that the atomic spins undergo short-range order at the hidden energy scale. In addition, it is reasonable to surmise that this is a bulk effect, i.e. the fraction of such spins is a non-negligible fraction of the total atomic spin population and, as discussed above, exists independent of the SG state.

How common is such a neutron signature of SRO in GF systems? We argued previously that the hidden energy scale was a generic feature of all GF magnets~\cite{Syzranov:HiddenEnergy} and, fortunately, neutron scattering has now been performed on a good handful of such $f>10$ materials. In Figure~\ref{fig:NeutronData}b-e, we show data 
for $NiGa_2S_4$~\cite{Stock:neutron,Nambu:NiGaS}, $Y_2Mo_2O_7$~\cite{Gardner:YMoOneutron,Greedan:YMoOglass}, $Fe_2TiO_5$~\cite{LaBarre:surfboards}
and Herbertsmithite~\cite{Helton:frustrationZnCuOHCl}
similar to that for $SrCr_{9p}Ga_{12-9p}O_{19}$~\cite{YangLee:SCGOjam}.
These data encompass different spin values ($S = 1/2, 1, 3/2, 5/2$), different lattice types (triangular, kagome, pyrochlore, hyperkagome), as well as effective spatial dimensionalities 2 and 3. We see that the growth of intensity at short wavelength is a generic feature of these systems, and always occurs in the vicinity of the SG transition, with the exception of Herbertsmithite, which has been argued to be 2D to a very good approximation (in contrast to the other lamellar materials, such as $SrCr_{9p}Ga_{12-9p}O_{19}$ and $NiGa_2S_4$, in which the interlayer coupling are more essential) and therefore will not exhibit SG freezing. Similarly to the other materials, the characteristic temperature $T^*\sim 5K$ at which Herbertsmithite  exhibits SRO is significantly smaller than the Weiss constant $\theta_W\approx 300K$.

We note that
for SG transitions in conventional, non-GF materials, neutron scattering also
shows an increase in diffuse scattering on cooling towards the glass transition temperature~\cite{BinderYoung:review,Mydosh:book}. In these systems, however, this magnetic scattering is caused by the degrees of freedom that undergo SG freezing. In GF systems, on the other hand, neutrons are scattered by the background spins.
Neutron scattering cannot, of course, differentiate between the two scenarios, but in GF systems, it clearly evidences the SRO established by the bulk of the 
spins.


\subsection{A scenario for the hidden energy scale}

\label{Sec:HiddenEnergyScenario}

The data summarized in Secs.~\ref{Sec:GlassTransitionHiddenScale}
and \ref{Sec:NeutronHiddenScale} illustrate that the hidden energy scale 
is a universal property of GF magnets and is a property of the clean system.
Recently, we proposed~\cite{PoppRamirezSyzranov} a scenario of the hidden energy scale. 

\begin{figure}[ht]
    \centering
    \includegraphics[width=0.7\linewidth]{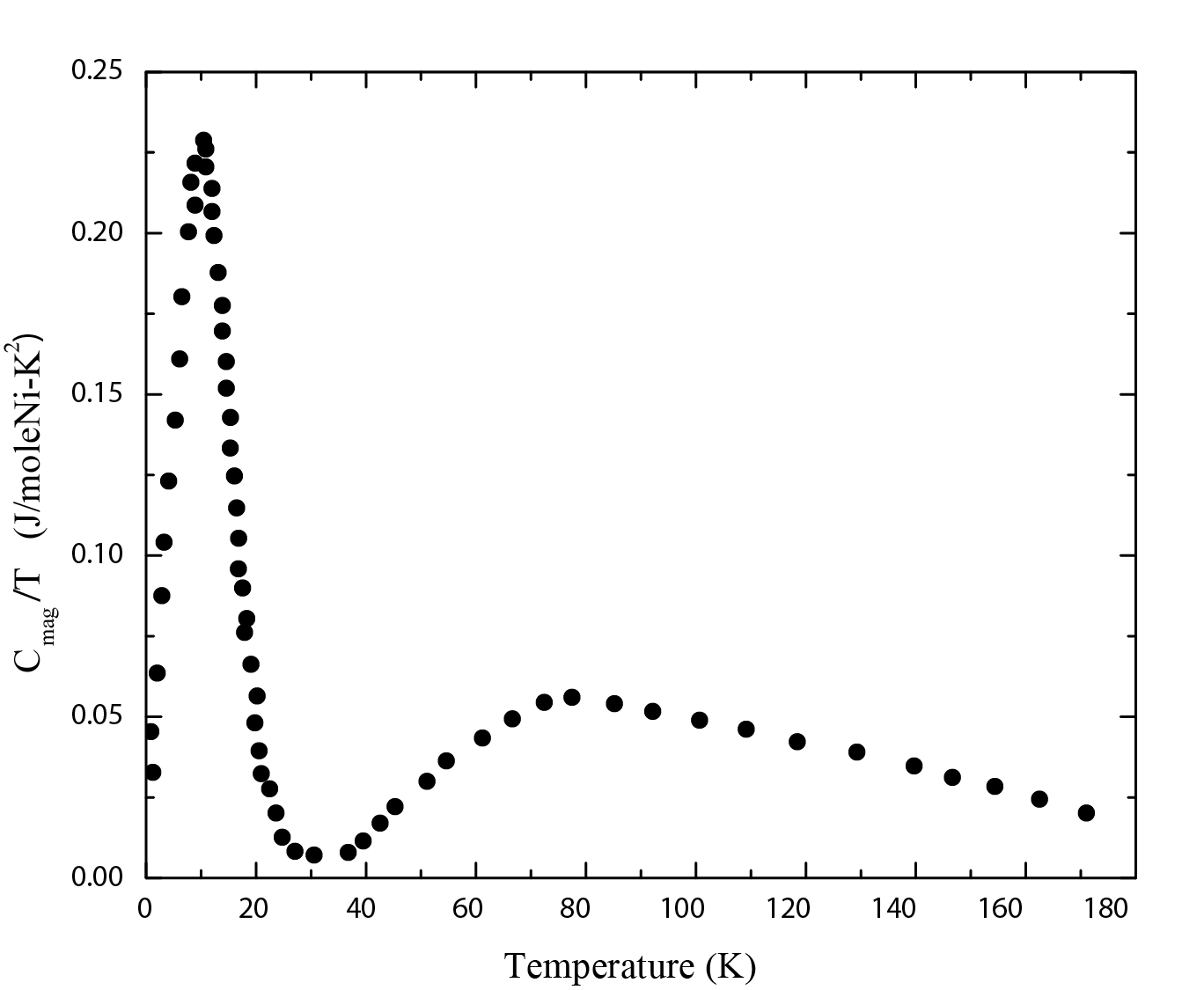}
    \caption{
    \label{fig:NiGaSTwoPeak}
    The $C(T)/T$ vs. $T$ dependence in $NiGa_2S_4$
    obtained by digitizing the data reported in Ref.~\cite{NakatshujiMaeno:NaGaSneutron}.
    }
\end{figure}

Essential to this scenario is the observation that the heat capacity $C(T)$
in GF frustrated materials, in which data in a sufficiently large temperature
range are available, display two peaks in the behavior of $C(T)$, 
as shown in Fig.~\ref{fig:NiGaSTwoPeak}.
This two-peak behavior is observed experimentally
in all materials for which the heat-capacity data are available in
a sufficiently broad temperature range, such
as $^3He$ atoms on graphite~\cite{Greywall:DoublePeakHe,Ishida:HeRingExchange},
$NiGa_2S_4$~\cite{NakatshujiMaeno:NaGaSneutron},
$NaYbO_2$~\cite{Bordelon:NaYbO:TwoPeak},
$NaYbSe_2$~\cite{RanjithBaenitz:NaYbSe:TwoPeak},
$Gd_3Ga_5O_{12}$~\cite{SchifferRamirez:GGG},
as well as in 
abundant numerical simulations~\cite{Elser:KHAF,ZengElser:KHAF,ElstnerYoung:kagome,NakamuraMiyashita:KHAF,TomczakRichter:KHAF,SindzingreMisguich:KHAF,MisguichBernu:KHAF,MisguichSindzingre:KHAF,IsodaNakano:XXZ,PrelovsekKokalj:THAF,SchnackSchulenberg:KHAF,ChenQu:THAF,Prelovsek:TriangularWilsonRatio,SekiYunoki:RingExchange,HutakKrokhmalskii:HKHAF,UlagaPrelovsek:TriangularKagomeTwoPeak,UlagaPrelovsek:latest}.

The temperature of the higher-energy peak is of the order of the 
Weiss constant $\theta_W$, while the 
low-temperature peak is located at significantly lower temperatures of the order
of the hidden energy scale $T^*$. An important insight into the 
nature of both peaks is provided by the numerical simulations~\cite{IsodaNakano:XXZ,UlagaPrelovsek:TriangularKagomeTwoPeak}\cite{PoppRamirezSyzranov} of 
a family of the 
$XXZ$ models described by the Hamiltonian
\begin{align}
    \cH=J\sum_{(ij)}\left(\hS_i^z\hS_j^z+\alpha\hS_i^x\hS_j^x+\alpha\hS_i^y\hS_j^y\right),
    \label{XXZHamiltonian}
\end{align}
with the anisotropy parameter $\alpha$ varying between $0$ (the Ising-model case) and $1$
(isotropic Heisenberg model). 
At small $\alpha\ll 1$, the 
states of the model are similar to those of the Ising model, and the
effect of the transverse couplings 
[$\propto \left(\hS_i^x\hS_j^x+\hS_i^y\hS_j^y\right)$] can be understood perturbatively in $\alpha$.
With increasing $\alpha$ from small values to values of order unity, 
the properties of the isotropic Heisenberg model can be connected to those 
near the Ising limit.

In Ref.~\cite{IsodaNakano:XXZ}, it was demonstrated that 
the $XXZ$ model on the kagome lattice exhibits
the two-peak structure of the heat capacity for any nonzero transverse coupling
[$\alpha> 0$ in Eq.~\eqref{XXZHamiltonian}]
but has only one peak in the Ising case ($\alpha =0$), suggesting that 
the transverse coupling is essential for the formation of the lower-temperature peak,
while the higher-temperature peak is of the Ising origin.

The ground states of the Ising model on common frustrating lattices 
have extensive degeneracy (i.e. degeneracy that scales exponentially
with the system size). 
Single spin-flip excitations have energies of the order $\theta_W\sim ZJ$,
where $Z$ and $J$ are the coordination number and the AF coupling constant, 
and give rise to one peak in the heat capacity $C(T)$.
In the presence of 
a weak transverse coupling ($0<\alpha\ll1$), the degeneracy of the Ising ground states is
lifted, with the respective low-energy excitations resulting in the formation of the 
lower-temperature peak~\cite{IsodaNakano:XXZ,UlagaPrelovsek:TriangularKagomeTwoPeak}\cite{PoppRamirezSyzranov}. 
The characteristic energy of that peak is
determined by the Ising instantons, 
the minimum sequences of pair-wise spin-exchange processes that connect 
similar Ising ground states to each other.

With increasing the transverse coupling, i.e. the value of the parameter $\alpha$,
the energies of the excitations that give rise to the lower peak of $C(T)$ increase.
However, the two peaks remain separated even for $\alpha\sim 1$~\cite{UlagaPrelovsek:TriangularKagomeTwoPeak}.
It has been estimated in Ref.~\cite{PoppRamirezSyzranov} that the ratio of the hidden
energy scale, i.e. the temperature of the first peak, to the characteristic temperature
$\theta_W$ of the second peak has the order of magnitude
\begin{align}
    \frac{T^*}{\theta_W}\sim \left(\frac{\alpha}{Z}\right)^{Z_1},
    \label{RatioTemperatures}
\end{align}
where $Z$ is the coordination number of the lattice and $Z_1$ is the size of the 
minimum Ising instanton, i.e. the minimum number of pairwise spin-exchange processes
that connect two Ising ground states.
As these numbers vary in the ranges $Z=4\ldots 6$ and $Z_1=1\ldots 3$ for common
frustrating lattices (triangular, kagome, pyrochlore, etc.), the ratio
\eqref{RatioTemperatures} remains
small even for $\alpha$ of order unity. 
This smallness ensures separation of the two peaks of the heat capacity $C(T)$ 
in frustrated magnets.

The estimate \eqref{RatioTemperatures} also imposes a constraint on the generalized
$f$-ratio, the figure of merit of frustration
given by Eq.~\eqref{fRatio} in which the temperature $T_c\sim T^*$ 
is the characteristic energy scale of the freezing of the magnetic degrees
of freedom that is not necessarily accompanied by a (magnetic-ordering or spin-glass) phase transition. However, in the presence of sufficient quenched disorder in 3D,
the freezing out of these degrees of freedom will lead to an SG transition at a 
temperature $T\sim T^*$.

We note that while the two-peak feature is a universal property of GF magnets, only one of those peaks is usually observed due to an insufficient temperature range accessed in the measurements. In GF materials with 3d elements, 
the higher-temperature peak lies at temperatures $\theta_W\gtrsim 100K$
and is rarely captured in experiments. In GF materials with
significantly lower $\theta_W$, such as rare-earth compounds,
the lower peak may lie {\it below} the lower limit of cryogenic apparatus ($\sim 50mK$)
and may thus be inaccessible in experiments.



\subsubsection*{Entropy associated with the hidden energy scale}

Because the states that lead to the formation of the lower-temperature peak in the heat capacity $C(T)$ are continuously connected to the ground states of the Ising model, the  entropy
\begin{align}
    S=\int_{\substack{\text{lower} \\ \text{peak}}} \frac{C(T)}{T} dT
    \label{EntropyLowPeak}
\end{align}
associated with the lower peak of $C(T)$
matches the respective Ising ground-state entropy for a particular lattice type.
If the lower-temperature peak is well separated from the higher-temperature one,
comparing the entropy under  it
with the Ising ground-state entropy on the same lattice
can be used, therefore, for verifying the described
mechanism of the hidden energy scale.

The values of the Ising ground-state entropy are known for common 2D lattices,
such as triangular~\cite{Wannier:Ising,HwangKim:TIAF,Kim:TIAF}) and kagome~\cite{KanoNaya:KIAF,SinghRigol:KIAF} lattices,
based on analytical arguments. For other lattices, e.g.
the pyrochlore lattice~\cite{Ramirez:iceEntropy,LauFreitas:SpinIce},
the value of the entropy can be found numerically.
We note that magnetism in many GF systems comes from higher spins, such
as spins-$1$ or spins-$3/2$. Strictly speaking, the entropy under the lower peak in these systems has to be compared with the ground-state entropy
of the respective higher-spin counterparts of the Ising model.
Often, however, the ground-state entropy of such models will match
the ground-state entropy of the conventional Ising model, corresponding to classical spins-$1/2$.

{\it A test case for the entropy}
associated with the hidden energy scale
is provided by $NiGa_2S_4$,
a layered, effectively 2D material with a triangular lattice and 
the heat capacity $C(T)$ shown in Fig.~\ref{fig:NiGaSTwoPeak}.
The value of the entropy~\eqref{EntropyLowPeak} under the lower peak
obtained from the digitized
data~\cite{NakatshujiMaeno:NaGaSneutron} shown in Fig.~\ref{fig:NiGaSTwoPeak}
is given by $S\approx 0.35$ per spin, which is indeed close to the value
of the Ising ground-state entropy $S_\text{Ising}\approx 0.32$ on the triangular lattice~\cite{Wannier:Ising,HwangKim:TIAF,Kim:TIAF}).


\subsection{Specific heat at the lowest temperatures}

In the previous subsections, we described manifestations of the hidden energy scale
in GF magnets, the type of excitations leading to that scale, and the total associated
entropy. In this subsection, we describe the temperature and magnetic-field dependencies
of the heat capacity $C(T)$, which provide further insight into the lowest-energy 
excitations in GF materials. 


\subsubsection{Temperature dependence: a test case}

Understanding the heat capacity $C(T)$ as a function of temperature in
GF systems with Heisenberg spins presents challenges,
many of which are illustrated in $SrCr_{9p}Ga_{12-9p}O_{19}$ (SCGO). This material, containing kagome bilayers of $Cr^{3+}$ $S = 3/2$ Heisenberg spins, exhibits an SG 
transition at $T_g\approx 2.7-4.2K$ for $0.6 < p < 0.98$~\cite{Ramirez:T2,Martinez:KagomeSCGO}. Specific heat measurements show 
the lower-temperature peak of $C(T)/T$ at $T\approx 6K$.
The area associated with that peak is given by $\sim 15\%$ of the total $R\ln 4$ entropy, per mole of $Cr$.


The family of the
SCGO materials exhibits a universal trend: 
the quadratic behavior $C(T)\propto T^2$ of the 
lowest-temperature ($T\ll T^*$) heat capacity, robust on 
varying the concentration $p$ of the magnetic $Cr$ atoms.
This temperature dependence is 
inconsistent with the linear heat capacity $C(T)\propto T$ 
generically expected to emerge in spin glasses
due to tunnelling between pairs of local energy minima separated by random potential barriers~\cite{AndersonHalperinVarma:GlassHeatCapacity}.

In a layered, i.e. effectively 2D material, the observed behavior of heat capacity
may come from any linearly dispersive excitations. We note that such excitations
in SCGO 
cannot be antiferromagnetic spin waves, since i) no magnetic Bragg peaks are seen in neutron scattering, as expected for an AF state~\cite{Broholm:SCGOcorrelationLength} and,  ii) the characteristic energy $\sim 6K$ of such excitations does not match 
the characteristic spin-spin interaction energy, which is estimated as the Weiss constant, approximately $500K$, in SCGO~\cite{Ramirez:T2SCGO}.
For a sufficiently large density of the linearly-dispersive 
excitations, their heat capacity may mask the conventional 
linear contribution common in spin glasses~\cite{AndersonHalperinVarma:GlassHeatCapacity}.

The quadratic behaviour $C(T)\propto T^2$
of the heat capacity has also been observed in the 
layered material $NiGa_2S_4$ described in Sec.~\ref{Sec:HiddenEnergyScenario}.
In this material, the SRO that develops at 
temperatures below the hidden energy scale $T^*$, has
a correlation region of $26 \AA$ in the plane of the layers,
and possesses a non-collinear spiral structure. The non-collinearity, along with the existence of static SRO, was argued to create the conditions for the so-called Halperin-Saslow modes~\cite{HalperinSaslowModes}, which can be thought of as linear-dispersing hydrodynamic excitations.
The theory~\cite{HalperinSaslowModes} of such modes was constructed for systems in which the ground state is comprised of non-collinear spins that are static but not long-range ordered, and while the theory was developed in the context of conventional spin glass, it should 
also apply to GF systems with SRO~\cite{PodolskyKim:HalperinSaslowNiGaS}.

Because the Halperin-Saslow modes are linearly dispersive, they can lead to the 
quadratic low-temperature dependence $C(T)\propto T^2$ in 
quasi-2D materials such as $NiGa_2S_4$ and other layered systems.
A conclusive identification of the mechanism
of this quadratic behaviour of the heat capacity requires further investigation.


\subsubsection{Magnetic-field (in-)dependence of the specific heat}   

Because the excitations giving rise to the lower-temperature peak in the heat capacity $C(T)$ are continuously connected to the ground states of the Ising model, the response of those states to a magnetic field depends on whether or not the respective ground states are magnetic.

For Ising spins on the pyrochlore lattice and the SCGO lattice~\cite{Ramirez:T2SCGO}, the magnetization in the ground states
is zero. Because in the presence of weak transverse coupling between spins, the low-energy excitations are given by superpositions of
such states and states obtained from them by permutations of spins, these
low-energy excitations will also be non-magnetic. On these lattices,
the heat capacity $C(T)$ at low temperatures $T\lesssim T^*$ will be insensitive to the 
magnetic field.

Other lattices, such as the triangular and kagome lattice, allow for nonzero magnetization of the ground states of the Ising model with nearest-neighbor interactions. On these lattices, the low-energy excitations will, in general, be magnetic. On the kagome lattice,
magnetic low-energy excitations have been observed numerically
in Refs.~\cite{WaldtmannEverts:KHAF,SindzingreMisguich:KHAF, IsodaNakano:XXZ,SchnackSchulenberg:KHAF,UlagaPrelovsek:TriangularKagomeTwoPeak,UlagaPrelovsek:latest,Prelovsek:TriangularWilsonRatio}.

Experimentally, it has been shown that $SrCr_{9p}Ga_{12-9p}O_{19}$, $NiGa_2S_4$, and also $Na_4Ir_3O_8$ all have peaks in $C(T)/T$ that are nearly field-independent~\cite{Ramirez:SCGOentropy,AndersonHalperinVarma:GlassHeatCapacity,Okamoto:NaIrOQSL} on Zeeman energy scales comparable to the hidden energy scale. 
While this field-independence is expected in SCGO and $Na_4Ir_3O_8$ based on the above 
argument, its microscopic mechanism in the (quasi-2D) compound $NiGa_2S_4$ with a triangular lattice calls for further investigation.
In this material, beyond-nearest-neighbor interactions or interlayer coupling 
might play a role. Such interactions are also required to prevent magnetic ordering 
on the triangular lattice~\cite{Bernu:triangularOrder,Capriotti:triangularOrder,White:traingularOrder}.


\section{Conclusions and further directions}

We have reviewed available data on the specific heat, magnetic susceptibility and neutron scattering in
geometrically frustrated (GF) materials. 
The available data for the intensity of neutron scattering as a function of temperature and for the glass transition temperature as a function
of impurity density 
suggest the existence of a characteristic energy scale, the  
``hidden energy scale'' $T^*$, which is of the order of $10K$
in the analyzed materials.

{\it Key manifestations of the hidden energy scale.}
The scale $T^*$ manifests itself in spin-glass (SG) freezing 
at temperatures $T<T^*$ in the cleanest 3D GF materials, i.e. in materials
with the smallest concentrations of the vacancy impurities, the most common form 
of quenched disorder in GF magnets. Cooling a material to temperatures 
below $T^*$ is also accompanied by an increase in the intensity of neutron scattering 
even in materials that do not exhibit the spin-glass transition due to the
low dimensionality. 

The GF materials for which specific heat $C(T)$ data are available 
in a sufficiently broad temperature range exhibit two peaks in the temperature dependence of the specific heat $C(T)$.
The higher-temperature peak is centered at temperatures of the order of the Weiss constant $\theta_W$, while the lower-temperature peak is located near the hidden energy scale $T^*$.
Such a structure in $C(T)$ is supported by
abundant numerical simulations of small clusters of spins
on geometrically frustrating lattices.

{\it The origin of the hidden energy scale and spin-glass transition.}
The SG freezing observed at a temperature of order $T^*$ in many rather clean GF materials
 raises the question of 
whether QSLs are achievable in GF systems or always get replaced by an SG state. This question, in part, motivated the current work.
While in our view, there is not a precise, commonly accepted definition of a QSL, an SG state is considered to be incompatible with a QSL.

In most GF systems, the main source of quenched disorder, which is necessary for the existence of an SG state, is vacancy defects. It has remained a puzzle for a while what causes the SG transitions in the purest GF materials, with vacancy defects virtually eliminated.
The existence of such SG transitions in rather clean GF systems seems to have created a school of thought that 
the SG transition is caused by a nearly immeasurable type of quenched disorder,
and the SRO observed in, e.g. neutron scattering, is caused by this glass transition.


By contrast, the available facts summarized in this review illustrate that the short-range order (SRO) that forms at temperatures $T$
below the hidden energy scale $T^*$ 
is a property of the disorder-free system.
The available facts also suggest that the formation of this SRO  
precipitates the SG transition in the presence of weak quenched disorder.

The hidden energy scale comes from the excitations that give rise to the lower-temperature peak in $C(T)$.
In a generic XXZ model on a GF lattice, such excitations are continuously connected to superpositions of Ising ground states when this model is deformed to the Ising model.
The total entropy of such 
excitations associated with the lower-temperature peak of $C(T)$
matches the entropy of the ground states
of the Ising model.
This correspondence is illustrated by the thermodynamic data in $NiGa_2S_4$
and can be used to verify the discussed origin of the hidden energy in other GF materials.

A GF material, thus, possesses a large density of excitations at the characteristic energy $T^*$, which are no longer activated when the material is cooled down to temperatures $T\lesssim T^*$.
In the presence of quenched disorder, the freezing out of such excitations leads to a spin-glass transition.
We emphasize that at such transitions,
the properties of the clean medium are in the driver's seat: while quenched disorder is required for SG freezing, the transition temperature will normally be determined by a crossover in the properties of the clean GF medium upon cooling.
Although an SG state can be avoided if quenched disorder is completely eliminated, currently available materials display an SG transition at a temperature of order $T^*$ for a broad range of disorder strengths.

Further theoretical advances are required to explain
various details of the 
heat capacity $C(T)$, for example, the observed quadratic temperature dependence
$C(T)\propto T^2$ at $T\ll T^*$ and the near-independence of $C(T)$ of the magnetic field.
Explaining these features will require
identifying the spin structure of the lowest-energy 
excitations, as well as evaluating their dispersion.

{\it Short-range order.}
The available neutron-scattering data further confirms the 
formation of short-range order near the hidden energy scale.
This type of magnetic order is rather unusual, distinct from
both the long-range magnetic order and spin-glass order, and creates
new exciting possibilities for understanding and manipulating it. 

For example, a characteristic feature of neutron scattering in spin ice is the set of pinch points in the momentum-dependent scattering. We expect that such structure will be revealed in strongly GF systems, but only in large-$q$ diffusive scattering. This structure should coincide with a build up of specific heat, already seen in a few systems. The two energy scales represented by the Weiss temperature and the hidden energy should also be systematically probed using specific heat. This will require measurements over a larger range of temperatures than is usually accessed.

{\it Magnetic susceptibility and impurities.}
As we discussed in this review, the magnetic susceptibility in GF magnets
follows the Curie-Weiss law, consistent with the mean-field description of 
magnetic materials, down to temperatures significantly lower than the Weiss constant
$\theta_W$. The reason for the persistence of the Curie-Weiss law in the lowest temperature range still remains to be investigated.

Quenched disorder has a profound effect on the magnetic susceptibility in GF materials.
The most common type of quenched disorder in GF magnets is vacancy defects. In terms of magnetic susceptibility, 
a vacancy defect is equivalent to an effective free spin, a ``quasispin''. Such quasispins lead to 
the Curie-tail $\propto N_\text{vac}/T$
contributions to $\chi(T)$ in the limit of small concentrations $N_\text{vac}$. At larger vacancy concentrations, the interactions between vacancy quasispins are essential.

While theoretical progress has been achieved in describing
quasispins in, e.g., Ising systems, in the vicinity of quantum criticality,
and in materials where vacancies do not disrupt short-range order, the values of 
quasispins and their interactions in generic quantum GF materials 
remain to be investigated. As the quasispin behavior 
has been investigated in terms of linear response of the magnetic susceptibility,
another question that awaits further study
is the response of defects to larger magnetic fields.
If quasispins can display quantum behavior, this may open new intriguing opportunities for storing and manipulating quantum information using quasispins. 
Encoded by numerous spins around a vacancy, the quantum state of a quasispin may be expected to be well protected from local noise.

{\it Vacancy defects and the spin-glass transition.} 
Another, counter-intuitive effect of vacancy defects on a GF material
is the decrease of the spin-glass transition temperature with increasing vacancy concentration. 
Conventional intuition and existing models of spin glasses suggest that increasing the strength of quenched disorder raises the temperature of the glass transition. Despite vacancies contribute to
the total disorder strength, adding them to a clean material has the opposite effect: they suppress the transition temperature.

As we discussed, in the vacancy-free limit, the glass transition takes place at a temperature of the order of the hidden energy 
scale $T^*$. Introducing dilute vacancies disrupts the glass order
and lowers the transition temperature.
While the effect of vacancies on a non-vacancy glass transition is readily captured by an effective field theory in terms of the glass order parameter~\cite{Syzranov:VacancyPhenomenological}, the described trend still awaits a microscopic description.

Finally, while we have concentrated on analyzing the most easily accessed measurements, now that the basic spin dynamics framework is established, other more targeted spectroscopies, such as muon spin relaxation and nuclear magnetic resonance, can also be brought to bear on describing the ground states and excitations in GF materials.
Of particular interest are the dynamics of excitations below the hidden energy scale,
which both NMR
and $\mu$SR have shown signs of~\cite{Takeya:NiGaSnmr,Limo:SCGOnmr} and may help to inform the microscopic theory of 
low-energy states.


\section{Data Availability}

The data used to generate Figs.~\ref{fig:ChiVsT}, \ref{fig:NeutronData} and \ref{fig:NiGaSTwoPeak} are available from the cited articles, except for the data for the magnetic susceptibility in $SrCr_{9p}Ga_{12-9p}O_{19}$ with $p=0.98$ and $p=0.67$. The latter data are available from the authors on request.
Figures~\ref{fig:HiddenEnergy} and \ref{fig:ChiTplots}
have been adapted with modification from the cited articles.


\section{Conflicts of Interest}

The authors declare that they have no conflicts of interest.


\section{Acknowledgements}

We are grateful to S.~Haravifard, P.~Popp, M.~Sedik, S.~Sun, and S.~Zhu for collaborations on the topics of this review.
We also appreciate useful discussions with R.~Colman, P.~Prelov\u{s}ek, B.~Sbierski, and M. Ulaga.
This work has been supported by the NSF grant DMR2218130.


%



\begin{thebibliography}{144}%
	\makeatletter
	\providecommand \@ifxundefined [1]{%
		\@ifx{#1\undefined}
	}%
	\providecommand \@ifnum [1]{%
		\ifnum #1\expandafter \@firstoftwo
		\else \expandafter \@secondoftwo
		\fi
	}%
	\providecommand \@ifx [1]{%
		\ifx #1\expandafter \@firstoftwo
		\else \expandafter \@secondoftwo
		\fi
	}%
	\providecommand \natexlab [1]{#1}%
	\providecommand \enquote  [1]{``#1''}%
	\providecommand \bibnamefont  [1]{#1}%
	\providecommand \bibfnamefont [1]{#1}%
	\providecommand \citenamefont [1]{#1}%
	\providecommand \href@noop [0]{\@secondoftwo}%
	\providecommand \href [0]{\begingroup \@sanitize@url \@href}%
	\providecommand \@href[1]{\@@startlink{#1}\@@href}%
	\providecommand \@@href[1]{\endgroup#1\@@endlink}%
	\providecommand \@sanitize@url [0]{\catcode `\\12\catcode `\$12\catcode
		`\&12\catcode `\#12\catcode `\^12\catcode `\_12\catcode `\%12\relax}%
	\providecommand \@@startlink[1]{}%
	\providecommand \@@endlink[0]{}%
	\providecommand \url  [0]{\begingroup\@sanitize@url \@url }%
	\providecommand \@url [1]{\endgroup\@href {#1}{\urlprefix }}%
	\providecommand \urlprefix  [0]{URL }%
	\providecommand \Eprint [0]{\href }%
	\providecommand \doibase [0]{https://doi.org/}%
	\providecommand \selectlanguage [0]{\@gobble}%
	\providecommand \bibinfo  [0]{\@secondoftwo}%
	\providecommand \bibfield  [0]{\@secondoftwo}%
	\providecommand \translation [1]{[#1]}%
	\providecommand \BibitemOpen [0]{}%
	\providecommand \bibitemStop [0]{}%
	\providecommand \bibitemNoStop [0]{.\EOS\space}%
	\providecommand \EOS [0]{\spacefactor3000\relax}%
	\providecommand \BibitemShut  [1]{\csname bibitem#1\endcsname}%
	\let\auto@bib@innerbib\@empty
	\bibitem [{\citenamefont {Anderson}(1973)}]{Anderson:firstQSL}%
	\BibitemOpen
	\bibfield  {author} {\bibinfo {author} {\bibfnamefont {P.}~\bibnamefont
			{Anderson}},\ }\bibfield  {title} {\bibinfo {title} {Resonating valence
			bonds: A new kind of insulator?},\ }\href
	{https://doi.org/https://doi.org/10.1016/0025-5408(73)90167-0} {\bibfield
		{journal} {\bibinfo  {journal} {Materials Research Bulletin}\ }\textbf
		{\bibinfo {volume} {8}},\ \bibinfo {pages} {153} (\bibinfo {year}
		{1973})}\BibitemShut {NoStop}%
	\bibitem [{\citenamefont {Bernu}\ \emph {et~al.}(1992)\citenamefont {Bernu},
		\citenamefont {Lhuillier},\ and\ \citenamefont
		{Pierre}}]{Bernu:triangularOrder}%
	\BibitemOpen
	\bibfield  {author} {\bibinfo {author} {\bibfnamefont {B.}~\bibnamefont
			{Bernu}}, \bibinfo {author} {\bibfnamefont {C.}~\bibnamefont {Lhuillier}},\
		and\ \bibinfo {author} {\bibfnamefont {L.}~\bibnamefont {Pierre}},\
	}\bibfield  {title} {\bibinfo {title} {{Signature of N\'eel order in exact
				spectra of quantum antiferromagnets on finite lattices}},\ }\href
	{https://doi.org/10.1103/PhysRevLett.69.2590} {\bibfield  {journal} {\bibinfo
			{journal} {Phys. Rev. Lett.}\ }\textbf {\bibinfo {volume} {69}},\ \bibinfo
		{pages} {2590} (\bibinfo {year} {1992})}\BibitemShut {NoStop}%
	\bibitem [{\citenamefont {Capriotti}\ \emph {et~al.}(1999)\citenamefont
		{Capriotti}, \citenamefont {Trumper},\ and\ \citenamefont
		{Sorella}}]{Capriotti:triangularOrder}%
	\BibitemOpen
	\bibfield  {author} {\bibinfo {author} {\bibfnamefont {L.}~\bibnamefont
			{Capriotti}}, \bibinfo {author} {\bibfnamefont {A.~E.}\ \bibnamefont
			{Trumper}},\ and\ \bibinfo {author} {\bibfnamefont {S.}~\bibnamefont
			{Sorella}},\ }\bibfield  {title} {\bibinfo {title} {{Long-Range N\'eel Order
				in the Triangular Heisenberg Model}},\ }\href
	{https://doi.org/10.1103/PhysRevLett.82.3899} {\bibfield  {journal} {\bibinfo
			{journal} {Phys. Rev. Lett.}\ }\textbf {\bibinfo {volume} {82}},\ \bibinfo
		{pages} {3899} (\bibinfo {year} {1999})}\BibitemShut {NoStop}%
	\bibitem [{\citenamefont {White}\ and\ \citenamefont
		{Chernyshev}(2007)}]{White:traingularOrder}%
	\BibitemOpen
	\bibfield  {author} {\bibinfo {author} {\bibfnamefont {S.~R.}\ \bibnamefont
			{White}}\ and\ \bibinfo {author} {\bibfnamefont {A.~L.}\ \bibnamefont
			{Chernyshev}},\ }\bibfield  {title} {\bibinfo {title} {Ne\'el order in square
			and triangular lattice heisenberg models},\ }\href
	{https://doi.org/10.1103/PhysRevLett.99.127004} {\bibfield  {journal}
		{\bibinfo  {journal} {Phys. Rev. Lett.}\ }\textbf {\bibinfo {volume} {99}},\
		\bibinfo {pages} {127004} (\bibinfo {year} {2007})}\BibitemShut {NoStop}%
	\bibitem [{\citenamefont {Savary}\ and\ \citenamefont
		{Balents}(2016)}]{SavaryBalents:review}%
	\BibitemOpen
	\bibfield  {author} {\bibinfo {author} {\bibfnamefont {L.}~\bibnamefont
			{Savary}}\ and\ \bibinfo {author} {\bibfnamefont {L.}~\bibnamefont
			{Balents}},\ }\bibfield  {title} {\bibinfo {title} {Quantum spin liquids: a
			review},\ }\href@noop {} {\bibfield  {journal} {\bibinfo  {journal} {Reports
				on Progress in Physics}\ }\textbf {\bibinfo {volume} {80}},\ \bibinfo {pages}
		{016502} (\bibinfo {year} {2016})}\BibitemShut {NoStop}%
	\bibitem [{\citenamefont {Knolle}\ and\ \citenamefont
		{Moessner}(2019)}]{KnolleMoessner:review}%
	\BibitemOpen
	\bibfield  {author} {\bibinfo {author} {\bibfnamefont {J.}~\bibnamefont
			{Knolle}}\ and\ \bibinfo {author} {\bibfnamefont {R.}~\bibnamefont
			{Moessner}},\ }\bibfield  {title} {\bibinfo {title} {A field guide to spin
			liquids},\ }\href
	{https://doi.org/https://doi.org/10.1146/annurev-conmatphys-031218-013401}
	{\bibfield  {journal} {\bibinfo  {journal} {Annual Review of Condensed Matter
				Physics}\ }\textbf {\bibinfo {volume} {10}},\ \bibinfo {pages} {451}
		(\bibinfo {year} {2019})}\BibitemShut {NoStop}%
	\bibitem [{\citenamefont {Broholm}\ \emph {et~al.}(2020)\citenamefont
		{Broholm}, \citenamefont {Cava}, \citenamefont {Kivelson}, \citenamefont
		{Nocera}, \citenamefont {Norman},\ and\ \citenamefont
		{Senthil}}]{Broholm:QSLreview}%
	\BibitemOpen
	\bibfield  {author} {\bibinfo {author} {\bibfnamefont {C.}~\bibnamefont
			{Broholm}}, \bibinfo {author} {\bibfnamefont {R.~J.}\ \bibnamefont {Cava}},
		\bibinfo {author} {\bibfnamefont {S.~A.}\ \bibnamefont {Kivelson}}, \bibinfo
		{author} {\bibfnamefont {D.~G.}\ \bibnamefont {Nocera}}, \bibinfo {author}
		{\bibfnamefont {M.~R.}\ \bibnamefont {Norman}},\ and\ \bibinfo {author}
		{\bibfnamefont {T.}~\bibnamefont {Senthil}},\ }\bibfield  {title} {\bibinfo
		{title} {Quantum spin liquids},\ }\href
	{https://doi.org/10.1126/science.aay0668} {\bibfield  {journal} {\bibinfo
			{journal} {Science}\ }\textbf {\bibinfo {volume} {367}},\ \bibinfo {pages}
		{eaay0668} (\bibinfo {year} {2020})},\ \Eprint
	{https://arxiv.org/abs/https://www.science.org/doi/pdf/10.1126/science.aay0668}
	{https://www.science.org/doi/pdf/10.1126/science.aay0668} \BibitemShut
	{NoStop}%
	\bibitem [{\citenamefont {Shastry}\ and\ \citenamefont
		{Sutherland}(1981)}]{ShastrySutherland}%
	\BibitemOpen
	\bibfield  {author} {\bibinfo {author} {\bibfnamefont {B.~S.}\ \bibnamefont
			{Shastry}}\ and\ \bibinfo {author} {\bibfnamefont {B.}~\bibnamefont
			{Sutherland}},\ }\bibfield  {title} {\bibinfo {title} {Exact ground state of
			a quantum mechanical antiferromagnet},\ }\href@noop {} {\bibfield  {journal}
		{\bibinfo  {journal} {Physica B+ C}\ }\textbf {\bibinfo {volume} {108}},\
		\bibinfo {pages} {1069} (\bibinfo {year} {1981})}\BibitemShut {NoStop}%
	\bibitem [{\citenamefont {Kitaev}(2006)}]{Kitaev:honeycob}%
	\BibitemOpen
	\bibfield  {author} {\bibinfo {author} {\bibfnamefont {A.}~\bibnamefont
			{Kitaev}},\ }\bibfield  {title} {\bibinfo {title} {Anyons in an exactly
			solved model and beyond},\ }\href
	{https://doi.org/https://doi.org/10.1016/j.aop.2005.10.005} {\bibfield
		{journal} {\bibinfo  {journal} {Annals of Physics}\ }\textbf {\bibinfo
			{volume} {321}},\ \bibinfo {pages} {2} (\bibinfo {year} {2006})},\ \bibinfo
	{note} {january Special Issue}\BibitemShut {NoStop}%
	\bibitem [{\citenamefont {Iga}\ \emph {et~al.}(2007)\citenamefont {Iga},
		\citenamefont {Shigekawa}, \citenamefont {Hasegawa}, \citenamefont
		{Michimura}, \citenamefont {Takabatake}, \citenamefont {Yoshii},
		\citenamefont {Yamamoto}, \citenamefont {Hagiwara},\ and\ \citenamefont
		{Kindo}}]{Iga:TmB4}%
	\BibitemOpen
	\bibfield  {author} {\bibinfo {author} {\bibfnamefont {F.}~\bibnamefont
			{Iga}}, \bibinfo {author} {\bibfnamefont {A.}~\bibnamefont {Shigekawa}},
		\bibinfo {author} {\bibfnamefont {Y.}~\bibnamefont {Hasegawa}}, \bibinfo
		{author} {\bibfnamefont {S.}~\bibnamefont {Michimura}}, \bibinfo {author}
		{\bibfnamefont {T.}~\bibnamefont {Takabatake}}, \bibinfo {author}
		{\bibfnamefont {S.}~\bibnamefont {Yoshii}}, \bibinfo {author} {\bibfnamefont
			{T.}~\bibnamefont {Yamamoto}}, \bibinfo {author} {\bibfnamefont
			{M.}~\bibnamefont {Hagiwara}},\ and\ \bibinfo {author} {\bibfnamefont
			{K.}~\bibnamefont {Kindo}},\ }\bibfield  {title} {\bibinfo {title} {{Highly
				anisotropic magnetic phase diagram of a 2-dimensional orthogonal dimer system
				$TmB_4$}},\ }\href
	{https://doi.org/https://doi.org/10.1016/j.jmmm.2006.10.476} {\bibfield
		{journal} {\bibinfo  {journal} {Journal of Magnetism and Magnetic Materials}\
		}\textbf {\bibinfo {volume} {310}},\ \bibinfo {pages} {e443} (\bibinfo {year}
		{2007})},\ \bibinfo {note} {proceedings of the 17th International Conference
		on Magnetism}\BibitemShut {NoStop}%
	\bibitem [{\citenamefont {Jackeli}\ and\ \citenamefont
		{Khaliullin}(2009)}]{Jackeli:SOMott}%
	\BibitemOpen
	\bibfield  {author} {\bibinfo {author} {\bibfnamefont {G.}~\bibnamefont
			{Jackeli}}\ and\ \bibinfo {author} {\bibfnamefont {G.}~\bibnamefont
			{Khaliullin}},\ }\bibfield  {title} {\bibinfo {title} {{Mott Insulators in
				the Strong Spin-Orbit Coupling Limit: From Heisenberg to a Quantum Compass
				and Kitaev Models}},\ }\href {https://doi.org/10.1103/PhysRevLett.102.017205}
	{\bibfield  {journal} {\bibinfo  {journal} {Phys. Rev. Lett.}\ }\textbf
		{\bibinfo {volume} {102}},\ \bibinfo {pages} {017205} (\bibinfo {year}
		{2009})}\BibitemShut {NoStop}%
	\bibitem [{\citenamefont {Johnson}\ \emph {et~al.}(2015)\citenamefont
		{Johnson}, \citenamefont {Williams}, \citenamefont {Haghighirad},
		\citenamefont {Singleton}, \citenamefont {Zapf}, \citenamefont {Manuel},
		\citenamefont {Mazin}, \citenamefont {Li}, \citenamefont {Jeschke},
		\citenamefont {Valent\'{\i}},\ and\ \citenamefont
		{Coldea}}]{Johnson:RuCl3ordering}%
	\BibitemOpen
	\bibfield  {author} {\bibinfo {author} {\bibfnamefont {R.~D.}\ \bibnamefont
			{Johnson}}, \bibinfo {author} {\bibfnamefont {S.~C.}\ \bibnamefont
			{Williams}}, \bibinfo {author} {\bibfnamefont {A.~A.}\ \bibnamefont
			{Haghighirad}}, \bibinfo {author} {\bibfnamefont {J.}~\bibnamefont
			{Singleton}}, \bibinfo {author} {\bibfnamefont {V.}~\bibnamefont {Zapf}},
		\bibinfo {author} {\bibfnamefont {P.}~\bibnamefont {Manuel}}, \bibinfo
		{author} {\bibfnamefont {I.~I.}\ \bibnamefont {Mazin}}, \bibinfo {author}
		{\bibfnamefont {Y.}~\bibnamefont {Li}}, \bibinfo {author} {\bibfnamefont
			{H.~O.}\ \bibnamefont {Jeschke}}, \bibinfo {author} {\bibfnamefont
			{R.}~\bibnamefont {Valent\'{\i}}},\ and\ \bibinfo {author} {\bibfnamefont
			{R.}~\bibnamefont {Coldea}},\ }\bibfield  {title} {\bibinfo {title}
		{{Monoclinic crystal structure of
				$\ensuremath{\alpha}\ensuremath{-}{\mathrm{RuCl}}_{3}$ and the zigzag
				antiferromagnetic ground state}},\ }\href
	{https://doi.org/10.1103/PhysRevB.92.235119} {\bibfield  {journal} {\bibinfo
			{journal} {Phys. Rev. B}\ }\textbf {\bibinfo {volume} {92}},\ \bibinfo
		{pages} {235119} (\bibinfo {year} {2015})}\BibitemShut {NoStop}%
	\bibitem [{\citenamefont {Kubota}\ \emph {et~al.}(2015)\citenamefont {Kubota},
		\citenamefont {Tanaka}, \citenamefont {Ono}, \citenamefont {Narumi},\ and\
		\citenamefont {Kindo}}]{Kubota:RuCl3ordering}%
	\BibitemOpen
	\bibfield  {author} {\bibinfo {author} {\bibfnamefont {Y.}~\bibnamefont
			{Kubota}}, \bibinfo {author} {\bibfnamefont {H.}~\bibnamefont {Tanaka}},
		\bibinfo {author} {\bibfnamefont {T.}~\bibnamefont {Ono}}, \bibinfo {author}
		{\bibfnamefont {Y.}~\bibnamefont {Narumi}},\ and\ \bibinfo {author}
		{\bibfnamefont {K.}~\bibnamefont {Kindo}},\ }\bibfield  {title} {\bibinfo
		{title} {{Successive magnetic phase transitions in
				$\ensuremath{\alpha}\ensuremath{-}{\mathrm{RuCl}}_{3}$: XY-like frustrated
				magnet on the honeycomb lattice}},\ }\href
	{https://doi.org/10.1103/PhysRevB.91.094422} {\bibfield  {journal} {\bibinfo
			{journal} {Phys. Rev. B}\ }\textbf {\bibinfo {volume} {91}},\ \bibinfo
		{pages} {094422} (\bibinfo {year} {2015})}\BibitemShut {NoStop}%
	\bibitem [{\citenamefont {Sears}\ \emph {et~al.}(2015)\citenamefont {Sears},
		\citenamefont {Songvilay}, \citenamefont {Plumb}, \citenamefont {Clancy},
		\citenamefont {Qiu}, \citenamefont {Zhao}, \citenamefont {Parshall},\ and\
		\citenamefont {Kim}}]{Sears:RuCl3magneticOrder}%
	\BibitemOpen
	\bibfield  {author} {\bibinfo {author} {\bibfnamefont {J.~A.}\ \bibnamefont
			{Sears}}, \bibinfo {author} {\bibfnamefont {M.}~\bibnamefont {Songvilay}},
		\bibinfo {author} {\bibfnamefont {K.~W.}\ \bibnamefont {Plumb}}, \bibinfo
		{author} {\bibfnamefont {J.~P.}\ \bibnamefont {Clancy}}, \bibinfo {author}
		{\bibfnamefont {Y.}~\bibnamefont {Qiu}}, \bibinfo {author} {\bibfnamefont
			{Y.}~\bibnamefont {Zhao}}, \bibinfo {author} {\bibfnamefont {D.}~\bibnamefont
			{Parshall}},\ and\ \bibinfo {author} {\bibfnamefont {Y.-J.}\ \bibnamefont
			{Kim}},\ }\bibfield  {title} {\bibinfo {title} {{Magnetic order in
				$\ensuremath{\alpha}\ensuremath{-}{\text{RuCl}}_{3}$: A honeycomb-lattice
				quantum magnet with strong spin-orbit coupling}},\ }\href
	{https://doi.org/10.1103/PhysRevB.91.144420} {\bibfield  {journal} {\bibinfo
			{journal} {Phys. Rev. B}\ }\textbf {\bibinfo {volume} {91}},\ \bibinfo
		{pages} {144420} (\bibinfo {year} {2015})}\BibitemShut {NoStop}%
	\bibitem [{\citenamefont {Banerjee}\ \emph {et~al.}(2017)\citenamefont
		{Banerjee}, \citenamefont {Yan}, \citenamefont {Knolle}, \citenamefont
		{Bridges}, \citenamefont {Stone}, \citenamefont {Lumsden}, \citenamefont
		{Mandrus}, \citenamefont {Tennant}, \citenamefont {Moessner},\ and\
		\citenamefont {Nagler}}]{Banerjee:RuCl3neutronScattering}%
	\BibitemOpen
	\bibfield  {author} {\bibinfo {author} {\bibfnamefont {A.}~\bibnamefont
			{Banerjee}}, \bibinfo {author} {\bibfnamefont {J.}~\bibnamefont {Yan}},
		\bibinfo {author} {\bibfnamefont {J.}~\bibnamefont {Knolle}}, \bibinfo
		{author} {\bibfnamefont {C.~A.}\ \bibnamefont {Bridges}}, \bibinfo {author}
		{\bibfnamefont {M.~B.}\ \bibnamefont {Stone}}, \bibinfo {author}
		{\bibfnamefont {M.~D.}\ \bibnamefont {Lumsden}}, \bibinfo {author}
		{\bibfnamefont {D.~G.}\ \bibnamefont {Mandrus}}, \bibinfo {author}
		{\bibfnamefont {D.~A.}\ \bibnamefont {Tennant}}, \bibinfo {author}
		{\bibfnamefont {R.}~\bibnamefont {Moessner}},\ and\ \bibinfo {author}
		{\bibfnamefont {S.~E.}\ \bibnamefont {Nagler}},\ }\bibfield  {title}
	{\bibinfo {title} {{Neutron scattering in the proximate quantum spin liquid
				$\alpha-RuCl_3$}},\ }\href {https://doi.org/10.1126/science.aah6015}
	{\bibfield  {journal} {\bibinfo  {journal} {Science}\ }\textbf {\bibinfo
			{volume} {356}},\ \bibinfo {pages} {1055} (\bibinfo {year} {2017})},\ \Eprint
	{https://arxiv.org/abs/https://www.science.org/doi/pdf/10.1126/science.aah6015}
	{https://www.science.org/doi/pdf/10.1126/science.aah6015} \BibitemShut
	{NoStop}%
	\bibitem [{\citenamefont {Kasahara}\ \emph {et~al.}(2018)\citenamefont
		{Kasahara}, \citenamefont {Ohnishi}, \citenamefont {Mizukami}, \citenamefont
		{Tanaka}, \citenamefont {Ma}, \citenamefont {Sugii}, \citenamefont {Kurita},
		\citenamefont {Tanaka}, \citenamefont {Nasu}, \citenamefont {Motome} \emph
		{et~al.}}]{Matsuda:RuCl3oscillations}%
	\BibitemOpen
	\bibfield  {author} {\bibinfo {author} {\bibfnamefont {Y.}~\bibnamefont
			{Kasahara}}, \bibinfo {author} {\bibfnamefont {T.}~\bibnamefont {Ohnishi}},
		\bibinfo {author} {\bibfnamefont {Y.}~\bibnamefont {Mizukami}}, \bibinfo
		{author} {\bibfnamefont {O.}~\bibnamefont {Tanaka}}, \bibinfo {author}
		{\bibfnamefont {S.}~\bibnamefont {Ma}}, \bibinfo {author} {\bibfnamefont
			{K.}~\bibnamefont {Sugii}}, \bibinfo {author} {\bibfnamefont
			{N.}~\bibnamefont {Kurita}}, \bibinfo {author} {\bibfnamefont
			{H.}~\bibnamefont {Tanaka}}, \bibinfo {author} {\bibfnamefont
			{J.}~\bibnamefont {Nasu}}, \bibinfo {author} {\bibfnamefont {Y.}~\bibnamefont
			{Motome}}, \emph {et~al.},\ }\bibfield  {title} {\bibinfo {title} {{Majorana
				quantization and half-integer thermal quantum Hall effect in a Kitaev spin
				liquid}},\ }\href@noop {} {\bibfield  {journal} {\bibinfo  {journal}
			{Nature}\ }\textbf {\bibinfo {volume} {559}},\ \bibinfo {pages} {227}
		(\bibinfo {year} {2018})}\BibitemShut {NoStop}%
	\bibitem [{\citenamefont {Czajka}\ \emph {et~al.}(2021)\citenamefont {Czajka},
		\citenamefont {Gao}, \citenamefont {Hirschberger}, \citenamefont
		{Lampen-Kelley}, \citenamefont {Banerjee}, \citenamefont {Yan}, \citenamefont
		{Mandrus}, \citenamefont {Nagler},\ and\ \citenamefont
		{Ong}}]{CzajkaOng:RuCl3oscillations}%
	\BibitemOpen
	\bibfield  {author} {\bibinfo {author} {\bibfnamefont {P.}~\bibnamefont
			{Czajka}}, \bibinfo {author} {\bibfnamefont {T.}~\bibnamefont {Gao}},
		\bibinfo {author} {\bibfnamefont {M.}~\bibnamefont {Hirschberger}}, \bibinfo
		{author} {\bibfnamefont {P.}~\bibnamefont {Lampen-Kelley}}, \bibinfo {author}
		{\bibfnamefont {A.}~\bibnamefont {Banerjee}}, \bibinfo {author}
		{\bibfnamefont {J.}~\bibnamefont {Yan}}, \bibinfo {author} {\bibfnamefont
			{D.~G.}\ \bibnamefont {Mandrus}}, \bibinfo {author} {\bibfnamefont {S.~E.}\
			\bibnamefont {Nagler}},\ and\ \bibinfo {author} {\bibfnamefont
			{N.}~\bibnamefont {Ong}},\ }\bibfield  {title} {\bibinfo {title}
		{{Oscillations of the thermal conductivity in the spin-liquid state of
				$\alpha$-RuCl3}},\ }\href@noop {} {\bibfield  {journal} {\bibinfo  {journal}
			{Nature Physics}\ }\textbf {\bibinfo {volume} {17}},\ \bibinfo {pages} {915}
		(\bibinfo {year} {2021})}\BibitemShut {NoStop}%
	\bibitem [{\citenamefont {Zheng}\ \emph {et~al.}(2017)\citenamefont {Zheng},
		\citenamefont {Ran}, \citenamefont {Li}, \citenamefont {Wang}, \citenamefont
		{Wang}, \citenamefont {Liu}, \citenamefont {Liu}, \citenamefont {Normand},
		\citenamefont {Wen},\ and\ \citenamefont {Yu}}]{ZhengYu:RuCl3QSL}%
	\BibitemOpen
	\bibfield  {author} {\bibinfo {author} {\bibfnamefont {J.}~\bibnamefont
			{Zheng}}, \bibinfo {author} {\bibfnamefont {K.}~\bibnamefont {Ran}}, \bibinfo
		{author} {\bibfnamefont {T.}~\bibnamefont {Li}}, \bibinfo {author}
		{\bibfnamefont {J.}~\bibnamefont {Wang}}, \bibinfo {author} {\bibfnamefont
			{P.}~\bibnamefont {Wang}}, \bibinfo {author} {\bibfnamefont {B.}~\bibnamefont
			{Liu}}, \bibinfo {author} {\bibfnamefont {Z.-X.}\ \bibnamefont {Liu}},
		\bibinfo {author} {\bibfnamefont {B.}~\bibnamefont {Normand}}, \bibinfo
		{author} {\bibfnamefont {J.}~\bibnamefont {Wen}},\ and\ \bibinfo {author}
		{\bibfnamefont {W.}~\bibnamefont {Yu}},\ }\bibfield  {title} {\bibinfo
		{title} {{Gapless Spin Excitations in the Field-Induced Quantum Spin Liquid
				Phase of $\ensuremath{\alpha}\text{\ensuremath{-}}{\mathrm{RuCl}}_{3}$}},\
	}\href {https://doi.org/10.1103/PhysRevLett.119.227208} {\bibfield  {journal}
		{\bibinfo  {journal} {Phys. Rev. Lett.}\ }\textbf {\bibinfo {volume} {119}},\
		\bibinfo {pages} {227208} (\bibinfo {year} {2017})}\BibitemShut {NoStop}%
	\bibitem [{\citenamefont {Bruin}\ \emph
		{et~al.}(2022{\natexlab{a}})\citenamefont {Bruin}, \citenamefont {Claus},
		\citenamefont {Matsumoto}, \citenamefont {Nuss}, \citenamefont {Laha},
		\citenamefont {Lotsch}, \citenamefont {Kurita}, \citenamefont {Tanaka},\ and\
		\citenamefont {Takagi}}]{BruinTakagi:RuCl3magneticTransitions}%
	\BibitemOpen
	\bibfield  {author} {\bibinfo {author} {\bibfnamefont {J.~A.~N.}\
			\bibnamefont {Bruin}}, \bibinfo {author} {\bibfnamefont {R.~R.}\ \bibnamefont
			{Claus}}, \bibinfo {author} {\bibfnamefont {Y.}~\bibnamefont {Matsumoto}},
		\bibinfo {author} {\bibfnamefont {J.}~\bibnamefont {Nuss}}, \bibinfo {author}
		{\bibfnamefont {S.}~\bibnamefont {Laha}}, \bibinfo {author} {\bibfnamefont
			{B.~V.}\ \bibnamefont {Lotsch}}, \bibinfo {author} {\bibfnamefont
			{N.}~\bibnamefont {Kurita}}, \bibinfo {author} {\bibfnamefont
			{H.}~\bibnamefont {Tanaka}},\ and\ \bibinfo {author} {\bibfnamefont
			{H.}~\bibnamefont {Takagi}},\ }\bibfield  {title} {\bibinfo {title} {{Origin
				of oscillatory structures in the magnetothermal conductivity of the putative
				Kitaev magnet $\alpha$-$RuCl_3$}},\ }\href
	{https://doi.org/10.1063/5.0101377} {\bibfield  {journal} {\bibinfo
			{journal} {APL Materials}\ }\textbf {\bibinfo {volume} {10}},\ \bibinfo
		{pages} {090703} (\bibinfo {year} {2022}{\natexlab{a}})},\ \Eprint
	{https://arxiv.org/abs/https://pubs.aip.org/aip/apm/article-pdf/doi/10.1063/5.0101377/19751661/090703\_1\_5.0101377.pdf}
	{https://pubs.aip.org/aip/apm/article-pdf/doi/10.1063/5.0101377/19751661/090703\_1\_5.0101377.pdf}
	\BibitemShut {NoStop}%
	\bibitem [{\citenamefont {Bruin}\ \emph
		{et~al.}(2022{\natexlab{b}})\citenamefont {Bruin}, \citenamefont {Claus},
		\citenamefont {Matsumoto}, \citenamefont {Kurita}, \citenamefont {Tanaka},\
		and\ \citenamefont {Takagi}}]{BruinTakagi:RuCl3oscillations}%
	\BibitemOpen
	\bibfield  {author} {\bibinfo {author} {\bibfnamefont {J.}~\bibnamefont
			{Bruin}}, \bibinfo {author} {\bibfnamefont {R.}~\bibnamefont {Claus}},
		\bibinfo {author} {\bibfnamefont {Y.}~\bibnamefont {Matsumoto}}, \bibinfo
		{author} {\bibfnamefont {N.}~\bibnamefont {Kurita}}, \bibinfo {author}
		{\bibfnamefont {H.}~\bibnamefont {Tanaka}},\ and\ \bibinfo {author}
		{\bibfnamefont {H.}~\bibnamefont {Takagi}},\ }\bibfield  {title} {\bibinfo
		{title} {{Robustness of the thermal Hall effect close to half-quantization in
				$\alpha$-$RuCl_3$}},\ }\href@noop {} {\bibfield  {journal} {\bibinfo
			{journal} {Nature Physics}\ }\textbf {\bibinfo {volume} {18}},\ \bibinfo
		{pages} {401} (\bibinfo {year} {2022}{\natexlab{b}})}\BibitemShut {NoStop}%
	\bibitem [{\citenamefont {Lefran\ifmmode~\mbox{\c{c}}\else \c{c}\fi{}ois}\
		\emph {et~al.}(2023)\citenamefont {Lefran\ifmmode~\mbox{\c{c}}\else
			\c{c}\fi{}ois}, \citenamefont {Baglo}, \citenamefont {Barth\'elemy},
		\citenamefont {Kim}, \citenamefont {Kim},\ and\ \citenamefont
		{Taillefer}}]{LefrancoiceTaillefer:RuCl3noOscillations}%
	\BibitemOpen
	\bibfield  {author} {\bibinfo {author} {\bibfnamefont {E.}~\bibnamefont
			{Lefran\ifmmode~\mbox{\c{c}}\else \c{c}\fi{}ois}}, \bibinfo {author}
		{\bibfnamefont {J.}~\bibnamefont {Baglo}}, \bibinfo {author} {\bibfnamefont
			{Q.}~\bibnamefont {Barth\'elemy}}, \bibinfo {author} {\bibfnamefont
			{S.}~\bibnamefont {Kim}}, \bibinfo {author} {\bibfnamefont {Y.-J.}\
			\bibnamefont {Kim}},\ and\ \bibinfo {author} {\bibfnamefont {L.}~\bibnamefont
			{Taillefer}},\ }\bibfield  {title} {\bibinfo {title} {{Oscillations in the
				magnetothermal conductivity of
				$\ensuremath{\alpha}\ensuremath{-}{\mathrm{RuCl}}_{3}$: Evidence of
				transition anomalies}},\ }\href {https://doi.org/10.1103/PhysRevB.107.064408}
	{\bibfield  {journal} {\bibinfo  {journal} {Phys. Rev. B}\ }\textbf {\bibinfo
			{volume} {107}},\ \bibinfo {pages} {064408} (\bibinfo {year}
		{2023})}\BibitemShut {NoStop}%
	\bibitem [{\citenamefont {Wannier}(1950)}]{Wannier:Ising}%
	\BibitemOpen
	\bibfield  {author} {\bibinfo {author} {\bibfnamefont {G.~H.}\ \bibnamefont
			{Wannier}},\ }\bibfield  {title} {\bibinfo {title} {Antiferromagnetism. the
			triangular ising net},\ }\href {https://doi.org/10.1103/PhysRev.79.357}
	{\bibfield  {journal} {\bibinfo  {journal} {Phys. Rev.}\ }\textbf {\bibinfo
			{volume} {79}},\ \bibinfo {pages} {357} (\bibinfo {year} {1950})}\BibitemShut
	{NoStop}%
	\bibitem [{\citenamefont {Anderson}(1956)}]{Anderson:OrderingFerrites}%
	\BibitemOpen
	\bibfield  {author} {\bibinfo {author} {\bibfnamefont {P.~W.}\ \bibnamefont
			{Anderson}},\ }\bibfield  {title} {\bibinfo {title} {Ordering and
			antiferromagnetism in ferrites},\ }\href
	{https://doi.org/10.1103/PhysRev.102.1008} {\bibfield  {journal} {\bibinfo
			{journal} {Phys. Rev.}\ }\textbf {\bibinfo {volume} {102}},\ \bibinfo {pages}
		{1008} (\bibinfo {year} {1956})}\BibitemShut {NoStop}%
	\bibitem [{\citenamefont {Ramirez}(1991)}]{Ramirez:FrustratedThermodynamic}%
	\BibitemOpen
	\bibfield  {author} {\bibinfo {author} {\bibfnamefont {A.}~\bibnamefont
			{Ramirez}},\ }\bibfield  {title} {\bibinfo {title} {Thermodynamic
			measurements on geometrically frustrated magnets},\ }\href@noop {} {\bibfield
		{journal} {\bibinfo  {journal} {Journal of applied physics}\ }\textbf
		{\bibinfo {volume} {70}},\ \bibinfo {pages} {5952} (\bibinfo {year}
		{1991})}\BibitemShut {NoStop}%
	\bibitem [{\citenamefont {Lacroix}\ \emph {et~al.}(2011)\citenamefont
		{Lacroix}, \citenamefont {Mendels},\ and\ \citenamefont
		{Mila}}]{Lacroix:book}%
	\BibitemOpen
	\bibfield  {author} {\bibinfo {author} {\bibfnamefont {C.}~\bibnamefont
			{Lacroix}}, \bibinfo {author} {\bibfnamefont {P.}~\bibnamefont {Mendels}},\
		and\ \bibinfo {author} {\bibfnamefont {F.}~\bibnamefont {Mila}},\ }\href@noop
	{} {\emph {\bibinfo {title} {Introduction to frustrated magnetism: materials,
				experiments, theory}}},\ Vol.\ \bibinfo {volume} {164}\ (\bibinfo
	{publisher} {Springer Science \& Business Media},\ \bibinfo {year}
	{2011})\BibitemShut {NoStop}%
	\bibitem [{\citenamefont {Chamorro}\ \emph {et~al.}(2020)\citenamefont
		{Chamorro}, \citenamefont {McQueen},\ and\ \citenamefont
		{Tran}}]{Chamorro:ChemReview}%
	\BibitemOpen
	\bibfield  {author} {\bibinfo {author} {\bibfnamefont {J.~R.}\ \bibnamefont
			{Chamorro}}, \bibinfo {author} {\bibfnamefont {T.~M.}\ \bibnamefont
			{McQueen}},\ and\ \bibinfo {author} {\bibfnamefont {T.~T.}\ \bibnamefont
			{Tran}},\ }\bibfield  {title} {\bibinfo {title} {Chemistry of quantum spin
			liquids},\ }\href@noop {} {\bibfield  {journal} {\bibinfo  {journal}
			{Chemical Reviews}\ }\textbf {\bibinfo {volume} {121}},\ \bibinfo {pages}
		{2898} (\bibinfo {year} {2020})}\BibitemShut {NoStop}%
	\bibitem [{\citenamefont {Zhou}\ \emph {et~al.}(2017)\citenamefont {Zhou},
		\citenamefont {Kanoda},\ and\ \citenamefont {Ng}}]{Zhou:QSLreview}%
	\BibitemOpen
	\bibfield  {author} {\bibinfo {author} {\bibfnamefont {Y.}~\bibnamefont
			{Zhou}}, \bibinfo {author} {\bibfnamefont {K.}~\bibnamefont {Kanoda}},\ and\
		\bibinfo {author} {\bibfnamefont {T.-K.}\ \bibnamefont {Ng}},\ }\bibfield
	{title} {\bibinfo {title} {Quantum spin liquid states},\ }\href
	{https://doi.org/10.1103/RevModPhys.89.025003} {\bibfield  {journal}
		{\bibinfo  {journal} {Rev. Mod. Phys.}\ }\textbf {\bibinfo {volume} {89}},\
		\bibinfo {pages} {025003} (\bibinfo {year} {2017})}\BibitemShut {NoStop}%
	\bibitem [{\citenamefont {Clark}\ and\ \citenamefont
		{Abdeldaim}(2021)}]{Clark:QSLreview}%
	\BibitemOpen
	\bibfield  {author} {\bibinfo {author} {\bibfnamefont {L.}~\bibnamefont
			{Clark}}\ and\ \bibinfo {author} {\bibfnamefont {A.~H.}\ \bibnamefont
			{Abdeldaim}},\ }\bibfield  {title} {\bibinfo {title} {Quantum spin liquids
			from a materials perspective},\ }\href
	{https://doi.org/https://doi.org/10.1146/annurev-matsci-080819-011453}
	{\bibfield  {journal} {\bibinfo  {journal} {Annual Review of Materials
				Research}\ }\textbf {\bibinfo {volume} {51}},\ \bibinfo {pages} {495}
		(\bibinfo {year} {2021})}\BibitemShut {NoStop}%
	\bibitem [{\citenamefont {Merchant}\ \emph {et~al.}(2014)\citenamefont
		{Merchant}, \citenamefont {Normand}, \citenamefont {Kr{\"a}mer},
		\citenamefont {Boehm}, \citenamefont {McMorrow},\ and\ \citenamefont
		{R{\"u}egg}}]{Merchant:dimerization}%
	\BibitemOpen
	\bibfield  {author} {\bibinfo {author} {\bibfnamefont {P.}~\bibnamefont
			{Merchant}}, \bibinfo {author} {\bibfnamefont {B.}~\bibnamefont {Normand}},
		\bibinfo {author} {\bibfnamefont {K.}~\bibnamefont {Kr{\"a}mer}}, \bibinfo
		{author} {\bibfnamefont {M.}~\bibnamefont {Boehm}}, \bibinfo {author}
		{\bibfnamefont {D.}~\bibnamefont {McMorrow}},\ and\ \bibinfo {author}
		{\bibfnamefont {C.}~\bibnamefont {R{\"u}egg}},\ }\bibfield  {title} {\bibinfo
		{title} {Quantum and classical criticality in a dimerized quantum
			antiferromagnet},\ }\href@noop {} {\bibfield  {journal} {\bibinfo  {journal}
			{Nature physics}\ }\textbf {\bibinfo {volume} {10}},\ \bibinfo {pages} {373}
		(\bibinfo {year} {2014})}\BibitemShut {NoStop}%
	\bibitem [{\citenamefont {Kimchi}\ \emph {et~al.}(2018)\citenamefont {Kimchi},
		\citenamefont {Sheckelton}, \citenamefont {McQueen},\ and\ \citenamefont
		{Lee}}]{Kimchi:disorder}%
	\BibitemOpen
	\bibfield  {author} {\bibinfo {author} {\bibfnamefont {I.}~\bibnamefont
			{Kimchi}}, \bibinfo {author} {\bibfnamefont {J.~P.}\ \bibnamefont
			{Sheckelton}}, \bibinfo {author} {\bibfnamefont {T.~M.}\ \bibnamefont
			{McQueen}},\ and\ \bibinfo {author} {\bibfnamefont {P.~A.}\ \bibnamefont
			{Lee}},\ }\bibfield  {title} {\bibinfo {title} {Scaling and data collapse
			from local moments in frustrated disordered quantum spin systems},\
	}\href@noop {} {\bibfield  {journal} {\bibinfo  {journal} {Nature
				communications}\ }\textbf {\bibinfo {volume} {9}},\ \bibinfo {pages} {4367}
		(\bibinfo {year} {2018})}\BibitemShut {NoStop}%
	\bibitem [{\citenamefont {Jongh}\ and\ \citenamefont
		{Miedema}(2001)}]{DeJongh:dimensionality}%
	\BibitemOpen
	\bibfield  {author} {\bibinfo {author} {\bibfnamefont {L.~J.~D.}\
			\bibnamefont {Jongh}}\ and\ \bibinfo {author} {\bibfnamefont {A.~R.}\
			\bibnamefont {Miedema}},\ }\bibfield  {title} {\bibinfo {title} {Experiments
			on simple magnetic model systems},\ }\href
	{https://doi.org/10.1080/00018730110101412} {\bibfield  {journal} {\bibinfo
			{journal} {Advances in Physics}\ }\textbf {\bibinfo {volume} {50}},\ \bibinfo
		{pages} {947} (\bibinfo {year} {2001})},\ \Eprint
	{https://arxiv.org/abs/https://doi.org/10.1080/00018730110101412}
	{https://doi.org/10.1080/00018730110101412} \BibitemShut {NoStop}%
	\bibitem [{\citenamefont {Takeya}\ \emph {et~al.}(2008)\citenamefont {Takeya},
		\citenamefont {Ishida}, \citenamefont {Kitagawa}, \citenamefont {Ihara},
		\citenamefont {Onuma}, \citenamefont {Maeno}, \citenamefont {Nambu},
		\citenamefont {Nakatsuji}, \citenamefont {MacLaughlin}, \citenamefont
		{Koda},\ and\ \citenamefont {Kadono}}]{Takeya:NiGaSnmr}%
	\BibitemOpen
	\bibfield  {author} {\bibinfo {author} {\bibfnamefont {H.}~\bibnamefont
			{Takeya}}, \bibinfo {author} {\bibfnamefont {K.}~\bibnamefont {Ishida}},
		\bibinfo {author} {\bibfnamefont {K.}~\bibnamefont {Kitagawa}}, \bibinfo
		{author} {\bibfnamefont {Y.}~\bibnamefont {Ihara}}, \bibinfo {author}
		{\bibfnamefont {K.}~\bibnamefont {Onuma}}, \bibinfo {author} {\bibfnamefont
			{Y.}~\bibnamefont {Maeno}}, \bibinfo {author} {\bibfnamefont
			{Y.}~\bibnamefont {Nambu}}, \bibinfo {author} {\bibfnamefont
			{S.}~\bibnamefont {Nakatsuji}}, \bibinfo {author} {\bibfnamefont {D.~E.}\
			\bibnamefont {MacLaughlin}}, \bibinfo {author} {\bibfnamefont
			{A.}~\bibnamefont {Koda}},\ and\ \bibinfo {author} {\bibfnamefont
			{R.}~\bibnamefont {Kadono}},\ }\bibfield  {title} {\bibinfo {title} {{Spin
				dynamics and spin freezing behavior in the two-dimensional antiferromagnet
				$\mathrm{Ni}{\mathrm{Ga}}_{2}{\mathrm{S}}_{4}$ revealed by Ga-NMR, NQR and
				$\ensuremath{\mu}\mathrm{SR}$ measurements}},\ }\href
	{https://doi.org/10.1103/PhysRevB.77.054429} {\bibfield  {journal} {\bibinfo
			{journal} {Phys. Rev. B}\ }\textbf {\bibinfo {volume} {77}},\ \bibinfo
		{pages} {054429} (\bibinfo {year} {2008})}\BibitemShut {NoStop}%
	\bibitem [{\citenamefont {Uemura}\ \emph {et~al.}(1994)\citenamefont {Uemura},
		\citenamefont {Keren}, \citenamefont {Kojima}, \citenamefont {Le},
		\citenamefont {Luke}, \citenamefont {Wu}, \citenamefont {Ajiro},
		\citenamefont {Asano}, \citenamefont {Kuriyama}, \citenamefont {Mekata},
		\citenamefont {Kikuchi},\ and\ \citenamefont {Kakurai}}]{Uemura:SCGOmuSR}%
	\BibitemOpen
	\bibfield  {author} {\bibinfo {author} {\bibfnamefont {Y.~J.}\ \bibnamefont
			{Uemura}}, \bibinfo {author} {\bibfnamefont {A.}~\bibnamefont {Keren}},
		\bibinfo {author} {\bibfnamefont {K.}~\bibnamefont {Kojima}}, \bibinfo
		{author} {\bibfnamefont {L.~P.}\ \bibnamefont {Le}}, \bibinfo {author}
		{\bibfnamefont {G.~M.}\ \bibnamefont {Luke}}, \bibinfo {author}
		{\bibfnamefont {W.~D.}\ \bibnamefont {Wu}}, \bibinfo {author} {\bibfnamefont
			{Y.}~\bibnamefont {Ajiro}}, \bibinfo {author} {\bibfnamefont
			{T.}~\bibnamefont {Asano}}, \bibinfo {author} {\bibfnamefont
			{Y.}~\bibnamefont {Kuriyama}}, \bibinfo {author} {\bibfnamefont
			{M.}~\bibnamefont {Mekata}}, \bibinfo {author} {\bibfnamefont
			{H.}~\bibnamefont {Kikuchi}},\ and\ \bibinfo {author} {\bibfnamefont
			{K.}~\bibnamefont {Kakurai}},\ }\bibfield  {title} {\bibinfo {title} {{Spin
				Fluctuations in Frustrated Kagom\'e Lattice System
				Sr${\mathrm{Cr}}_{8}$${\mathrm{Ga}}_{4}$${\mathrm{O}}_{19}$ Studied by Muon
				Spin Relaxation}},\ }\href {https://doi.org/10.1103/PhysRevLett.73.3306}
	{\bibfield  {journal} {\bibinfo  {journal} {Phys. Rev. Lett.}\ }\textbf
		{\bibinfo {volume} {73}},\ \bibinfo {pages} {3306} (\bibinfo {year}
		{1994})}\BibitemShut {NoStop}%
	\bibitem [{\citenamefont {Bono}\ \emph {et~al.}(2004)\citenamefont {Bono},
		\citenamefont {Mendels}, \citenamefont {Collin}, \citenamefont {Blanchard},
		\citenamefont {Bert}, \citenamefont {Amato}, \citenamefont {Baines},\ and\
		\citenamefont {Hillier}}]{Bono:KagomeMuSR}%
	\BibitemOpen
	\bibfield  {author} {\bibinfo {author} {\bibfnamefont {D.}~\bibnamefont
			{Bono}}, \bibinfo {author} {\bibfnamefont {P.}~\bibnamefont {Mendels}},
		\bibinfo {author} {\bibfnamefont {G.}~\bibnamefont {Collin}}, \bibinfo
		{author} {\bibfnamefont {N.}~\bibnamefont {Blanchard}}, \bibinfo {author}
		{\bibfnamefont {F.}~\bibnamefont {Bert}}, \bibinfo {author} {\bibfnamefont
			{A.}~\bibnamefont {Amato}}, \bibinfo {author} {\bibfnamefont
			{C.}~\bibnamefont {Baines}},\ and\ \bibinfo {author} {\bibfnamefont {A.~D.}\
			\bibnamefont {Hillier}},\ }\bibfield  {title} {\bibinfo {title}
		{$\ensuremath{\mu}\mathrm{S}\mathrm{R}$ study of the quantum dynamics in the
			frustrated $s=\frac{3}{2}$ kagom\'e bilayers},\ }\href
	{https://doi.org/10.1103/PhysRevLett.93.187201} {\bibfield  {journal}
		{\bibinfo  {journal} {Phys. Rev. Lett.}\ }\textbf {\bibinfo {volume} {93}},\
		\bibinfo {pages} {187201} (\bibinfo {year} {2004})}\BibitemShut {NoStop}%
	\bibitem [{\citenamefont {{Popp}}\ \emph {et~al.}(2024)\citenamefont {{Popp}},
		\citenamefont {{Ramirez}},\ and\ \citenamefont
		{{Syzranov}}}]{PoppRamirezSyzranov}%
	\BibitemOpen
	\bibfield  {author} {\bibinfo {author} {\bibfnamefont {P.}~\bibnamefont
			{{Popp}}}, \bibinfo {author} {\bibfnamefont {A.~P.}\ \bibnamefont
			{{Ramirez}}},\ and\ \bibinfo {author} {\bibfnamefont {S.}~\bibnamefont
			{{Syzranov}}},\ }\bibfield  {title} {\bibinfo {title} {{Origin of the hidden
				energy scale and the $f$-ratio in geometrically frustrated magnets}},\ }\href
	{https://doi.org/10.48550/arXiv.2406.12966} {\bibfield  {journal} {\bibinfo
			{journal} {arXiv e-prints}\ ,\ \bibinfo {eid} {arXiv:2406.12966}} (\bibinfo
		{year} {2024})},\ \Eprint {https://arxiv.org/abs/2406.12966}
	{arXiv:2406.12966 [cond-mat.str-el]} \BibitemShut {NoStop}%
	\bibitem [{\citenamefont {Helton}\ \emph {et~al.}(2007)\citenamefont {Helton},
		\citenamefont {Matan}, \citenamefont {Shores}, \citenamefont {Nytko},
		\citenamefont {Bartlett}, \citenamefont {Yoshida}, \citenamefont {Takano},
		\citenamefont {Suslov}, \citenamefont {Qiu}, \citenamefont {Chung},
		\citenamefont {Nocera},\ and\ \citenamefont
		{Lee}}]{Helton:frustrationZnCuOHCl}%
	\BibitemOpen
	\bibfield  {author} {\bibinfo {author} {\bibfnamefont {J.~S.}\ \bibnamefont
			{Helton}}, \bibinfo {author} {\bibfnamefont {K.}~\bibnamefont {Matan}},
		\bibinfo {author} {\bibfnamefont {M.~P.}\ \bibnamefont {Shores}}, \bibinfo
		{author} {\bibfnamefont {E.~A.}\ \bibnamefont {Nytko}}, \bibinfo {author}
		{\bibfnamefont {B.~M.}\ \bibnamefont {Bartlett}}, \bibinfo {author}
		{\bibfnamefont {Y.}~\bibnamefont {Yoshida}}, \bibinfo {author} {\bibfnamefont
			{Y.}~\bibnamefont {Takano}}, \bibinfo {author} {\bibfnamefont
			{A.}~\bibnamefont {Suslov}}, \bibinfo {author} {\bibfnamefont
			{Y.}~\bibnamefont {Qiu}}, \bibinfo {author} {\bibfnamefont {J.-H.}\
			\bibnamefont {Chung}}, \bibinfo {author} {\bibfnamefont {D.~G.}\ \bibnamefont
			{Nocera}},\ and\ \bibinfo {author} {\bibfnamefont {Y.~S.}\ \bibnamefont
			{Lee}},\ }\bibfield  {title} {\bibinfo {title} {{Spin Dynamics of the
				Spin-$1/2$ Kagome Lattice Antiferromagnet
				${\mathrm{ZnCu}}_{3}(\mathrm{OH}{)}_{6}{\mathrm{Cl}}_{2}$}},\ }\href
	{https://doi.org/10.1103/PhysRevLett.98.107204} {\bibfield  {journal}
		{\bibinfo  {journal} {Phys. Rev. Lett.}\ }\textbf {\bibinfo {volume} {98}},\
		\bibinfo {pages} {107204} (\bibinfo {year} {2007})}\BibitemShut {NoStop}%
	\bibitem [{\citenamefont {Lafontaine}\ \emph {et~al.}(1990)\citenamefont
		{Lafontaine}, \citenamefont {{Le Bail}},\ and\ \citenamefont
		{Férey}}]{LaFontaine:DFTCu3V2O7(OH)2}%
	\BibitemOpen
	\bibfield  {author} {\bibinfo {author} {\bibfnamefont {M.}~\bibnamefont
			{Lafontaine}}, \bibinfo {author} {\bibfnamefont {A.}~\bibnamefont {{Le
					Bail}}},\ and\ \bibinfo {author} {\bibfnamefont {G.}~\bibnamefont {Férey}},\
	}\bibfield  {title} {\bibinfo {title} {{Copper-containing minerals—I.
				$Cu_3V_2O_7(OH)_2$, $2H_2O$: The synthetichomolog of volborthite; crystal
				structure determination from X-ray and neutron data; structural
				correlations}},\ }\href
	{https://doi.org/https://doi.org/10.1016/S0022-4596(05)80078-7} {\bibfield
		{journal} {\bibinfo  {journal} {Journal of Solid State Chemistry}\ }\textbf
		{\bibinfo {volume} {85}},\ \bibinfo {pages} {220} (\bibinfo {year}
		{1990})}\BibitemShut {NoStop}%
	\bibitem [{\citenamefont {Hiroi}\ \emph {et~al.}(2001)\citenamefont {Hiroi},
		\citenamefont {Hanawa}, \citenamefont {Kobayashi}, \citenamefont {Nohara},
		\citenamefont {Takagi}, \citenamefont {Kato},\ and\ \citenamefont
		{Takigawa}}]{HiroiTakigawa:Cu3V2O7(OH)2H2O}%
	\BibitemOpen
	\bibfield  {author} {\bibinfo {author} {\bibfnamefont {Z.}~\bibnamefont
			{Hiroi}}, \bibinfo {author} {\bibfnamefont {M.}~\bibnamefont {Hanawa}},
		\bibinfo {author} {\bibfnamefont {N.}~\bibnamefont {Kobayashi}}, \bibinfo
		{author} {\bibfnamefont {M.}~\bibnamefont {Nohara}}, \bibinfo {author}
		{\bibfnamefont {H.}~\bibnamefont {Takagi}}, \bibinfo {author} {\bibfnamefont
			{Y.}~\bibnamefont {Kato}},\ and\ \bibinfo {author} {\bibfnamefont
			{M.}~\bibnamefont {Takigawa}},\ }\bibfield  {title} {\bibinfo {title}
		{{Spin-1/2 Kagom{\'e}-Like Lattice in Volborthite $Cu_3 V_2 O_7
				(OH)_2{\textperiodcentered} 2H_2 O$}},\ }\href
	{https://doi.org/10.1143/jpsj.70.3377} {\bibfield  {journal} {\bibinfo
			{journal} {Journal of the Physical Society of Japan}\ }\textbf {\bibinfo
			{volume} {70}},\ \bibinfo {pages} {3377} (\bibinfo {year}
		{2001})}\BibitemShut {NoStop}%
	\bibitem [{\citenamefont {Yoshida}\ \emph {et~al.}(2009)\citenamefont
		{Yoshida}, \citenamefont {Okamoto}, \citenamefont {Tayama}, \citenamefont
		{Sakakibara}, \citenamefont {Tokunaga}, \citenamefont {Matsuo}, \citenamefont
		{Narumi}, \citenamefont {Kindo}, \citenamefont {Yoshida}, \citenamefont
		{Takigawa},\ and\ \citenamefont
		{Hiroi}}]{Hiroyuki:magnetizationStepsVolborthite}%
	\BibitemOpen
	\bibfield  {author} {\bibinfo {author} {\bibfnamefont {H.}~\bibnamefont
			{Yoshida}}, \bibinfo {author} {\bibfnamefont {Y.}~\bibnamefont {Okamoto}},
		\bibinfo {author} {\bibfnamefont {T.}~\bibnamefont {Tayama}}, \bibinfo
		{author} {\bibfnamefont {T.}~\bibnamefont {Sakakibara}}, \bibinfo {author}
		{\bibfnamefont {M.}~\bibnamefont {Tokunaga}}, \bibinfo {author}
		{\bibfnamefont {A.}~\bibnamefont {Matsuo}}, \bibinfo {author} {\bibfnamefont
			{Y.}~\bibnamefont {Narumi}}, \bibinfo {author} {\bibfnamefont
			{K.}~\bibnamefont {Kindo}}, \bibinfo {author} {\bibfnamefont
			{M.}~\bibnamefont {Yoshida}}, \bibinfo {author} {\bibfnamefont
			{M.}~\bibnamefont {Takigawa}},\ and\ \bibinfo {author} {\bibfnamefont
			{Z.}~\bibnamefont {Hiroi}},\ }\bibfield  {title} {\bibinfo {title}
		{Magnetization “steps” on a kagome lattice in volborthite},\ }\href
	{https://doi.org/10.1143/JPSJ.78.043704} {\bibfield  {journal} {\bibinfo
			{journal} {Journal of the Physical Society of Japan}\ }\textbf {\bibinfo
			{volume} {78}},\ \bibinfo {pages} {043704} (\bibinfo {year} {2009})},\
	\Eprint {https://arxiv.org/abs/https://doi.org/10.1143/JPSJ.78.043704}
	{https://doi.org/10.1143/JPSJ.78.043704} \BibitemShut {NoStop}%
	\bibitem [{\citenamefont {Yoshida}\ \emph {et~al.}(2013)\citenamefont
		{Yoshida}, \citenamefont {Okamoto}, \citenamefont {Takigawa},\ and\
		\citenamefont {Hiroi}}]{MakotoTakigawa:susceptibilityVesigniete}%
	\BibitemOpen
	\bibfield  {author} {\bibinfo {author} {\bibfnamefont {M.}~\bibnamefont
			{Yoshida}}, \bibinfo {author} {\bibfnamefont {Y.}~\bibnamefont {Okamoto}},
		\bibinfo {author} {\bibfnamefont {M.}~\bibnamefont {Takigawa}},\ and\
		\bibinfo {author} {\bibfnamefont {Z.}~\bibnamefont {Hiroi}},\ }\bibfield
	{title} {\bibinfo {title} {Magnetic order in the spin-1/2 kagome
			antiferromagnet vesignieite},\ }\href
	{https://doi.org/10.7566/JPSJ.82.013702} {\bibfield  {journal} {\bibinfo
			{journal} {Journal of the Physical Society of Japan}\ }\textbf {\bibinfo
			{volume} {82}},\ \bibinfo {pages} {013702} (\bibinfo {year} {2013})},\
	\Eprint {https://arxiv.org/abs/https://doi.org/10.7566/JPSJ.82.013702}
	{https://doi.org/10.7566/JPSJ.82.013702} \BibitemShut {NoStop}%
	\bibitem [{\citenamefont {Janson}\ \emph {et~al.}(2008)\citenamefont {Janson},
		\citenamefont {Richter},\ and\ \citenamefont
		{Rosner}}]{Janson:kapellasiteSimulation}%
	\BibitemOpen
	\bibfield  {author} {\bibinfo {author} {\bibfnamefont {O.}~\bibnamefont
			{Janson}}, \bibinfo {author} {\bibfnamefont {J.}~\bibnamefont {Richter}},\
		and\ \bibinfo {author} {\bibfnamefont {H.}~\bibnamefont {Rosner}},\
	}\bibfield  {title} {\bibinfo {title} {{Modified Kagome Physics in the
				Natural Spin-$1/2$ Kagome Lattice Systems: Kapellasite
				${\mathrm{Cu}}_{3}\mathrm{Zn}(\mathrm{OH}{)}_{6}{\mathrm{Cl}}_{2}$ and
				Haydeeite
				${\mathrm{Cu}}_{3}\mathrm{Mg}(\mathrm{OH}{)}_{6}{\mathrm{Cl}}_{2}$}},\ }\href
	{https://doi.org/10.1103/PhysRevLett.101.106403} {\bibfield  {journal}
		{\bibinfo  {journal} {Phys. Rev. Lett.}\ }\textbf {\bibinfo {volume} {101}},\
		\bibinfo {pages} {106403} (\bibinfo {year} {2008})}\BibitemShut {NoStop}%
	\bibitem [{\citenamefont {Colman}\ \emph {et~al.}(2010)\citenamefont {Colman},
		\citenamefont {Sinclair},\ and\ \citenamefont
		{Wills}}]{Colman:haydeeiteSynthesis}%
	\BibitemOpen
	\bibfield  {author} {\bibinfo {author} {\bibfnamefont {R.}~\bibnamefont
			{Colman}}, \bibinfo {author} {\bibfnamefont {A.}~\bibnamefont {Sinclair}},\
		and\ \bibinfo {author} {\bibfnamefont {A.}~\bibnamefont {Wills}},\ }\bibfield
	{title} {\bibinfo {title} {{Comparisons between Haydeeite,
				$\alpha-Cu_3Mg(OD)_6Cl_2$, and Kapellasite, $\alpha-Cu_3Zn(OD)_6Cl_2$,
				Isostructural S = 1/2 Kagome Magnets}},\ }\href
	{https://doi.org/10.1021/cm101594c} {\bibfield  {journal} {\bibinfo
			{journal} {Chemistry of Materials}\ }\textbf {\bibinfo {volume} {22}},\
		\bibinfo {pages} {5774} (\bibinfo {year} {2010})},\ \Eprint
	{https://arxiv.org/abs/https://doi.org/10.1021/cm101594c}
	{https://doi.org/10.1021/cm101594c} \BibitemShut {NoStop}%
	\bibitem [{\citenamefont {Han}\ \emph {et~al.}(2014)\citenamefont {Han},
		\citenamefont {Singleton},\ and\ \citenamefont {Schlueter}}]{Han:Barlowite}%
	\BibitemOpen
	\bibfield  {author} {\bibinfo {author} {\bibfnamefont {T.-H.}\ \bibnamefont
			{Han}}, \bibinfo {author} {\bibfnamefont {J.}~\bibnamefont {Singleton}},\
		and\ \bibinfo {author} {\bibfnamefont {J.~A.}\ \bibnamefont {Schlueter}},\
	}\bibfield  {title} {\bibinfo {title} {Barlowite: A spin-$1/2$
			antiferromagnet with a geometrically perfect kagome motif},\ }\href
	{https://doi.org/10.1103/PhysRevLett.113.227203} {\bibfield  {journal}
		{\bibinfo  {journal} {Phys. Rev. Lett.}\ }\textbf {\bibinfo {volume} {113}},\
		\bibinfo {pages} {227203} (\bibinfo {year} {2014})}\BibitemShut {NoStop}%
	\bibitem [{\citenamefont {Yue}\ \emph {et~al.}(2018)\citenamefont {Yue},
		\citenamefont {Ouyang}, \citenamefont {Wang}, \citenamefont {Wang},
		\citenamefont {Xia},\ and\ \citenamefont {He}}]{Yue:Barlowite}%
	\BibitemOpen
	\bibfield  {author} {\bibinfo {author} {\bibfnamefont {X.~Y.}\ \bibnamefont
			{Yue}}, \bibinfo {author} {\bibfnamefont {Z.~W.}\ \bibnamefont {Ouyang}},
		\bibinfo {author} {\bibfnamefont {J.~F.}\ \bibnamefont {Wang}}, \bibinfo
		{author} {\bibfnamefont {Z.~X.}\ \bibnamefont {Wang}}, \bibinfo {author}
		{\bibfnamefont {Z.~C.}\ \bibnamefont {Xia}},\ and\ \bibinfo {author}
		{\bibfnamefont {Z.~Z.}\ \bibnamefont {He}},\ }\bibfield  {title} {\bibinfo
		{title} {Magnetization and esr studies on
			$\mathrm{C}{\mathrm{u}}_{4}{(\mathrm{OH})}_{6}\mathrm{FCl}$: An
			antiferromagnet with a kagome lattice},\ }\href
	{https://doi.org/10.1103/PhysRevB.97.054417} {\bibfield  {journal} {\bibinfo
			{journal} {Phys. Rev. B}\ }\textbf {\bibinfo {volume} {97}},\ \bibinfo
		{pages} {054417} (\bibinfo {year} {2018})}\BibitemShut {NoStop}%
	\bibitem [{\citenamefont {Greedan}\ \emph {et~al.}(1986)\citenamefont
		{Greedan}, \citenamefont {Sato}, \citenamefont {Yan},\ and\ \citenamefont
		{Razavi}}]{Greedan:YMoOglass}%
	\BibitemOpen
	\bibfield  {author} {\bibinfo {author} {\bibfnamefont {J.}~\bibnamefont
			{Greedan}}, \bibinfo {author} {\bibfnamefont {M.}~\bibnamefont {Sato}},
		\bibinfo {author} {\bibfnamefont {X.}~\bibnamefont {Yan}},\ and\ \bibinfo
		{author} {\bibfnamefont {F.}~\bibnamefont {Razavi}},\ }\bibfield  {title}
	{\bibinfo {title} {{Spin-glass-like behavior in $Y_2Mo_2O_7$, a concentrated,
				crystalline system with negligible apparent disorder}},\ }\href
	{https://doi.org/https://doi.org/10.1016/0038-1098(86)90652-6} {\bibfield
		{journal} {\bibinfo  {journal} {Solid State Communications}\ }\textbf
		{\bibinfo {volume} {59}},\ \bibinfo {pages} {895} (\bibinfo {year}
		{1986})}\BibitemShut {NoStop}%
	\bibitem [{\citenamefont {Thygesen}\ \emph {et~al.}(2017)\citenamefont
		{Thygesen}, \citenamefont {Paddison}, \citenamefont {Zhang}, \citenamefont
		{Beyer}, \citenamefont {Chapman}, \citenamefont {Playford}, \citenamefont
		{Tucker}, \citenamefont {Keen}, \citenamefont {Hayward},\ and\ \citenamefont
		{Goodwin}}]{Thygesen:YMOdisplacements}%
	\BibitemOpen
	\bibfield  {author} {\bibinfo {author} {\bibfnamefont {P.~M.~M.}\
			\bibnamefont {Thygesen}}, \bibinfo {author} {\bibfnamefont {J.~A.~M.}\
			\bibnamefont {Paddison}}, \bibinfo {author} {\bibfnamefont {R.}~\bibnamefont
			{Zhang}}, \bibinfo {author} {\bibfnamefont {K.~A.}\ \bibnamefont {Beyer}},
		\bibinfo {author} {\bibfnamefont {K.~W.}\ \bibnamefont {Chapman}}, \bibinfo
		{author} {\bibfnamefont {H.~Y.}\ \bibnamefont {Playford}}, \bibinfo {author}
		{\bibfnamefont {M.~G.}\ \bibnamefont {Tucker}}, \bibinfo {author}
		{\bibfnamefont {D.~A.}\ \bibnamefont {Keen}}, \bibinfo {author}
		{\bibfnamefont {M.~A.}\ \bibnamefont {Hayward}},\ and\ \bibinfo {author}
		{\bibfnamefont {A.~L.}\ \bibnamefont {Goodwin}},\ }\bibfield  {title}
	{\bibinfo {title} {{Orbital Dimer Model for the Spin-Glass State in
				${\mathrm{Y}}_{2}{\mathrm{Mo}}_{2}{\mathrm{O}}_{7}$}},\ }\href
	{https://doi.org/10.1103/PhysRevLett.118.067201} {\bibfield  {journal}
		{\bibinfo  {journal} {Phys. Rev. Lett.}\ }\textbf {\bibinfo {volume} {118}},\
		\bibinfo {pages} {067201} (\bibinfo {year} {2017})}\BibitemShut {NoStop}%
	\bibitem [{\citenamefont {LaBarre}\ \emph {et~al.}(2021)\citenamefont
		{LaBarre}, \citenamefont {Phelan}, \citenamefont {Xin}, \citenamefont {Ye},
		\citenamefont {Besara}, \citenamefont {Siegrist}, \citenamefont {Syzranov},
		\citenamefont {Rosenkranz},\ and\ \citenamefont
		{Ramirez}}]{LaBarre:surfboards}%
	\BibitemOpen
	\bibfield  {author} {\bibinfo {author} {\bibfnamefont {P.~G.}\ \bibnamefont
			{LaBarre}}, \bibinfo {author} {\bibfnamefont {D.}~\bibnamefont {Phelan}},
		\bibinfo {author} {\bibfnamefont {Y.}~\bibnamefont {Xin}}, \bibinfo {author}
		{\bibfnamefont {F.}~\bibnamefont {Ye}}, \bibinfo {author} {\bibfnamefont
			{T.}~\bibnamefont {Besara}}, \bibinfo {author} {\bibfnamefont
			{T.}~\bibnamefont {Siegrist}}, \bibinfo {author} {\bibfnamefont {S.~V.}\
			\bibnamefont {Syzranov}}, \bibinfo {author} {\bibfnamefont {S.}~\bibnamefont
			{Rosenkranz}},\ and\ \bibinfo {author} {\bibfnamefont {A.~P.}\ \bibnamefont
			{Ramirez}},\ }\bibfield  {title} {\bibinfo {title} {{Fluctuation-induced
				interactions and the spin-glass transition in
				${\mathrm{Fe}}_{2}{\mathrm{TiO}}_{5}$}},\ }\href
	{https://doi.org/10.1103/PhysRevB.103.L220404} {\bibfield  {journal}
		{\bibinfo  {journal} {Phys. Rev. B}\ }\textbf {\bibinfo {volume} {103}},\
		\bibinfo {pages} {L220404} (\bibinfo {year} {2021})}\BibitemShut {NoStop}%
	\bibitem [{\citenamefont {Kancko}\ \emph {et~al.}(2023)\citenamefont {Kancko},
		\citenamefont {Giester},\ and\ \citenamefont
		{Colman}}]{KanckoColman:mixingNaCdCoF}%
	\BibitemOpen
	\bibfield  {author} {\bibinfo {author} {\bibfnamefont {A.}~\bibnamefont
			{Kancko}}, \bibinfo {author} {\bibfnamefont {G.}~\bibnamefont {Giester}},\
		and\ \bibinfo {author} {\bibfnamefont {R.~H.}\ \bibnamefont {Colman}},\
	}\bibfield  {title} {\bibinfo {title} {{Structural and spin-glass properties
				of single crystal $J_{eff} = 1/2$ pyrochlore antiferromagnet $NaCdCo_2F_7$:
				correlating Tf with magnetic-bond-disorder}},\ }\href
	{https://doi.org/10.1088/1402-4896/acdeb7} {\bibfield  {journal} {\bibinfo
			{journal} {Physica Scripta}\ }\textbf {\bibinfo {volume} {98}},\ \bibinfo
		{pages} {075947} (\bibinfo {year} {2023})}\BibitemShut {NoStop}%
	\bibitem [{\citenamefont {Goodenough}(1963)}]{Goodenough:book}%
	\BibitemOpen
	\bibfield  {author} {\bibinfo {author} {\bibfnamefont {J.~B.}\ \bibnamefont
			{Goodenough}},\ }\href@noop {} {\emph {\bibinfo {title} {Magnetism and the
				chemical bond}}}\ (\bibinfo  {publisher} {R. E. Krieger Pub. Co.},\ \bibinfo
	{year} {1963})\BibitemShut {NoStop}%
	\bibitem [{\citenamefont {Martinho}\ \emph {et~al.}(2001)\citenamefont
		{Martinho}, \citenamefont {Moreno}, \citenamefont {Sanjurjo}, \citenamefont
		{Rettori}, \citenamefont {Garc\'{\i}a-Adeva}, \citenamefont {Huber},
		\citenamefont {Oseroff}, \citenamefont {Ratcliff}, \citenamefont {Cheong},
		\citenamefont {Pagliuso}, \citenamefont {Sarrao},\ and\ \citenamefont
		{Martins}}]{Martinho:ZnCrOsubstitution}%
	\BibitemOpen
	\bibfield  {author} {\bibinfo {author} {\bibfnamefont {H.}~\bibnamefont
			{Martinho}}, \bibinfo {author} {\bibfnamefont {N.~O.}\ \bibnamefont
			{Moreno}}, \bibinfo {author} {\bibfnamefont {J.~A.}\ \bibnamefont
			{Sanjurjo}}, \bibinfo {author} {\bibfnamefont {C.}~\bibnamefont {Rettori}},
		\bibinfo {author} {\bibfnamefont {A.~J.}\ \bibnamefont {Garc\'{\i}a-Adeva}},
		\bibinfo {author} {\bibfnamefont {D.~L.}\ \bibnamefont {Huber}}, \bibinfo
		{author} {\bibfnamefont {S.~B.}\ \bibnamefont {Oseroff}}, \bibinfo {author}
		{\bibfnamefont {W.}~\bibnamefont {Ratcliff}}, \bibinfo {author}
		{\bibfnamefont {S.-W.}\ \bibnamefont {Cheong}}, \bibinfo {author}
		{\bibfnamefont {P.~G.}\ \bibnamefont {Pagliuso}}, \bibinfo {author}
		{\bibfnamefont {J.~L.}\ \bibnamefont {Sarrao}},\ and\ \bibinfo {author}
		{\bibfnamefont {G.~B.}\ \bibnamefont {Martins}},\ }\bibfield  {title}
	{\bibinfo {title} {{Magnetic properties of the frustrated antiferromagnetic
				spinel ${\mathrm{ZnCr}}_{2}{\mathrm{O}}_{4}$ and the spin-glass
				${\mathrm{Zn}}_{1\ensuremath{-}x}{\mathrm{Cd}}_{x}{\mathrm{Cr}}_{2}{\mathrm{O}}_{4}$
				$(x=0.05,0.10)$}},\ }\href {https://doi.org/10.1103/PhysRevB.64.024408}
	{\bibfield  {journal} {\bibinfo  {journal} {Phys. Rev. B}\ }\textbf {\bibinfo
			{volume} {64}},\ \bibinfo {pages} {024408} (\bibinfo {year}
		{2001})}\BibitemShut {NoStop}%
	\bibitem [{\citenamefont {Booth}\ \emph {et~al.}(2000)\citenamefont {Booth},
		\citenamefont {Gardner}, \citenamefont {Kwei}, \citenamefont {Heffner},
		\citenamefont {Bridges},\ and\ \citenamefont
		{Subramanian}}]{BoothSubramanian:YMOdisorder}%
	\BibitemOpen
	\bibfield  {author} {\bibinfo {author} {\bibfnamefont {C.~H.}\ \bibnamefont
			{Booth}}, \bibinfo {author} {\bibfnamefont {J.~S.}\ \bibnamefont {Gardner}},
		\bibinfo {author} {\bibfnamefont {G.~H.}\ \bibnamefont {Kwei}}, \bibinfo
		{author} {\bibfnamefont {R.~H.}\ \bibnamefont {Heffner}}, \bibinfo {author}
		{\bibfnamefont {F.}~\bibnamefont {Bridges}},\ and\ \bibinfo {author}
		{\bibfnamefont {M.~A.}\ \bibnamefont {Subramanian}},\ }\bibfield  {title}
	{\bibinfo {title} {Local lattice disorder in the geometrically frustrated
			spin-glass pyrochlore ${\mathrm{y}}_{2}{\mathrm{mo}}_{2}{\mathrm{o}}_{7}$},\
	}\href {https://doi.org/10.1103/PhysRevB.62.R755} {\bibfield  {journal}
		{\bibinfo  {journal} {Phys. Rev. B}\ }\textbf {\bibinfo {volume} {62}},\
		\bibinfo {pages} {R755} (\bibinfo {year} {2000})}\BibitemShut {NoStop}%
	\bibitem [{\citenamefont {Blanc}\ \emph {et~al.}(2018)\citenamefont {Blanc},
		\citenamefont {Trinh}, \citenamefont {Dong}, \citenamefont {Bai},
		\citenamefont {Aczel}, \citenamefont {Mourigal}, \citenamefont {Balents},
		\citenamefont {Siegrist},\ and\ \citenamefont
		{Ramirez}}]{BlancRamirez:SaltEffectiveD}%
	\BibitemOpen
	\bibfield  {author} {\bibinfo {author} {\bibfnamefont {N.}~\bibnamefont
			{Blanc}}, \bibinfo {author} {\bibfnamefont {J.}~\bibnamefont {Trinh}},
		\bibinfo {author} {\bibfnamefont {L.}~\bibnamefont {Dong}}, \bibinfo {author}
		{\bibfnamefont {X.}~\bibnamefont {Bai}}, \bibinfo {author} {\bibfnamefont
			{A.~A.}\ \bibnamefont {Aczel}}, \bibinfo {author} {\bibfnamefont
			{M.}~\bibnamefont {Mourigal}}, \bibinfo {author} {\bibfnamefont
			{L.}~\bibnamefont {Balents}}, \bibinfo {author} {\bibfnamefont
			{T.}~\bibnamefont {Siegrist}},\ and\ \bibinfo {author} {\bibfnamefont
			{A.}~\bibnamefont {Ramirez}},\ }\bibfield  {title} {\bibinfo {title} {Quantum
			criticality among entangled spin chains},\ }\href@noop {} {\bibfield
		{journal} {\bibinfo  {journal} {Nature Physics}\ }\textbf {\bibinfo {volume}
			{14}},\ \bibinfo {pages} {273} (\bibinfo {year} {2018})}\BibitemShut
	{NoStop}%
	\bibitem [{\citenamefont {Fernandez}\ \emph {et~al.}(2016)\citenamefont
		{Fernandez}, \citenamefont {Marinari}, \citenamefont {Martin-Mayor},
		\citenamefont {Parisi},\ and\ \citenamefont
		{Ruiz-Lorenzo}}]{FernandezParisi:Ising2Dtransition}%
	\BibitemOpen
	\bibfield  {author} {\bibinfo {author} {\bibfnamefont {L.~A.}\ \bibnamefont
			{Fernandez}}, \bibinfo {author} {\bibfnamefont {E.}~\bibnamefont {Marinari}},
		\bibinfo {author} {\bibfnamefont {V.}~\bibnamefont {Martin-Mayor}}, \bibinfo
		{author} {\bibfnamefont {G.}~\bibnamefont {Parisi}},\ and\ \bibinfo {author}
		{\bibfnamefont {J.~J.}\ \bibnamefont {Ruiz-Lorenzo}},\ }\bibfield  {title}
	{\bibinfo {title} {Universal critical behavior of the two-dimensional ising
			spin glass},\ }\href {https://doi.org/10.1103/PhysRevB.94.024402} {\bibfield
		{journal} {\bibinfo  {journal} {Phys. Rev. B}\ }\textbf {\bibinfo {volume}
			{94}},\ \bibinfo {pages} {024402} (\bibinfo {year} {2016})}\BibitemShut
	{NoStop}%
	\bibitem [{\citenamefont {Dekker}\ \emph {et~al.}(1988)\citenamefont {Dekker},
		\citenamefont {Arts}, \citenamefont {de~Wijn}, \citenamefont {van
			Duyneveldt},\ and\ \citenamefont {Mydosh}}]{DekkerMydosh:SGrelaxation}%
	\BibitemOpen
	\bibfield  {author} {\bibinfo {author} {\bibfnamefont {C.}~\bibnamefont
			{Dekker}}, \bibinfo {author} {\bibfnamefont {A.~F.~M.}\ \bibnamefont {Arts}},
		\bibinfo {author} {\bibfnamefont {H.~W.}\ \bibnamefont {de~Wijn}}, \bibinfo
		{author} {\bibfnamefont {A.~J.}\ \bibnamefont {van Duyneveldt}},\ and\
		\bibinfo {author} {\bibfnamefont {J.~A.}\ \bibnamefont {Mydosh}},\ }\bibfield
	{title} {\bibinfo {title} {{Activated Dynamics in the Two-Dimensional Ising
				Spin-Glass
				${\mathrm{Rb}}_{2}{\mathrm{Cu}}_{1\ensuremath{-}x}{\mathrm{Co}}_{x}{\mathrm{F}}_{4}$}},\
	}\href {https://doi.org/10.1103/PhysRevLett.61.1780} {\bibfield  {journal}
		{\bibinfo  {journal} {Phys. Rev. Lett.}\ }\textbf {\bibinfo {volume} {61}},\
		\bibinfo {pages} {1780} (\bibinfo {year} {1988})}\BibitemShut {NoStop}%
	\bibitem [{\citenamefont {Shimizu}\ \emph {et~al.}(2003)\citenamefont
		{Shimizu}, \citenamefont {Miyagawa}, \citenamefont {Kanoda}, \citenamefont
		{Maesato},\ and\ \citenamefont {Saito}}]{Shimizu:QSLclaim}%
	\BibitemOpen
	\bibfield  {author} {\bibinfo {author} {\bibfnamefont {Y.}~\bibnamefont
			{Shimizu}}, \bibinfo {author} {\bibfnamefont {K.}~\bibnamefont {Miyagawa}},
		\bibinfo {author} {\bibfnamefont {K.}~\bibnamefont {Kanoda}}, \bibinfo
		{author} {\bibfnamefont {M.}~\bibnamefont {Maesato}},\ and\ \bibinfo {author}
		{\bibfnamefont {G.}~\bibnamefont {Saito}},\ }\bibfield  {title} {\bibinfo
		{title} {Spin liquid state in an organic mott insulator with a triangular
			lattice},\ }\href {https://doi.org/10.1103/PhysRevLett.91.107001} {\bibfield
		{journal} {\bibinfo  {journal} {Phys. Rev. Lett.}\ }\textbf {\bibinfo
			{volume} {91}},\ \bibinfo {pages} {107001} (\bibinfo {year}
		{2003})}\BibitemShut {NoStop}%
	\bibitem [{\citenamefont {Ma}\ \emph {et~al.}(2018)\citenamefont {Ma},
		\citenamefont {Wang}, \citenamefont {Dong}, \citenamefont {Zhang},
		\citenamefont {Li}, \citenamefont {Zheng}, \citenamefont {Yu}, \citenamefont
		{Wang}, \citenamefont {Che}, \citenamefont {Ran}, \citenamefont {Bao},
		\citenamefont {Cai}, \citenamefont {\ifmmode~\check{C}\else
			\v{C}\fi{}erm\'ak}, \citenamefont {Schneidewind}, \citenamefont {Yano},
		\citenamefont {Gardner}, \citenamefont {Lu}, \citenamefont {Yu},
		\citenamefont {Liu}, \citenamefont {Li}, \citenamefont {Li},\ and\
		\citenamefont {Wen}}]{Ma:QSLclaim}%
	\BibitemOpen
	\bibfield  {author} {\bibinfo {author} {\bibfnamefont {Z.}~\bibnamefont
			{Ma}}, \bibinfo {author} {\bibfnamefont {J.}~\bibnamefont {Wang}}, \bibinfo
		{author} {\bibfnamefont {Z.-Y.}\ \bibnamefont {Dong}}, \bibinfo {author}
		{\bibfnamefont {J.}~\bibnamefont {Zhang}}, \bibinfo {author} {\bibfnamefont
			{S.}~\bibnamefont {Li}}, \bibinfo {author} {\bibfnamefont {S.-H.}\
			\bibnamefont {Zheng}}, \bibinfo {author} {\bibfnamefont {Y.}~\bibnamefont
			{Yu}}, \bibinfo {author} {\bibfnamefont {W.}~\bibnamefont {Wang}}, \bibinfo
		{author} {\bibfnamefont {L.}~\bibnamefont {Che}}, \bibinfo {author}
		{\bibfnamefont {K.}~\bibnamefont {Ran}}, \bibinfo {author} {\bibfnamefont
			{S.}~\bibnamefont {Bao}}, \bibinfo {author} {\bibfnamefont {Z.}~\bibnamefont
			{Cai}}, \bibinfo {author} {\bibfnamefont {P.}~\bibnamefont
			{\ifmmode~\check{C}\else \v{C}\fi{}erm\'ak}}, \bibinfo {author}
		{\bibfnamefont {A.}~\bibnamefont {Schneidewind}}, \bibinfo {author}
		{\bibfnamefont {S.}~\bibnamefont {Yano}}, \bibinfo {author} {\bibfnamefont
			{J.~S.}\ \bibnamefont {Gardner}}, \bibinfo {author} {\bibfnamefont
			{X.}~\bibnamefont {Lu}}, \bibinfo {author} {\bibfnamefont {S.-L.}\
			\bibnamefont {Yu}}, \bibinfo {author} {\bibfnamefont {J.-M.}\ \bibnamefont
			{Liu}}, \bibinfo {author} {\bibfnamefont {S.}~\bibnamefont {Li}}, \bibinfo
		{author} {\bibfnamefont {J.-X.}\ \bibnamefont {Li}},\ and\ \bibinfo {author}
		{\bibfnamefont {J.}~\bibnamefont {Wen}},\ }\bibfield  {title} {\bibinfo
		{title} {{Spin-Glass Ground State in a Triangular-Lattice Compound
				${\mathrm{YbZnGaO}}_{4}$}},\ }\href
	{https://doi.org/10.1103/PhysRevLett.120.087201} {\bibfield  {journal}
		{\bibinfo  {journal} {Phys. Rev. Lett.}\ }\textbf {\bibinfo {volume} {120}},\
		\bibinfo {pages} {087201} (\bibinfo {year} {2018})}\BibitemShut {NoStop}%
	\bibitem [{\citenamefont {Kitagawa}\ \emph {et~al.}(2018)\citenamefont
		{Kitagawa}, \citenamefont {Takayama}, \citenamefont {Matsumoto},
		\citenamefont {Kato}, \citenamefont {Takano}, \citenamefont {Kishimoto},
		\citenamefont {Bette}, \citenamefont {Dinnebier}, \citenamefont {Jackeli},\
		and\ \citenamefont {Takagi}}]{Kitagawa:QSLclaim}%
	\BibitemOpen
	\bibfield  {author} {\bibinfo {author} {\bibfnamefont {K.}~\bibnamefont
			{Kitagawa}}, \bibinfo {author} {\bibfnamefont {T.}~\bibnamefont {Takayama}},
		\bibinfo {author} {\bibfnamefont {Y.}~\bibnamefont {Matsumoto}}, \bibinfo
		{author} {\bibfnamefont {A.}~\bibnamefont {Kato}}, \bibinfo {author}
		{\bibfnamefont {R.}~\bibnamefont {Takano}}, \bibinfo {author} {\bibfnamefont
			{Y.}~\bibnamefont {Kishimoto}}, \bibinfo {author} {\bibfnamefont
			{S.}~\bibnamefont {Bette}}, \bibinfo {author} {\bibfnamefont
			{R.}~\bibnamefont {Dinnebier}}, \bibinfo {author} {\bibfnamefont
			{G.}~\bibnamefont {Jackeli}},\ and\ \bibinfo {author} {\bibfnamefont
			{H.}~\bibnamefont {Takagi}},\ }\bibfield  {title} {\bibinfo {title} {{A
				spin--orbital-entangled quantum liquid on a honeycomb lattice}},\ }\href@noop
	{} {\bibfield  {journal} {\bibinfo  {journal} {Nature}\ }\textbf {\bibinfo
			{volume} {554}},\ \bibinfo {pages} {341} (\bibinfo {year}
		{2018})}\BibitemShut {NoStop}%
	\bibitem [{\citenamefont {Fujihala}\ \emph {et~al.}(2012)\citenamefont
		{Fujihala}, \citenamefont {Zheng}, \citenamefont {Oohara}, \citenamefont
		{Morodomi}, \citenamefont {Kawae}, \citenamefont {Matsuo},\ and\
		\citenamefont {Kindo}}]{Fujihala:SpinIce}%
	\BibitemOpen
	\bibfield  {author} {\bibinfo {author} {\bibfnamefont {M.}~\bibnamefont
			{Fujihala}}, \bibinfo {author} {\bibfnamefont {X.~G.}\ \bibnamefont {Zheng}},
		\bibinfo {author} {\bibfnamefont {Y.}~\bibnamefont {Oohara}}, \bibinfo
		{author} {\bibfnamefont {H.}~\bibnamefont {Morodomi}}, \bibinfo {author}
		{\bibfnamefont {T.}~\bibnamefont {Kawae}}, \bibinfo {author} {\bibfnamefont
			{A.}~\bibnamefont {Matsuo}},\ and\ \bibinfo {author} {\bibfnamefont
			{K.}~\bibnamefont {Kindo}},\ }\bibfield  {title} {\bibinfo {title}
		{{Short-range correlations and persistent spin fluctuations in the
				undistorted kagome lattice Ising antiferromagnet
				Co${}_{3}$Mg(OH)${}_{6}$Cl${}_{2}$}},\ }\href
	{https://doi.org/10.1103/PhysRevB.85.012402} {\bibfield  {journal} {\bibinfo
			{journal} {Phys. Rev. B}\ }\textbf {\bibinfo {volume} {85}},\ \bibinfo
		{pages} {012402} (\bibinfo {year} {2012})}\BibitemShut {NoStop}%
	\bibitem [{\citenamefont {Zhong}\ \emph {et~al.}(2020)\citenamefont {Zhong},
		\citenamefont {Guo},\ and\ \citenamefont {Cava}}]{Zhong:LayeredTriangular}%
	\BibitemOpen
	\bibfield  {author} {\bibinfo {author} {\bibfnamefont {R.}~\bibnamefont
			{Zhong}}, \bibinfo {author} {\bibfnamefont {S.}~\bibnamefont {Guo}},\ and\
		\bibinfo {author} {\bibfnamefont {R.~J.}\ \bibnamefont {Cava}},\ }\bibfield
	{title} {\bibinfo {title} {{Frustrated magnetism in the layered triangular
				lattice materials ${\mathrm{K}}_{2}\mathrm{Co}{({\mathrm{SeO}}_{3})}_{2}$ and
				${\mathrm{Rb}}_{2}\mathrm{Co}{({\mathrm{SeO}}_{3})}_{2}$}},\ }\href
	{https://doi.org/10.1103/PhysRevMaterials.4.084406} {\bibfield  {journal}
		{\bibinfo  {journal} {Phys. Rev. Mater.}\ }\textbf {\bibinfo {volume} {4}},\
		\bibinfo {pages} {084406} (\bibinfo {year} {2020})}\BibitemShut {NoStop}%
	\bibitem [{\citenamefont {Harris}\ \emph {et~al.}(1997)\citenamefont {Harris},
		\citenamefont {Bramwell}, \citenamefont {McMorrow}, \citenamefont {Zeiske},\
		and\ \citenamefont {Godfrey}}]{Harris:HoTiOice}%
	\BibitemOpen
	\bibfield  {author} {\bibinfo {author} {\bibfnamefont {M.~J.}\ \bibnamefont
			{Harris}}, \bibinfo {author} {\bibfnamefont {S.~T.}\ \bibnamefont
			{Bramwell}}, \bibinfo {author} {\bibfnamefont {D.~F.}\ \bibnamefont
			{McMorrow}}, \bibinfo {author} {\bibfnamefont {T.}~\bibnamefont {Zeiske}},\
		and\ \bibinfo {author} {\bibfnamefont {K.~W.}\ \bibnamefont {Godfrey}},\
	}\bibfield  {title} {\bibinfo {title} {{Geometrical Frustration in the
				Ferromagnetic Pyrochlore ${\mathrm{Ho}}_{2}{\mathrm{Ti}}_{2}{O}_{7}$}},\
	}\href {https://doi.org/10.1103/PhysRevLett.79.2554} {\bibfield  {journal}
		{\bibinfo  {journal} {Phys. Rev. Lett.}\ }\textbf {\bibinfo {volume} {79}},\
		\bibinfo {pages} {2554} (\bibinfo {year} {1997})}\BibitemShut {NoStop}%
	\bibitem [{\citenamefont {Ramirez}\ \emph {et~al.}(1999)\citenamefont
		{Ramirez}, \citenamefont {Hayashi}, \citenamefont {Cava}, \citenamefont
		{Siddharthan},\ and\ \citenamefont {Shastry}}]{Ramirez:iceEntropy}%
	\BibitemOpen
	\bibfield  {author} {\bibinfo {author} {\bibfnamefont {A.~P.}\ \bibnamefont
			{Ramirez}}, \bibinfo {author} {\bibfnamefont {A.}~\bibnamefont {Hayashi}},
		\bibinfo {author} {\bibfnamefont {R.~J.}\ \bibnamefont {Cava}}, \bibinfo
		{author} {\bibfnamefont {R.}~\bibnamefont {Siddharthan}},\ and\ \bibinfo
		{author} {\bibfnamefont {B.}~\bibnamefont {Shastry}},\ }\bibfield  {title}
	{\bibinfo {title} {Zero-point entropy in ‘spin ice’},\ }\href@noop {}
	{\bibfield  {journal} {\bibinfo  {journal} {Nature}\ }\textbf {\bibinfo
			{volume} {399}},\ \bibinfo {pages} {333} (\bibinfo {year}
		{1999})}\BibitemShut {NoStop}%
	\bibitem [{\citenamefont {Snyder}\ \emph {et~al.}(2004)\citenamefont {Snyder},
		\citenamefont {Ueland}, \citenamefont {Mizel}, \citenamefont {Slusky},
		\citenamefont {Karunadasa}, \citenamefont {Cava},\ and\ \citenamefont
		{Schiffer}}]{SnyderSchiffer:ices}%
	\BibitemOpen
	\bibfield  {author} {\bibinfo {author} {\bibfnamefont {J.}~\bibnamefont
			{Snyder}}, \bibinfo {author} {\bibfnamefont {B.~G.}\ \bibnamefont {Ueland}},
		\bibinfo {author} {\bibfnamefont {A.}~\bibnamefont {Mizel}}, \bibinfo
		{author} {\bibfnamefont {J.~S.}\ \bibnamefont {Slusky}}, \bibinfo {author}
		{\bibfnamefont {H.}~\bibnamefont {Karunadasa}}, \bibinfo {author}
		{\bibfnamefont {R.~J.}\ \bibnamefont {Cava}},\ and\ \bibinfo {author}
		{\bibfnamefont {P.}~\bibnamefont {Schiffer}},\ }\bibfield  {title} {\bibinfo
		{title} {{Quantum and thermal spin relaxation in the diluted spin ice
				${\mathrm{Dy}}_{2\ensuremath{-}x}{M}_{x}{\mathrm{Ti}}_{2}{\mathrm{O}}_{7}$
				$(M=\mathrm{Lu},\mathrm{Y})$}},\ }\href
	{https://doi.org/10.1103/PhysRevB.70.184431} {\bibfield  {journal} {\bibinfo
			{journal} {Phys. Rev. B}\ }\textbf {\bibinfo {volume} {70}},\ \bibinfo
		{pages} {184431} (\bibinfo {year} {2004})}\BibitemShut {NoStop}%
	\bibitem [{\citenamefont {den Hertog}\ and\ \citenamefont
		{Gingras}(2000)}]{DenHertogGingras:spinIce}%
	\BibitemOpen
	\bibfield  {author} {\bibinfo {author} {\bibfnamefont {B.~C.}\ \bibnamefont
			{den Hertog}}\ and\ \bibinfo {author} {\bibfnamefont {M.~J.~P.}\ \bibnamefont
			{Gingras}},\ }\bibfield  {title} {\bibinfo {title} {Dipolar interactions and
			origin of spin ice in ising pyrochlore magnets},\ }\href
	{https://doi.org/10.1103/PhysRevLett.84.3430} {\bibfield  {journal} {\bibinfo
			{journal} {Phys. Rev. Lett.}\ }\textbf {\bibinfo {volume} {84}},\ \bibinfo
		{pages} {3430} (\bibinfo {year} {2000})}\BibitemShut {NoStop}%
	\bibitem [{\citenamefont {Nirmala}\ \emph {et~al.}(2017)\citenamefont
		{Nirmala}, \citenamefont {Jang}, \citenamefont {Sim}, \citenamefont {Cho},
		\citenamefont {Lee}, \citenamefont {Yang}, \citenamefont {Lee}, \citenamefont
		{Ibberson}, \citenamefont {Kakurai}, \citenamefont {Matsuda} \emph
		{et~al.}}]{Nirmala:SGinCuAlO}%
	\BibitemOpen
	\bibfield  {author} {\bibinfo {author} {\bibfnamefont {R.}~\bibnamefont
			{Nirmala}}, \bibinfo {author} {\bibfnamefont {K.-H.}\ \bibnamefont {Jang}},
		\bibinfo {author} {\bibfnamefont {H.}~\bibnamefont {Sim}}, \bibinfo {author}
		{\bibfnamefont {H.}~\bibnamefont {Cho}}, \bibinfo {author} {\bibfnamefont
			{J.}~\bibnamefont {Lee}}, \bibinfo {author} {\bibfnamefont {N.-G.}\
			\bibnamefont {Yang}}, \bibinfo {author} {\bibfnamefont {S.}~\bibnamefont
			{Lee}}, \bibinfo {author} {\bibfnamefont {R.}~\bibnamefont {Ibberson}},
		\bibinfo {author} {\bibfnamefont {K.}~\bibnamefont {Kakurai}}, \bibinfo
		{author} {\bibfnamefont {M.}~\bibnamefont {Matsuda}}, \emph {et~al.},\
	}\bibfield  {title} {\bibinfo {title} {{Spin glass behavior in frustrated
				quantum spin system $CuAl_2O_4$ with a possible orbital liquid state}},\
	}\href@noop {} {\bibfield  {journal} {\bibinfo  {journal} {Journal of
				Physics: Condensed Matter}\ }\textbf {\bibinfo {volume} {29}},\ \bibinfo
		{pages} {13LT01} (\bibinfo {year} {2017})}\BibitemShut {NoStop}%
	\bibitem [{\citenamefont {Fenner}\ \emph {et~al.}(2009)\citenamefont {Fenner},
		\citenamefont {Wills}, \citenamefont {Bramwell}, \citenamefont {Dahlberg},\
		and\ \citenamefont {Schiffer}}]{Fenner:EntropyCuGlasses}%
	\BibitemOpen
	\bibfield  {author} {\bibinfo {author} {\bibfnamefont {L.~A.}\ \bibnamefont
			{Fenner}}, \bibinfo {author} {\bibfnamefont {A.~S.}\ \bibnamefont {Wills}},
		\bibinfo {author} {\bibfnamefont {S.~T.}\ \bibnamefont {Bramwell}}, \bibinfo
		{author} {\bibfnamefont {M.}~\bibnamefont {Dahlberg}},\ and\ \bibinfo
		{author} {\bibfnamefont {P.}~\bibnamefont {Schiffer}},\ }\bibfield  {title}
	{\bibinfo {title} {{Zero-point entropy of the spinel spin glasses $CuGa_2O_4$
				and $CuAl_2O_4$}},\ }\href {https://doi.org/10.1088/1742-6596/145/1/012029}
	{\bibfield  {journal} {\bibinfo  {journal} {Journal of Physics: Conference
				Series}\ }\textbf {\bibinfo {volume} {145}},\ \bibinfo {pages} {012029}
		(\bibinfo {year} {2009})}\BibitemShut {NoStop}%
	\bibitem [{\citenamefont {Cooke}\ \emph {et~al.}(1959)\citenamefont {Cooke},
		\citenamefont {Lazenby}, \citenamefont {McKim}, \citenamefont {Owen},
		\citenamefont {Wolf},\ and\ \citenamefont {Bleaney}}]{Cooke:KIrCl}%
	\BibitemOpen
	\bibfield  {author} {\bibinfo {author} {\bibfnamefont {A.~H.}\ \bibnamefont
			{Cooke}}, \bibinfo {author} {\bibfnamefont {R.}~\bibnamefont {Lazenby}},
		\bibinfo {author} {\bibfnamefont {F.~R.}\ \bibnamefont {McKim}}, \bibinfo
		{author} {\bibfnamefont {J.}~\bibnamefont {Owen}}, \bibinfo {author}
		{\bibfnamefont {W.~P.}\ \bibnamefont {Wolf}},\ and\ \bibinfo {author}
		{\bibfnamefont {B.}~\bibnamefont {Bleaney}},\ }\bibfield  {title} {\bibinfo
		{title} {{Exchange interactions in antiferromagnetic salts of iridium II.
				Magnetic susceptibility measurements}},\ }\href
	{https://doi.org/10.1098/rspa.1959.0053} {\bibfield  {journal} {\bibinfo
			{journal} {Proceedings of the Royal Society of London. Series A. Mathematical
				and Physical Sciences}\ }\textbf {\bibinfo {volume} {250}},\ \bibinfo {pages}
		{97} (\bibinfo {year} {1959})},\ \Eprint
	{https://arxiv.org/abs/https://royalsocietypublishing.org/doi/pdf/10.1098/rspa.1959.0053}
	{https://royalsocietypublishing.org/doi/pdf/10.1098/rspa.1959.0053}
	\BibitemShut {NoStop}%
	\bibitem [{\citenamefont {LaForge}\ \emph {et~al.}(2013)\citenamefont
		{LaForge}, \citenamefont {Pulido}, \citenamefont {Cava}, \citenamefont
		{Chan},\ and\ \citenamefont {Ramirez}}]{LaForge:quasispin}%
	\BibitemOpen
	\bibfield  {author} {\bibinfo {author} {\bibfnamefont {A.~D.}\ \bibnamefont
			{LaForge}}, \bibinfo {author} {\bibfnamefont {S.~H.}\ \bibnamefont {Pulido}},
		\bibinfo {author} {\bibfnamefont {R.~J.}\ \bibnamefont {Cava}}, \bibinfo
		{author} {\bibfnamefont {B.~C.}\ \bibnamefont {Chan}},\ and\ \bibinfo
		{author} {\bibfnamefont {A.~P.}\ \bibnamefont {Ramirez}},\ }\bibfield
	{title} {\bibinfo {title} {Quasispin glass in a geometrically frustrated
			magnet},\ }\href {https://doi.org/10.1103/PhysRevLett.110.017203} {\bibfield
		{journal} {\bibinfo  {journal} {Phys. Rev. Lett.}\ }\textbf {\bibinfo
			{volume} {110}},\ \bibinfo {pages} {017203} (\bibinfo {year}
		{2013})}\BibitemShut {NoStop}%
	\bibitem [{\citenamefont {Okamoto}\ \emph {et~al.}(2007)\citenamefont
		{Okamoto}, \citenamefont {Nohara}, \citenamefont {Aruga-Katori},\ and\
		\citenamefont {Takagi}}]{Okamoto:NaIrOQSL}%
	\BibitemOpen
	\bibfield  {author} {\bibinfo {author} {\bibfnamefont {Y.}~\bibnamefont
			{Okamoto}}, \bibinfo {author} {\bibfnamefont {M.}~\bibnamefont {Nohara}},
		\bibinfo {author} {\bibfnamefont {H.}~\bibnamefont {Aruga-Katori}},\ and\
		\bibinfo {author} {\bibfnamefont {H.}~\bibnamefont {Takagi}},\ }\bibfield
	{title} {\bibinfo {title} {{Spin-Liquid State in the $S=1/2$ Hyperkagome
				Antiferromagnet ${\mathrm{Na}}_{4}{\mathrm{Ir}}_{3}{\mathrm{O}}_{8}$}},\
	}\href {https://doi.org/10.1103/PhysRevLett.99.137207} {\bibfield  {journal}
		{\bibinfo  {journal} {Phys. Rev. Lett.}\ }\textbf {\bibinfo {volume} {99}},\
		\bibinfo {pages} {137207} (\bibinfo {year} {2007})}\BibitemShut {NoStop}%
	\bibitem [{\citenamefont {Ramirez}\ \emph {et~al.}(2002)\citenamefont
		{Ramirez}, \citenamefont {Shastry}, \citenamefont {Hayashi}, \citenamefont
		{Krajewski}, \citenamefont {Huse},\ and\ \citenamefont
		{Cava}}]{Ramirez:GdTiO}%
	\BibitemOpen
	\bibfield  {author} {\bibinfo {author} {\bibfnamefont {A.~P.}\ \bibnamefont
			{Ramirez}}, \bibinfo {author} {\bibfnamefont {B.~S.}\ \bibnamefont
			{Shastry}}, \bibinfo {author} {\bibfnamefont {A.}~\bibnamefont {Hayashi}},
		\bibinfo {author} {\bibfnamefont {J.~J.}\ \bibnamefont {Krajewski}}, \bibinfo
		{author} {\bibfnamefont {D.~A.}\ \bibnamefont {Huse}},\ and\ \bibinfo
		{author} {\bibfnamefont {R.~J.}\ \bibnamefont {Cava}},\ }\bibfield  {title}
	{\bibinfo {title} {{Multiple Field-Induced Phase Transitions in the
				Geometrically Frustrated Dipolar Magnet:
				${\mathrm{G}\mathrm{d}}_{2}{\mathrm{T}\mathrm{i}}_{2}{\mathrm{O}}_{7}$}},\
	}\href {https://doi.org/10.1103/PhysRevLett.89.067202} {\bibfield  {journal}
		{\bibinfo  {journal} {Phys. Rev. Lett.}\ }\textbf {\bibinfo {volume} {89}},\
		\bibinfo {pages} {067202} (\bibinfo {year} {2002})}\BibitemShut {NoStop}%
	\bibitem [{SCG()}]{SCGO:unpublished}%
	\BibitemOpen
	\href@noop {} {}\bibinfo {note} {{The data for $SrCr_{9p}Ga_{12-9p}O_{19}$
			with $p=0.98$ and $p=0.67$ is available on request.}}\BibitemShut {Stop}%
	\bibitem [{\citenamefont {Ramirez}\ \emph {et~al.}(1993)\citenamefont
		{Ramirez}, \citenamefont {Broholm}, \citenamefont {Lee}, \citenamefont
		{Collins}, \citenamefont {Heller}, \citenamefont {Kloc},\ and\ \citenamefont
		{Bucher}}]{Ramirez:KCrOHSO}%
	\BibitemOpen
	\bibfield  {author} {\bibinfo {author} {\bibfnamefont {A.~P.}\ \bibnamefont
			{Ramirez}}, \bibinfo {author} {\bibfnamefont {C.}~\bibnamefont {Broholm}},
		\bibinfo {author} {\bibfnamefont {S.~H.}\ \bibnamefont {Lee}}, \bibinfo
		{author} {\bibfnamefont {M.~F.}\ \bibnamefont {Collins}}, \bibinfo {author}
		{\bibfnamefont {L.}~\bibnamefont {Heller}}, \bibinfo {author} {\bibfnamefont
			{C.}~\bibnamefont {Kloc}},\ and\ \bibinfo {author} {\bibfnamefont
			{E.}~\bibnamefont {Bucher}},\ }\bibfield  {title} {\bibinfo {title}
		{{Magnetism in a chromium jarosite Kagomé lattice KCr3 (OH)6 (SO4)2
				(abstract)}},\ }\href {https://doi.org/10.1063/1.353631} {\bibfield
		{journal} {\bibinfo  {journal} {Journal of Applied Physics}\ }\textbf
		{\bibinfo {volume} {73}},\ \bibinfo {pages} {5658} (\bibinfo {year}
		{1993})},\ \Eprint
	{https://arxiv.org/abs/https://pubs.aip.org/aip/jap/article-pdf/73/10/5658/18657207/5658\_1\_online.pdf}
	{https://pubs.aip.org/aip/jap/article-pdf/73/10/5658/18657207/5658\_1\_online.pdf}
	\BibitemShut {NoStop}%
	\bibitem [{\citenamefont {Nakatsuji}\ \emph {et~al.}(2005)\citenamefont
		{Nakatsuji}, \citenamefont {Nambu}, \citenamefont {Tonomura}, \citenamefont
		{Sakai}, \citenamefont {Jonas}, \citenamefont {Broholm}, \citenamefont
		{Tsunetsugu}, \citenamefont {Qiu},\ and\ \citenamefont
		{Maeno}}]{NakatshujiMaeno:NaGaSneutron}%
	\BibitemOpen
	\bibfield  {author} {\bibinfo {author} {\bibfnamefont {S.}~\bibnamefont
			{Nakatsuji}}, \bibinfo {author} {\bibfnamefont {Y.}~\bibnamefont {Nambu}},
		\bibinfo {author} {\bibfnamefont {H.}~\bibnamefont {Tonomura}}, \bibinfo
		{author} {\bibfnamefont {O.}~\bibnamefont {Sakai}}, \bibinfo {author}
		{\bibfnamefont {S.}~\bibnamefont {Jonas}}, \bibinfo {author} {\bibfnamefont
			{C.}~\bibnamefont {Broholm}}, \bibinfo {author} {\bibfnamefont
			{H.}~\bibnamefont {Tsunetsugu}}, \bibinfo {author} {\bibfnamefont
			{Y.}~\bibnamefont {Qiu}},\ and\ \bibinfo {author} {\bibfnamefont
			{Y.}~\bibnamefont {Maeno}},\ }\bibfield  {title} {\bibinfo {title} {Spin
			disorder on a triangular lattice},\ }\href
	{https://doi.org/10.1126/science.1114727} {\bibfield  {journal} {\bibinfo
			{journal} {Science}\ }\textbf {\bibinfo {volume} {309}},\ \bibinfo {pages}
		{1697} (\bibinfo {year} {2005})},\ \Eprint
	{https://arxiv.org/abs/https://www.science.org/doi/pdf/10.1126/science.1114727}
	{https://www.science.org/doi/pdf/10.1126/science.1114727} \BibitemShut
	{NoStop}%
	\bibitem [{\citenamefont {Hagemann}\ \emph {et~al.}(2001)\citenamefont
		{Hagemann}, \citenamefont {Huang}, \citenamefont {Gao}, \citenamefont
		{Ramirez},\ and\ \citenamefont {Cava}}]{HagemannCava:BaSnGaZnCrO}%
	\BibitemOpen
	\bibfield  {author} {\bibinfo {author} {\bibfnamefont {I.~S.}\ \bibnamefont
			{Hagemann}}, \bibinfo {author} {\bibfnamefont {Q.}~\bibnamefont {Huang}},
		\bibinfo {author} {\bibfnamefont {X.~P.~A.}\ \bibnamefont {Gao}}, \bibinfo
		{author} {\bibfnamefont {A.~P.}\ \bibnamefont {Ramirez}},\ and\ \bibinfo
		{author} {\bibfnamefont {R.~J.}\ \bibnamefont {Cava}},\ }\bibfield  {title}
	{\bibinfo {title} {{Geometric Magnetic Frustration in
				${\mathrm{Ba}}_{2}{\mathrm{Sn}}_{2}{\mathrm{Ga}}_{3}{\mathrm{ZnCr}}_{7}{O}_{22}$:
				A Two-Dimensional Spinel Based Kagom\'e Lattice}},\ }\href
	{https://doi.org/10.1103/PhysRevLett.86.894} {\bibfield  {journal} {\bibinfo
			{journal} {Phys. Rev. Lett.}\ }\textbf {\bibinfo {volume} {86}},\ \bibinfo
		{pages} {894} (\bibinfo {year} {2001})}\BibitemShut {NoStop}%
	\bibitem [{\citenamefont {Nambu}\ and\ \citenamefont
		{Nakatsuji}(2011)}]{Nambu:NiGaS}%
	\BibitemOpen
	\bibfield  {author} {\bibinfo {author} {\bibfnamefont {Y.}~\bibnamefont
			{Nambu}}\ and\ \bibinfo {author} {\bibfnamefont {S.}~\bibnamefont
			{Nakatsuji}},\ }\bibfield  {title} {\bibinfo {title} {{Two-dimensional
				magnetism and spin-size effect in the S = 1 triangular antiferromagnet
				$NiGa_2S_4$}},\ }\href {https://doi.org/10.1088/0953-8984/23/16/164202}
	{\bibfield  {journal} {\bibinfo  {journal} {Journal of Physics: Condensed
				Matter}\ }\textbf {\bibinfo {volume} {23}},\ \bibinfo {pages} {164202}
		(\bibinfo {year} {2011})}\BibitemShut {NoStop}%
	\bibitem [{\citenamefont {Binder}\ and\ \citenamefont
		{Young}(1986)}]{BinderYoung:review}%
	\BibitemOpen
	\bibfield  {author} {\bibinfo {author} {\bibfnamefont {K.}~\bibnamefont
			{Binder}}\ and\ \bibinfo {author} {\bibfnamefont {A.~P.}\ \bibnamefont
			{Young}},\ }\bibfield  {title} {\bibinfo {title} {Spin glasses: Experimental
			facts, theoretical concepts, and open questions},\ }\href
	{https://doi.org/10.1103/RevModPhys.58.801} {\bibfield  {journal} {\bibinfo
			{journal} {Rev. Mod. Phys.}\ }\textbf {\bibinfo {volume} {58}},\ \bibinfo
		{pages} {801} (\bibinfo {year} {1986})}\BibitemShut {NoStop}%
	\bibitem [{\citenamefont {Ramirez}\ \emph {et~al.}(1990)\citenamefont
		{Ramirez}, \citenamefont {Espinosa},\ and\ \citenamefont
		{Cooper}}]{Ramirez:T2}%
	\BibitemOpen
	\bibfield  {author} {\bibinfo {author} {\bibfnamefont {A.~P.}\ \bibnamefont
			{Ramirez}}, \bibinfo {author} {\bibfnamefont {G.~P.}\ \bibnamefont
			{Espinosa}},\ and\ \bibinfo {author} {\bibfnamefont {A.~S.}\ \bibnamefont
			{Cooper}},\ }\bibfield  {title} {\bibinfo {title} {{Strong frustration and
				dilution-enhanced order in a quasi-2D spin glass}},\ }\href
	{https://doi.org/10.1103/PhysRevLett.64.2070} {\bibfield  {journal} {\bibinfo
			{journal} {Phys. Rev. Lett.}\ }\textbf {\bibinfo {volume} {64}},\ \bibinfo
		{pages} {2070} (\bibinfo {year} {1990})}\BibitemShut {NoStop}%
	\bibitem [{\citenamefont {Gingras}\ \emph {et~al.}(1996)\citenamefont
		{Gingras}, \citenamefont {Stager}, \citenamefont {Gaulin}, \citenamefont
		{Raju},\ and\ \citenamefont {Greedan}}]{Gingras:YMoOsusceptiblityGlass}%
	\BibitemOpen
	\bibfield  {author} {\bibinfo {author} {\bibfnamefont {M.}~\bibnamefont
			{Gingras}}, \bibinfo {author} {\bibfnamefont {C.}~\bibnamefont {Stager}},
		\bibinfo {author} {\bibfnamefont {B.}~\bibnamefont {Gaulin}}, \bibinfo
		{author} {\bibfnamefont {N.}~\bibnamefont {Raju}},\ and\ \bibinfo {author}
		{\bibfnamefont {J.}~\bibnamefont {Greedan}},\ }\bibfield  {title} {\bibinfo
		{title} {{Nonlinear susceptibility measurements at the spin-glass transition
				of the pyrochlore antiferromagnet $Y_2Mo_2O_7$}},\ }\href@noop {} {\bibfield
		{journal} {\bibinfo  {journal} {Journal of applied physics}\ }\textbf
		{\bibinfo {volume} {79}},\ \bibinfo {pages} {6170} (\bibinfo {year}
		{1996})}\BibitemShut {NoStop}%
	\bibitem [{\citenamefont {Schiffer}\ and\ \citenamefont
		{Daruka}(1997)}]{Schiffer:TwoPopulationModel}%
	\BibitemOpen
	\bibfield  {author} {\bibinfo {author} {\bibfnamefont {P.}~\bibnamefont
			{Schiffer}}\ and\ \bibinfo {author} {\bibfnamefont {I.}~\bibnamefont
			{Daruka}},\ }\bibfield  {title} {\bibinfo {title} {Two-population model for
			anomalous low-temperature magnetism in geometrically frustrated magnets},\
	}\href {https://doi.org/10.1103/PhysRevB.56.13712} {\bibfield  {journal}
		{\bibinfo  {journal} {Phys. Rev. B}\ }\textbf {\bibinfo {volume} {56}},\
		\bibinfo {pages} {13712} (\bibinfo {year} {1997})}\BibitemShut {NoStop}%
	\bibitem [{\citenamefont {Sandvik}\ \emph {et~al.}(1997)\citenamefont
		{Sandvik}, \citenamefont {Dagotto},\ and\ \citenamefont
		{Scalapino}}]{SandvikDagottoScalapino}%
	\BibitemOpen
	\bibfield  {author} {\bibinfo {author} {\bibfnamefont {A.~W.}\ \bibnamefont
			{Sandvik}}, \bibinfo {author} {\bibfnamefont {E.}~\bibnamefont {Dagotto}},\
		and\ \bibinfo {author} {\bibfnamefont {D.~J.}\ \bibnamefont {Scalapino}},\
	}\bibfield  {title} {\bibinfo {title} {Nonmagnetic impurities in spin-gapped
			and gapless {Heisenberg} antiferromagnets},\ }\href
	{https://doi.org/10.1103/PhysRevB.56.11701} {\bibfield  {journal} {\bibinfo
			{journal} {Phys. Rev. B}\ }\textbf {\bibinfo {volume} {56}},\ \bibinfo
		{pages} {11701} (\bibinfo {year} {1997})}\BibitemShut {NoStop}%
	\bibitem [{\citenamefont {Sachdev}\ \emph {et~al.}(1999)\citenamefont
		{Sachdev}, \citenamefont {Buragohain},\ and\ \citenamefont
		{Vojta}}]{SachdevBuragonhainVojta}%
	\BibitemOpen
	\bibfield  {author} {\bibinfo {author} {\bibfnamefont {S.}~\bibnamefont
			{Sachdev}}, \bibinfo {author} {\bibfnamefont {C.}~\bibnamefont
			{Buragohain}},\ and\ \bibinfo {author} {\bibfnamefont {M.}~\bibnamefont
			{Vojta}},\ }\bibfield  {title} {\bibinfo {title} {Quantum impurity in a
			nearly critical two-dimensional antiferromagnet},\ }\href
	{https://doi.org/10.1126/science.286.5449.2479} {\bibfield  {journal}
		{\bibinfo  {journal} {Science}\ }\textbf {\bibinfo {volume} {286}},\ \bibinfo
		{pages} {2479} (\bibinfo {year} {1999})}\BibitemShut {NoStop}%
	\bibitem [{\citenamefont {H\"oglund}\ and\ \citenamefont
		{Sandvik}(2003)}]{HoglundSandvik03}%
	\BibitemOpen
	\bibfield  {author} {\bibinfo {author} {\bibfnamefont {K.~H.}\ \bibnamefont
			{H\"oglund}}\ and\ \bibinfo {author} {\bibfnamefont {A.~W.}\ \bibnamefont
			{Sandvik}},\ }\bibfield  {title} {\bibinfo {title} {{Susceptibility of the 2D
				Spin-$\frac{1}{2}$ {Heisenberg} Antiferromagnet with an Impurity}},\ }\href
	{https://doi.org/10.1103/PhysRevLett.91.077204} {\bibfield  {journal}
		{\bibinfo  {journal} {Phys. Rev. Lett.}\ }\textbf {\bibinfo {volume} {91}},\
		\bibinfo {pages} {077204} (\bibinfo {year} {2003})}\BibitemShut {NoStop}%
	\bibitem [{\citenamefont {Maryasin}\ and\ \citenamefont
		{Zhitomirsky}(2013)}]{MaryasinZhitomirsky:VacancyPhaseDiagram}%
	\BibitemOpen
	\bibfield  {author} {\bibinfo {author} {\bibfnamefont {V.~S.}\ \bibnamefont
			{Maryasin}}\ and\ \bibinfo {author} {\bibfnamefont {M.~E.}\ \bibnamefont
			{Zhitomirsky}},\ }\bibfield  {title} {\bibinfo {title} {Triangular
			antiferromagnet with nonmagnetic impurities},\ }\href
	{https://doi.org/10.1103/PhysRevLett.111.247201} {\bibfield  {journal}
		{\bibinfo  {journal} {Phys. Rev. Lett.}\ }\textbf {\bibinfo {volume} {111}},\
		\bibinfo {pages} {247201} (\bibinfo {year} {2013})}\BibitemShut {NoStop}%
	\bibitem [{\citenamefont {Maryasin}\ and\ \citenamefont
		{Zhitomirsky}(2015)}]{Maryasin_2015}%
	\BibitemOpen
	\bibfield  {author} {\bibinfo {author} {\bibfnamefont {V.~S.}\ \bibnamefont
			{Maryasin}}\ and\ \bibinfo {author} {\bibfnamefont {M.~E.}\ \bibnamefont
			{Zhitomirsky}},\ }\bibfield  {title} {\bibinfo {title} {Collective impurity
			effects in the heisenberg triangular antiferromagnet},\ }\href
	{https://doi.org/10.1088/1742-6596/592/1/012112} {\bibfield  {journal}
		{\bibinfo  {journal} {Journal of Physics: Conference Series}\ }\textbf
		{\bibinfo {volume} {592}},\ \bibinfo {pages} {012112} (\bibinfo {year}
		{2015})}\BibitemShut {NoStop}%
	\bibitem [{\citenamefont {Sun}\ \emph {et~al.}(2023)\citenamefont {Sun},
		\citenamefont {Ramirez},\ and\ \citenamefont
		{Syzranov}}]{SunRamirezSyzranov:1Dquasispin}%
	\BibitemOpen
	\bibfield  {author} {\bibinfo {author} {\bibfnamefont {S.}~\bibnamefont
			{Sun}}, \bibinfo {author} {\bibfnamefont {A.~P.}\ \bibnamefont {Ramirez}},\
		and\ \bibinfo {author} {\bibfnamefont {S.}~\bibnamefont {Syzranov}},\
	}\bibfield  {title} {\bibinfo {title} {Quasispins of vacancy defects in ising
			chains with nearest- and next-to-nearest-neighbor interactions},\ }\href
	{https://doi.org/10.1103/PhysRevB.108.174436} {\bibfield  {journal} {\bibinfo
			{journal} {Phys. Rev. B}\ }\textbf {\bibinfo {volume} {108}},\ \bibinfo
		{pages} {174436} (\bibinfo {year} {2023})}\BibitemShut {NoStop}%
	\bibitem [{\citenamefont {Katsura}\ and\ \citenamefont
		{Tsujiyama}(1966)}]{KatsuraTsujiyama}%
	\BibitemOpen
	\bibfield  {author} {\bibinfo {author} {\bibfnamefont {S.}~\bibnamefont
			{Katsura}}\ and\ \bibinfo {author} {\bibfnamefont {B.}~\bibnamefont
			{Tsujiyama}},\ }\bibinfo {title} {Ferro- and antiferromagnetism of dilute
		{Ising} model},\ in\ \href {https://doi.org/10.6028/NBS.MP.273} {\emph
		{\bibinfo {booktitle} {Critical phenomena}}}\ (\bibinfo  {publisher}
	{National Institute of Standards and Technology, Gaithersburg, MD},\ \bibinfo
	{year} {1966})\ pp.\ \bibinfo {pages} {219 -- 225}\BibitemShut {NoStop}%
	\bibitem [{\citenamefont {Wortis}(1974)}]{Wortis1974}%
	\BibitemOpen
	\bibfield  {author} {\bibinfo {author} {\bibfnamefont {M.}~\bibnamefont
			{Wortis}},\ }\bibfield  {title} {\bibinfo {title} {Griffiths singularities in
			the randomly dilute one-dimensional ising model},\ }\href
	{https://doi.org/10.1103/PhysRevB.10.4665} {\bibfield  {journal} {\bibinfo
			{journal} {Phys. Rev. B}\ }\textbf {\bibinfo {volume} {10}},\ \bibinfo
		{pages} {4665} (\bibinfo {year} {1974})}\BibitemShut {NoStop}%
	\bibitem [{\citenamefont {Bogani}\ \emph {et~al.}(2005)\citenamefont {Bogani},
		\citenamefont {Sessoli}, \citenamefont {Pini}, \citenamefont {Rettori},
		\citenamefont {Novak}, \citenamefont {Rosa}, \citenamefont {Massi},
		\citenamefont {Fedi}, \citenamefont {Giuntini}, \citenamefont {Caneschi},\
		and\ \citenamefont {Gatteschi}}]{Bogani}%
	\BibitemOpen
	\bibfield  {author} {\bibinfo {author} {\bibfnamefont {L.}~\bibnamefont
			{Bogani}}, \bibinfo {author} {\bibfnamefont {R.}~\bibnamefont {Sessoli}},
		\bibinfo {author} {\bibfnamefont {M.~G.}\ \bibnamefont {Pini}}, \bibinfo
		{author} {\bibfnamefont {A.}~\bibnamefont {Rettori}}, \bibinfo {author}
		{\bibfnamefont {M.~A.}\ \bibnamefont {Novak}}, \bibinfo {author}
		{\bibfnamefont {P.}~\bibnamefont {Rosa}}, \bibinfo {author} {\bibfnamefont
			{M.}~\bibnamefont {Massi}}, \bibinfo {author} {\bibfnamefont {M.~E.}\
			\bibnamefont {Fedi}}, \bibinfo {author} {\bibfnamefont {L.}~\bibnamefont
			{Giuntini}}, \bibinfo {author} {\bibfnamefont {A.}~\bibnamefont {Caneschi}},\
		and\ \bibinfo {author} {\bibfnamefont {D.}~\bibnamefont {Gatteschi}},\
	}\bibfield  {title} {\bibinfo {title} {Finite-size effects on the static
			properties of a single-chain magnet},\ }\href
	{https://doi.org/10.1103/PhysRevB.72.064406} {\bibfield  {journal} {\bibinfo
			{journal} {Phys. Rev. B}\ }\textbf {\bibinfo {volume} {72}},\ \bibinfo
		{pages} {064406} (\bibinfo {year} {2005})}\BibitemShut {NoStop}%
	\bibitem [{\citenamefont {Goupalov}\ and\ \citenamefont
		{Mattis}(2007)}]{GoupalovMattis}%
	\BibitemOpen
	\bibfield  {author} {\bibinfo {author} {\bibfnamefont {S.~V.}\ \bibnamefont
			{Goupalov}}\ and\ \bibinfo {author} {\bibfnamefont {D.~C.}\ \bibnamefont
			{Mattis}},\ }\bibfield  {title} {\bibinfo {title} {Magnetic susceptibilities
			of finite {Ising} chains in the presence of defect sites},\ }\href
	{https://doi.org/10.1103/PhysRevB.76.224412} {\bibfield  {journal} {\bibinfo
			{journal} {Phys. Rev. B}\ }\textbf {\bibinfo {volume} {76}},\ \bibinfo
		{pages} {224412} (\bibinfo {year} {2007})}\BibitemShut {NoStop}%
	\bibitem [{\citenamefont {Val’kov}\ and\ \citenamefont
		{Shustin}(2016)}]{ValkovShustin}%
	\BibitemOpen
	\bibfield  {author} {\bibinfo {author} {\bibfnamefont {V.}~\bibnamefont
			{Val’kov}}\ and\ \bibinfo {author} {\bibfnamefont {M.}~\bibnamefont
			{Shustin}},\ }\bibfield  {title} {\bibinfo {title} {Effect of intersite
			repulsion on the correlation functions and thermodynamics of an {Ising} chain
			with annealed magnetic disorder},\ }\href
	{https://doi.org/10.3103/S1062873816110290} {\bibfield  {journal} {\bibinfo
			{journal} {Bulletin of the Russian Academy of Sciences: Physics}\ }\textbf
		{\bibinfo {volume} {80}},\ \bibinfo {pages} {1335} (\bibinfo {year}
		{2016})}\BibitemShut {NoStop}%
	\bibitem [{\citenamefont {{Sedik}}\ \emph {et~al.}(2024)\citenamefont
		{{Sedik}}, \citenamefont {{Sun}}, \citenamefont {{Ramirez}},\ and\
		\citenamefont {{Syzranov}}}]{Sedik:QuasispinInteraction}%
	\BibitemOpen
	\bibfield  {author} {\bibinfo {author} {\bibfnamefont {M.}~\bibnamefont
			{{Sedik}}}, \bibinfo {author} {\bibfnamefont {S.}~\bibnamefont {{Sun}}},
		\bibinfo {author} {\bibfnamefont {A.~P.}\ \bibnamefont {{Ramirez}}},\ and\
		\bibinfo {author} {\bibfnamefont {S.}~\bibnamefont {{Syzranov}}},\ }\bibfield
	{title} {\bibinfo {title} {{Quasispins of vacancy defects and their
				interactions in disordered antiferromagnets}},\ }\href
	{https://doi.org/10.48550/arXiv.2404.05845} {\bibfield  {journal} {\bibinfo
			{journal} {arXiv e-prints}\ ,\ \bibinfo {eid} {arXiv:2404.05845}} (\bibinfo
		{year} {2024})},\ \Eprint {https://arxiv.org/abs/2404.05845}
	{arXiv:2404.05845 [cond-mat.dis-nn]} \BibitemShut {NoStop}%
	\bibitem [{\citenamefont {Wollny}\ \emph {et~al.}(2012)\citenamefont {Wollny},
		\citenamefont {Andrade},\ and\ \citenamefont
		{Vojta}}]{WollnyVojta:vacancies}%
	\BibitemOpen
	\bibfield  {author} {\bibinfo {author} {\bibfnamefont {A.}~\bibnamefont
			{Wollny}}, \bibinfo {author} {\bibfnamefont {E.~C.}\ \bibnamefont
			{Andrade}},\ and\ \bibinfo {author} {\bibfnamefont {M.}~\bibnamefont
			{Vojta}},\ }\bibfield  {title} {\bibinfo {title} {Singular field response and
			singular screening of vacancies in antiferromagnets},\ }\href
	{https://doi.org/10.1103/PhysRevLett.109.177203} {\bibfield  {journal}
		{\bibinfo  {journal} {Phys. Rev. Lett.}\ }\textbf {\bibinfo {volume} {109}},\
		\bibinfo {pages} {177203} (\bibinfo {year} {2012})}\BibitemShut {NoStop}%
	\bibitem [{\citenamefont {Wollny}\ \emph {et~al.}(2011)\citenamefont {Wollny},
		\citenamefont {Fritz},\ and\ \citenamefont {Vojta}}]{WollnyFritzVojta}%
	\BibitemOpen
	\bibfield  {author} {\bibinfo {author} {\bibfnamefont {A.}~\bibnamefont
			{Wollny}}, \bibinfo {author} {\bibfnamefont {L.}~\bibnamefont {Fritz}},\ and\
		\bibinfo {author} {\bibfnamefont {M.}~\bibnamefont {Vojta}},\ }\bibfield
	{title} {\bibinfo {title} {Fractional impurity moments in two-dimensional
			noncollinear magnets},\ }\href
	{https://doi.org/10.1103/PhysRevLett.107.137204} {\bibfield  {journal}
		{\bibinfo  {journal} {Phys. Rev. Lett.}\ }\textbf {\bibinfo {volume} {107}},\
		\bibinfo {pages} {137204} (\bibinfo {year} {2011})}\BibitemShut {NoStop}%
	\bibitem [{\citenamefont {H\"oglund}\ and\ \citenamefont
		{Sandvik}(2007)}]{HoglundSandvik:FractionalSpin}%
	\BibitemOpen
	\bibfield  {author} {\bibinfo {author} {\bibfnamefont {K.~H.}\ \bibnamefont
			{H\"oglund}}\ and\ \bibinfo {author} {\bibfnamefont {A.~W.}\ \bibnamefont
			{Sandvik}},\ }\bibfield  {title} {\bibinfo {title} {{Anomalous Curie Response
				of Impurities in Quantum-Critical Spin-$1/2$ Heisenberg Antiferromagnets}},\
	}\href {https://doi.org/10.1103/PhysRevLett.99.027205} {\bibfield  {journal}
		{\bibinfo  {journal} {Phys. Rev. Lett.}\ }\textbf {\bibinfo {volume} {99}},\
		\bibinfo {pages} {027205} (\bibinfo {year} {2007})}\BibitemShut {NoStop}%
	\bibitem [{\citenamefont {Syzranov}\ and\ \citenamefont
		{Ramirez}(2022)}]{Syzranov:HiddenEnergy}%
	\BibitemOpen
	\bibfield  {author} {\bibinfo {author} {\bibfnamefont {S.~V.}\ \bibnamefont
			{Syzranov}}\ and\ \bibinfo {author} {\bibfnamefont {A.~P.}\ \bibnamefont
			{Ramirez}},\ }\bibfield  {title} {\bibinfo {title} {Eminuscent phase in
			frustrated magnets: a challenge to quantum spin liquids},\ }\href
	{https://www.nature.com/articles/s41467-022-30739-0} {\bibfield  {journal}
		{\bibinfo  {journal} {Nature Communications}\ }\textbf {\bibinfo {volume}
			{13}},\ \bibinfo {pages} {2993} (\bibinfo {year} {2022})}\BibitemShut
	{NoStop}%
	\bibitem [{\citenamefont {Dotsenko}(1993)}]{Dotsenko:GlassReview}%
	\BibitemOpen
	\bibfield  {author} {\bibinfo {author} {\bibfnamefont {V.~S.}\ \bibnamefont
			{Dotsenko}},\ }\bibfield  {title} {\bibinfo {title} {Physics of the
			spin-glass state},\ }\href@noop {} {\bibfield  {journal} {\bibinfo  {journal}
			{Physics-Uspekhi}\ }\textbf {\bibinfo {volume} {36}},\ \bibinfo {pages} {455}
		(\bibinfo {year} {1993})}\BibitemShut {NoStop}%
	\bibitem [{\citenamefont {M{\'e}zard}\ \emph {et~al.}(1987)\citenamefont
		{M{\'e}zard}, \citenamefont {Parisi},\ and\ \citenamefont
		{Virasoro}}]{Mezard:GlassReviewBook}%
	\BibitemOpen
	\bibfield  {author} {\bibinfo {author} {\bibfnamefont {M.}~\bibnamefont
			{M{\'e}zard}}, \bibinfo {author} {\bibfnamefont {G.}~\bibnamefont {Parisi}},\
		and\ \bibinfo {author} {\bibfnamefont {M.~A.}\ \bibnamefont {Virasoro}},\
	}\href@noop {} {\emph {\bibinfo {title} {Spin glass theory and beyond: An
				Introduction to the Replica Method and Its Applications}}},\ Vol.~\bibinfo
	{volume} {9}\ (\bibinfo  {publisher} {World Scientific Publishing Company},\
	\bibinfo {year} {1987})\BibitemShut {NoStop}%
	\bibitem [{Note1()}]{Note1}%
	\BibitemOpen
	\bibinfo {note} {With the exception of the compounds
		$Fe_{3-3x}Ga_{3x-1}Ti_5$, in which a significantly higher glass-transition
		temperature may be attributable to the formation of antiferromagnetic
		clusters~\cite {LaBarre:surfboards,Phelan:surfboards}}\BibitemShut {NoStop}%
	\bibitem [{\citenamefont {Maletta}\ and\ \citenamefont
		{Felsch}(1979)}]{Maletta:EuSrS}%
	\BibitemOpen
	\bibfield  {author} {\bibinfo {author} {\bibfnamefont {H.}~\bibnamefont
			{Maletta}}\ and\ \bibinfo {author} {\bibfnamefont {W.}~\bibnamefont
			{Felsch}},\ }\bibfield  {title} {\bibinfo {title} {{Insulating spin-glass
				system ${\mathrm{Eu}}_{x}{\mathrm{Sr}}_{1\ensuremath{-}x}\mathrm{S}$}},\
	}\href {https://doi.org/10.1103/PhysRevB.20.1245} {\bibfield  {journal}
		{\bibinfo  {journal} {Phys. Rev. B}\ }\textbf {\bibinfo {volume} {20}},\
		\bibinfo {pages} {1245} (\bibinfo {year} {1979})}\BibitemShut {NoStop}%
	\bibitem [{\citenamefont {Syzranov}(2022)}]{Syzranov:VacancyPhenomenological}%
	\BibitemOpen
	\bibfield  {author} {\bibinfo {author} {\bibfnamefont {S.}~\bibnamefont
			{Syzranov}},\ }\bibfield  {title} {\bibinfo {title} {Effect of vacancy
			defects on geometrically frustrated magnets},\ }\href
	{https://doi.org/10.1103/PhysRevB.106.L140202} {\bibfield  {journal}
		{\bibinfo  {journal} {Phys. Rev. B}\ }\textbf {\bibinfo {volume} {106}},\
		\bibinfo {pages} {L140202} (\bibinfo {year} {2022})}\BibitemShut {NoStop}%
	\bibitem [{\citenamefont {Henley}(2010)}]{Henley:CoulombReview}%
	\BibitemOpen
	\bibfield  {author} {\bibinfo {author} {\bibfnamefont {C.~L.}\ \bibnamefont
			{Henley}},\ }\bibfield  {title} {\bibinfo {title} {{The “Coulomb Phase”
				in Frustrated Systems}},\ }\href
	{https://doi.org/10.1146/annurev-conmatphys-070909-104138} {\bibfield
		{journal} {\bibinfo  {journal} {Annual Review of Condensed Matter Physics}\
		}\textbf {\bibinfo {volume} {1}},\ \bibinfo {pages} {179} (\bibinfo {year}
		{2010})},\ \Eprint
	{https://arxiv.org/abs/https://doi.org/10.1146/annurev-conmatphys-070909-104138}
	{https://doi.org/10.1146/annurev-conmatphys-070909-104138} \BibitemShut
	{NoStop}%
	\bibitem [{\citenamefont {Tabata}\ \emph {et~al.}(2006)\citenamefont {Tabata},
		\citenamefont {Kadowaki}, \citenamefont {Matsuhira}, \citenamefont {Hiroi},
		\citenamefont {Aso}, \citenamefont {Ressouche},\ and\ \citenamefont
		{F\aa{}k}}]{Tabata:KagomeIce}%
	\BibitemOpen
	\bibfield  {author} {\bibinfo {author} {\bibfnamefont {Y.}~\bibnamefont
			{Tabata}}, \bibinfo {author} {\bibfnamefont {H.}~\bibnamefont {Kadowaki}},
		\bibinfo {author} {\bibfnamefont {K.}~\bibnamefont {Matsuhira}}, \bibinfo
		{author} {\bibfnamefont {Z.}~\bibnamefont {Hiroi}}, \bibinfo {author}
		{\bibfnamefont {N.}~\bibnamefont {Aso}}, \bibinfo {author} {\bibfnamefont
			{E.}~\bibnamefont {Ressouche}},\ and\ \bibinfo {author} {\bibfnamefont
			{B.}~\bibnamefont {F\aa{}k}},\ }\bibfield  {title} {\bibinfo {title}
		{{Kagom\'e Ice State in the Dipolar Spin Ice
				${\mathrm{Dy}}_{2}{\mathrm{Ti}}_{2}{\mathrm{O}}_{7}$}},\ }\href
	{https://doi.org/10.1103/PhysRevLett.97.257205} {\bibfield  {journal}
		{\bibinfo  {journal} {Phys. Rev. Lett.}\ }\textbf {\bibinfo {volume} {97}},\
		\bibinfo {pages} {257205} (\bibinfo {year} {2006})}\BibitemShut {NoStop}%
	\bibitem [{\citenamefont {Fennell}\ \emph {et~al.}(2009)\citenamefont
		{Fennell}, \citenamefont {Deen}, \citenamefont {Wildes}, \citenamefont
		{Schmalzl}, \citenamefont {Prabhakaran}, \citenamefont {Boothroyd},
		\citenamefont {Aldus}, \citenamefont {McMorrow},\ and\ \citenamefont
		{Bramwell}}]{Fennell:CoulombHoTiO}%
	\BibitemOpen
	\bibfield  {author} {\bibinfo {author} {\bibfnamefont {T.}~\bibnamefont
			{Fennell}}, \bibinfo {author} {\bibfnamefont {P.~P.}\ \bibnamefont {Deen}},
		\bibinfo {author} {\bibfnamefont {A.~R.}\ \bibnamefont {Wildes}}, \bibinfo
		{author} {\bibfnamefont {K.}~\bibnamefont {Schmalzl}}, \bibinfo {author}
		{\bibfnamefont {D.}~\bibnamefont {Prabhakaran}}, \bibinfo {author}
		{\bibfnamefont {A.~T.}\ \bibnamefont {Boothroyd}}, \bibinfo {author}
		{\bibfnamefont {R.~J.}\ \bibnamefont {Aldus}}, \bibinfo {author}
		{\bibfnamefont {D.~F.}\ \bibnamefont {McMorrow}},\ and\ \bibinfo {author}
		{\bibfnamefont {S.~T.}\ \bibnamefont {Bramwell}},\ }\bibfield  {title}
	{\bibinfo {title} {{Magnetic Coulomb Phase in the Spin Ice $Ho_2Ti_2O_7$}},\
	}\href {https://doi.org/10.1126/science.1177582} {\bibfield  {journal}
		{\bibinfo  {journal} {Science}\ }\textbf {\bibinfo {volume} {326}},\ \bibinfo
		{pages} {415} (\bibinfo {year} {2009})},\ \Eprint
	{https://arxiv.org/abs/https://www.science.org/doi/pdf/10.1126/science.1177582}
	{https://www.science.org/doi/pdf/10.1126/science.1177582} \BibitemShut
	{NoStop}%
	\bibitem [{\citenamefont {Broholm}\ \emph
		{et~al.}(1990{\natexlab{a}})\citenamefont {Broholm}, \citenamefont {Aeppli},
		\citenamefont {Espinosa},\ and\ \citenamefont
		{Cooper}}]{Broholm:SCGOneutronScattering}%
	\BibitemOpen
	\bibfield  {author} {\bibinfo {author} {\bibfnamefont {C.}~\bibnamefont
			{Broholm}}, \bibinfo {author} {\bibfnamefont {G.}~\bibnamefont {Aeppli}},
		\bibinfo {author} {\bibfnamefont {G.~P.}\ \bibnamefont {Espinosa}},\ and\
		\bibinfo {author} {\bibfnamefont {A.~S.}\ \bibnamefont {Cooper}},\ }\bibfield
	{title} {\bibinfo {title} {Antiferromagnetic fluctuations and short-range
			order in a kagom\'e lattice},\ }\href
	{https://doi.org/10.1103/PhysRevLett.65.3173} {\bibfield  {journal} {\bibinfo
			{journal} {Phys. Rev. Lett.}\ }\textbf {\bibinfo {volume} {65}},\ \bibinfo
		{pages} {3173} (\bibinfo {year} {1990}{\natexlab{a}})}\BibitemShut {NoStop}%
	\bibitem [{\citenamefont {Yang}\ \emph {et~al.}(2015)\citenamefont {Yang},
		\citenamefont {Samarakoon}, \citenamefont {Dissanayake}, \citenamefont
		{Ueda}, \citenamefont {Klich}, \citenamefont {Iida}, \citenamefont
		{Pajerowski}, \citenamefont {Butch}, \citenamefont {Huang}, \citenamefont
		{Copley},\ and\ \citenamefont {Lee}}]{YangLee:SCGOjam}%
	\BibitemOpen
	\bibfield  {author} {\bibinfo {author} {\bibfnamefont {J.}~\bibnamefont
			{Yang}}, \bibinfo {author} {\bibfnamefont {A.}~\bibnamefont {Samarakoon}},
		\bibinfo {author} {\bibfnamefont {S.}~\bibnamefont {Dissanayake}}, \bibinfo
		{author} {\bibfnamefont {H.}~\bibnamefont {Ueda}}, \bibinfo {author}
		{\bibfnamefont {I.}~\bibnamefont {Klich}}, \bibinfo {author} {\bibfnamefont
			{K.}~\bibnamefont {Iida}}, \bibinfo {author} {\bibfnamefont {D.}~\bibnamefont
			{Pajerowski}}, \bibinfo {author} {\bibfnamefont {N.~P.}\ \bibnamefont
			{Butch}}, \bibinfo {author} {\bibfnamefont {Q.}~\bibnamefont {Huang}},
		\bibinfo {author} {\bibfnamefont {J.~R.~D.}\ \bibnamefont {Copley}},\ and\
		\bibinfo {author} {\bibfnamefont {S.-H.}\ \bibnamefont {Lee}},\ }\bibfield
	{title} {\bibinfo {title} {Spin jam induced by quantum fluctuations in a
			frustrated magnet},\ }\href {https://doi.org/10.1073/pnas.1503126112}
	{\bibfield  {journal} {\bibinfo  {journal} {Proceedings of the National
				Academy of Sciences}\ }\textbf {\bibinfo {volume} {112}},\ \bibinfo {pages}
		{11519} (\bibinfo {year} {2015})},\ \Eprint
	{https://arxiv.org/abs/https://www.pnas.org/doi/pdf/10.1073/pnas.1503126112}
	{https://www.pnas.org/doi/pdf/10.1073/pnas.1503126112} \BibitemShut {NoStop}%
	\bibitem [{\citenamefont {Stock}\ \emph {et~al.}(2010)\citenamefont {Stock},
		\citenamefont {Jonas}, \citenamefont {Broholm}, \citenamefont {Nakatsuji},
		\citenamefont {Nambu}, \citenamefont {Onuma}, \citenamefont {Maeno},\ and\
		\citenamefont {Chung}}]{Stock:neutron}%
	\BibitemOpen
	\bibfield  {author} {\bibinfo {author} {\bibfnamefont {C.}~\bibnamefont
			{Stock}}, \bibinfo {author} {\bibfnamefont {S.}~\bibnamefont {Jonas}},
		\bibinfo {author} {\bibfnamefont {C.}~\bibnamefont {Broholm}}, \bibinfo
		{author} {\bibfnamefont {S.}~\bibnamefont {Nakatsuji}}, \bibinfo {author}
		{\bibfnamefont {Y.}~\bibnamefont {Nambu}}, \bibinfo {author} {\bibfnamefont
			{K.}~\bibnamefont {Onuma}}, \bibinfo {author} {\bibfnamefont
			{Y.}~\bibnamefont {Maeno}},\ and\ \bibinfo {author} {\bibfnamefont {J.-H.}\
			\bibnamefont {Chung}},\ }\bibfield  {title} {\bibinfo {title}
		{{Neutron-Scattering Measurement of Incommensurate Short-Range Order in
				Single Crystals of the $S=1$ Triangular Antiferromagnet
				${\mathrm{NiGa}}_{2}{\mathrm{S}}_{4}$}},\ }\href
	{https://doi.org/10.1103/PhysRevLett.105.037402} {\bibfield  {journal}
		{\bibinfo  {journal} {Phys. Rev. Lett.}\ }\textbf {\bibinfo {volume} {105}},\
		\bibinfo {pages} {037402} (\bibinfo {year} {2010})}\BibitemShut {NoStop}%
	\bibitem [{\citenamefont {Gardner}\ \emph {et~al.}(1999)\citenamefont
		{Gardner}, \citenamefont {Gaulin}, \citenamefont {Lee}, \citenamefont
		{Broholm}, \citenamefont {Raju},\ and\ \citenamefont
		{Greedan}}]{Gardner:YMoOneutron}%
	\BibitemOpen
	\bibfield  {author} {\bibinfo {author} {\bibfnamefont {J.~S.}\ \bibnamefont
			{Gardner}}, \bibinfo {author} {\bibfnamefont {B.~D.}\ \bibnamefont {Gaulin}},
		\bibinfo {author} {\bibfnamefont {S.-H.}\ \bibnamefont {Lee}}, \bibinfo
		{author} {\bibfnamefont {C.}~\bibnamefont {Broholm}}, \bibinfo {author}
		{\bibfnamefont {N.~P.}\ \bibnamefont {Raju}},\ and\ \bibinfo {author}
		{\bibfnamefont {J.~E.}\ \bibnamefont {Greedan}},\ }\bibfield  {title}
	{\bibinfo {title} {{Glassy Statics and Dynamics in the Chemically Ordered
				Pyrochlore Antiferromagnet ${Y}_{2}{\mathrm{Mo}}_{2}{O}_{7}$}},\ }\href
	{https://doi.org/10.1103/PhysRevLett.83.211} {\bibfield  {journal} {\bibinfo
			{journal} {Phys. Rev. Lett.}\ }\textbf {\bibinfo {volume} {83}},\ \bibinfo
		{pages} {211} (\bibinfo {year} {1999})}\BibitemShut {NoStop}%
	\bibitem [{\citenamefont {Mydosh}(1993)}]{Mydosh:book}%
	\BibitemOpen
	\bibfield  {author} {\bibinfo {author} {\bibfnamefont {J.~A.}\ \bibnamefont
			{Mydosh}},\ }\href@noop {} {\emph {\bibinfo {title} {Spin glasses: an
				experimental introduction}}}\ (\bibinfo  {publisher} {CRC Press},\ \bibinfo
	{year} {1993})\BibitemShut {NoStop}%
	\bibitem [{\citenamefont {Greywall}\ and\ \citenamefont
		{Busch}(1989)}]{Greywall:DoublePeakHe}%
	\BibitemOpen
	\bibfield  {author} {\bibinfo {author} {\bibfnamefont {D.~S.}\ \bibnamefont
			{Greywall}}\ and\ \bibinfo {author} {\bibfnamefont {P.~A.}\ \bibnamefont
			{Busch}},\ }\bibfield  {title} {\bibinfo {title} {Heat capacity of
			$^{3}\mathrm{He}$ adsorbed on graphite at millikelvin temperatures and near
			third-layer promotion},\ }\href {https://doi.org/10.1103/PhysRevLett.62.1868}
	{\bibfield  {journal} {\bibinfo  {journal} {Phys. Rev. Lett.}\ }\textbf
		{\bibinfo {volume} {62}},\ \bibinfo {pages} {1868} (\bibinfo {year}
		{1989})}\BibitemShut {NoStop}%
	\bibitem [{\citenamefont {Ishida}\ \emph {et~al.}(1997)\citenamefont {Ishida},
		\citenamefont {Morishita}, \citenamefont {Yawata},\ and\ \citenamefont
		{Fukuyama}}]{Ishida:HeRingExchange}%
	\BibitemOpen
	\bibfield  {author} {\bibinfo {author} {\bibfnamefont {K.}~\bibnamefont
			{Ishida}}, \bibinfo {author} {\bibfnamefont {M.}~\bibnamefont {Morishita}},
		\bibinfo {author} {\bibfnamefont {K.}~\bibnamefont {Yawata}},\ and\ \bibinfo
		{author} {\bibfnamefont {H.}~\bibnamefont {Fukuyama}},\ }\bibfield  {title}
	{\bibinfo {title} {Low temperature heat-capacity anomalies in two-dimensional
			solid {$^{3}\mathrm{He}$}},\ }\href
	{https://doi.org/10.1103/PhysRevLett.79.3451} {\bibfield  {journal} {\bibinfo
			{journal} {Phys. Rev. Lett.}\ }\textbf {\bibinfo {volume} {79}},\ \bibinfo
		{pages} {3451} (\bibinfo {year} {1997})}\BibitemShut {NoStop}%
	\bibitem [{\citenamefont {Bordelon}\ \emph {et~al.}(2019)\citenamefont
		{Bordelon}, \citenamefont {Kenney}, \citenamefont {Liu}, \citenamefont
		{Hogan}, \citenamefont {Posthuma}, \citenamefont {Kavand}, \citenamefont
		{Lyu}, \citenamefont {Sherwin}, \citenamefont {Butch}, \citenamefont {Brown}
		\emph {et~al.}}]{Bordelon:NaYbO:TwoPeak}%
	\BibitemOpen
	\bibfield  {author} {\bibinfo {author} {\bibfnamefont {M.~M.}\ \bibnamefont
			{Bordelon}}, \bibinfo {author} {\bibfnamefont {E.}~\bibnamefont {Kenney}},
		\bibinfo {author} {\bibfnamefont {C.}~\bibnamefont {Liu}}, \bibinfo {author}
		{\bibfnamefont {T.}~\bibnamefont {Hogan}}, \bibinfo {author} {\bibfnamefont
			{L.}~\bibnamefont {Posthuma}}, \bibinfo {author} {\bibfnamefont
			{M.}~\bibnamefont {Kavand}}, \bibinfo {author} {\bibfnamefont
			{Y.}~\bibnamefont {Lyu}}, \bibinfo {author} {\bibfnamefont {M.}~\bibnamefont
			{Sherwin}}, \bibinfo {author} {\bibfnamefont {N.~P.}\ \bibnamefont {Butch}},
		\bibinfo {author} {\bibfnamefont {C.}~\bibnamefont {Brown}}, \emph {et~al.},\
	}\bibfield  {title} {\bibinfo {title} {{Field-tunable quantum disordered
				ground state in the triangular-lattice antiferromagnet NaYbO2}},\ }\href@noop
	{} {\bibfield  {journal} {\bibinfo  {journal} {Nature Physics}\ }\textbf
		{\bibinfo {volume} {15}},\ \bibinfo {pages} {1058} (\bibinfo {year}
		{2019})}\BibitemShut {NoStop}%
	\bibitem [{\citenamefont {Ranjith}\ \emph {et~al.}(2019)\citenamefont
		{Ranjith}, \citenamefont {Luther}, \citenamefont {Reimann}, \citenamefont
		{Schmidt}, \citenamefont {Schlender}, \citenamefont {Sichelschmidt},
		\citenamefont {Yasuoka}, \citenamefont {Strydom}, \citenamefont {Skourski},
		\citenamefont {Wosnitza}, \citenamefont {K\"uhne}, \citenamefont {Doert},\
		and\ \citenamefont {Baenitz}}]{RanjithBaenitz:NaYbSe:TwoPeak}%
	\BibitemOpen
	\bibfield  {author} {\bibinfo {author} {\bibfnamefont {K.~M.}\ \bibnamefont
			{Ranjith}}, \bibinfo {author} {\bibfnamefont {S.}~\bibnamefont {Luther}},
		\bibinfo {author} {\bibfnamefont {T.}~\bibnamefont {Reimann}}, \bibinfo
		{author} {\bibfnamefont {B.}~\bibnamefont {Schmidt}}, \bibinfo {author}
		{\bibfnamefont {P.}~\bibnamefont {Schlender}}, \bibinfo {author}
		{\bibfnamefont {J.}~\bibnamefont {Sichelschmidt}}, \bibinfo {author}
		{\bibfnamefont {H.}~\bibnamefont {Yasuoka}}, \bibinfo {author} {\bibfnamefont
			{A.~M.}\ \bibnamefont {Strydom}}, \bibinfo {author} {\bibfnamefont
			{Y.}~\bibnamefont {Skourski}}, \bibinfo {author} {\bibfnamefont
			{J.}~\bibnamefont {Wosnitza}}, \bibinfo {author} {\bibfnamefont
			{H.}~\bibnamefont {K\"uhne}}, \bibinfo {author} {\bibfnamefont
			{T.}~\bibnamefont {Doert}},\ and\ \bibinfo {author} {\bibfnamefont
			{M.}~\bibnamefont {Baenitz}},\ }\bibfield  {title} {\bibinfo {title}
		{{Anisotropic field-induced ordering in the triangular-lattice quantum spin
				liquid ${\mathrm{NaYbSe}}_{2}$}},\ }\href
	{https://doi.org/10.1103/PhysRevB.100.224417} {\bibfield  {journal} {\bibinfo
			{journal} {Phys. Rev. B}\ }\textbf {\bibinfo {volume} {100}},\ \bibinfo
		{pages} {224417} (\bibinfo {year} {2019})}\BibitemShut {NoStop}%
	\bibitem [{\citenamefont {Schiffer}\ \emph {et~al.}(1995)\citenamefont
		{Schiffer}, \citenamefont {Ramirez}, \citenamefont {Huse}, \citenamefont
		{Gammel}, \citenamefont {Yaron}, \citenamefont {Bishop},\ and\ \citenamefont
		{Valentino}}]{SchifferRamirez:GGG}%
	\BibitemOpen
	\bibfield  {author} {\bibinfo {author} {\bibfnamefont {P.}~\bibnamefont
			{Schiffer}}, \bibinfo {author} {\bibfnamefont {A.~P.}\ \bibnamefont
			{Ramirez}}, \bibinfo {author} {\bibfnamefont {D.~A.}\ \bibnamefont {Huse}},
		\bibinfo {author} {\bibfnamefont {P.~L.}\ \bibnamefont {Gammel}}, \bibinfo
		{author} {\bibfnamefont {U.}~\bibnamefont {Yaron}}, \bibinfo {author}
		{\bibfnamefont {D.~J.}\ \bibnamefont {Bishop}},\ and\ \bibinfo {author}
		{\bibfnamefont {A.~J.}\ \bibnamefont {Valentino}},\ }\bibfield  {title}
	{\bibinfo {title} {Frustration induced spin freezing in a site-ordered
			magnet: Gadolinium gallium garnet},\ }\href
	{https://doi.org/10.1103/PhysRevLett.74.2379} {\bibfield  {journal} {\bibinfo
			{journal} {Phys. Rev. Lett.}\ }\textbf {\bibinfo {volume} {74}},\ \bibinfo
		{pages} {2379} (\bibinfo {year} {1995})}\BibitemShut {NoStop}%
	\bibitem [{\citenamefont {Elser}(1989)}]{Elser:KHAF}%
	\BibitemOpen
	\bibfield  {author} {\bibinfo {author} {\bibfnamefont {V.}~\bibnamefont
			{Elser}},\ }\bibfield  {title} {\bibinfo {title} {Nuclear antiferromagnetism
			in a registered $^{3}\mathrm{He}$ solid},\ }\href
	{https://doi.org/10.1103/PhysRevLett.62.2405} {\bibfield  {journal} {\bibinfo
			{journal} {Phys. Rev. Lett.}\ }\textbf {\bibinfo {volume} {62}},\ \bibinfo
		{pages} {2405} (\bibinfo {year} {1989})}\BibitemShut {NoStop}%
	\bibitem [{\citenamefont {Zeng}\ and\ \citenamefont
		{Elser}(1990)}]{ZengElser:KHAF}%
	\BibitemOpen
	\bibfield  {author} {\bibinfo {author} {\bibfnamefont {C.}~\bibnamefont
			{Zeng}}\ and\ \bibinfo {author} {\bibfnamefont {V.}~\bibnamefont {Elser}},\
	}\bibfield  {title} {\bibinfo {title} {Numerical studies of
			antiferromagnetism on a kagom\'e net},\ }\href
	{https://doi.org/10.1103/PhysRevB.42.8436} {\bibfield  {journal} {\bibinfo
			{journal} {Phys. Rev. B}\ }\textbf {\bibinfo {volume} {42}},\ \bibinfo
		{pages} {8436} (\bibinfo {year} {1990})}\BibitemShut {NoStop}%
	\bibitem [{\citenamefont {Elstner}\ and\ \citenamefont
		{Young}(1994)}]{ElstnerYoung:kagome}%
	\BibitemOpen
	\bibfield  {author} {\bibinfo {author} {\bibfnamefont {N.}~\bibnamefont
			{Elstner}}\ and\ \bibinfo {author} {\bibfnamefont {A.~P.}\ \bibnamefont
			{Young}},\ }\bibfield  {title} {\bibinfo {title} {Spin-1/2 {H}eisenberg
			antiferromagnet on the kagome\ifmmode\acute\else\textasciiacute\fi{} lattice:
			High-temperature expansion and exact-diagonalization studies},\ }\href
	{https://doi.org/10.1103/PhysRevB.50.6871} {\bibfield  {journal} {\bibinfo
			{journal} {Phys. Rev. B}\ }\textbf {\bibinfo {volume} {50}},\ \bibinfo
		{pages} {6871} (\bibinfo {year} {1994})}\BibitemShut {NoStop}%
	\bibitem [{\citenamefont {Nakamura}\ and\ \citenamefont
		{Miyashita}(1995)}]{NakamuraMiyashita:KHAF}%
	\BibitemOpen
	\bibfield  {author} {\bibinfo {author} {\bibfnamefont {T.}~\bibnamefont
			{Nakamura}}\ and\ \bibinfo {author} {\bibfnamefont {S.}~\bibnamefont
			{Miyashita}},\ }\bibfield  {title} {\bibinfo {title} {Thermodynamic
			properties of the quantum {H}eisenberg antiferromagnet on the kagom\'e
			lattice},\ }\href {https://doi.org/10.1103/PhysRevB.52.9174} {\bibfield
		{journal} {\bibinfo  {journal} {Phys. Rev. B}\ }\textbf {\bibinfo {volume}
			{52}},\ \bibinfo {pages} {9174} (\bibinfo {year} {1995})}\BibitemShut
	{NoStop}%
	\bibitem [{\citenamefont {Tomczak}\ and\ \citenamefont
		{Richter}(1996)}]{TomczakRichter:KHAF}%
	\BibitemOpen
	\bibfield  {author} {\bibinfo {author} {\bibfnamefont {P.}~\bibnamefont
			{Tomczak}}\ and\ \bibinfo {author} {\bibfnamefont {J.}~\bibnamefont
			{Richter}},\ }\bibfield  {title} {\bibinfo {title} {Thermodynamical
			properties of the {H}eisenberg antiferromagnet on the kagom\'e lattice},\
	}\href {https://doi.org/10.1103/PhysRevB.54.9004} {\bibfield  {journal}
		{\bibinfo  {journal} {Phys. Rev. B}\ }\textbf {\bibinfo {volume} {54}},\
		\bibinfo {pages} {9004} (\bibinfo {year} {1996})}\BibitemShut {NoStop}%
	\bibitem [{\citenamefont {Sindzingre}\ \emph {et~al.}(2000)\citenamefont
		{Sindzingre}, \citenamefont {Misguich}, \citenamefont {Lhuillier},
		\citenamefont {Bernu}, \citenamefont {Pierre}, \citenamefont {Waldtmann},\
		and\ \citenamefont {Everts}}]{SindzingreMisguich:KHAF}%
	\BibitemOpen
	\bibfield  {author} {\bibinfo {author} {\bibfnamefont {P.}~\bibnamefont
			{Sindzingre}}, \bibinfo {author} {\bibfnamefont {G.}~\bibnamefont
			{Misguich}}, \bibinfo {author} {\bibfnamefont {C.}~\bibnamefont {Lhuillier}},
		\bibinfo {author} {\bibfnamefont {B.}~\bibnamefont {Bernu}}, \bibinfo
		{author} {\bibfnamefont {L.}~\bibnamefont {Pierre}}, \bibinfo {author}
		{\bibfnamefont {C.}~\bibnamefont {Waldtmann}},\ and\ \bibinfo {author}
		{\bibfnamefont {H.-U.}\ \bibnamefont {Everts}},\ }\bibfield  {title}
	{\bibinfo {title} {Magnetothermodynamics of the spin- $\frac{1}{2}$ kagom\'e
			antiferromagnet},\ }\href {https://doi.org/10.1103/PhysRevLett.84.2953}
	{\bibfield  {journal} {\bibinfo  {journal} {Phys. Rev. Lett.}\ }\textbf
		{\bibinfo {volume} {84}},\ \bibinfo {pages} {2953} (\bibinfo {year}
		{2000})}\BibitemShut {NoStop}%
	\bibitem [{\citenamefont {Misguich}\ and\ \citenamefont
		{Bernu}(2005)}]{MisguichBernu:KHAF}%
	\BibitemOpen
	\bibfield  {author} {\bibinfo {author} {\bibfnamefont {G.}~\bibnamefont
			{Misguich}}\ and\ \bibinfo {author} {\bibfnamefont {B.}~\bibnamefont
			{Bernu}},\ }\bibfield  {title} {\bibinfo {title} {Specific heat of the
			{$S=\frac{1}{2}$} {H}eisenberg model on the kagome lattice: High-temperature
			series expansion analysis},\ }\href
	{https://doi.org/10.1103/PhysRevB.71.014417} {\bibfield  {journal} {\bibinfo
			{journal} {Phys. Rev. B}\ }\textbf {\bibinfo {volume} {71}},\ \bibinfo
		{pages} {014417} (\bibinfo {year} {2005})}\BibitemShut {NoStop}%
	\bibitem [{\citenamefont {Misguich}\ and\ \citenamefont
		{Sindzingre}(2007)}]{MisguichSindzingre:KHAF}%
	\BibitemOpen
	\bibfield  {author} {\bibinfo {author} {\bibfnamefont {G.}~\bibnamefont
			{Misguich}}\ and\ \bibinfo {author} {\bibfnamefont {P.}~\bibnamefont
			{Sindzingre}},\ }\bibfield  {title} {\bibinfo {title} {Magnetic
			susceptibility and specific heat of the spin- ${\frac{1}{2}}$ {H}eisenberg
			model on the kagome lattice and experimental data on
			{$\mathrm{ZnCu}_3\mathrm{(OH)}_6\mathrm{Cl}_2$}},\ }\href
	{https://doi.org/10.1140/epjb/e2007-00301-6} {\bibfield  {journal} {\bibinfo
			{journal} {The European Physical Journal B}\ }\textbf {\bibinfo {volume}
			{59}},\ \bibinfo {pages} {305–309} (\bibinfo {year} {2007})}\BibitemShut
	{NoStop}%
	\bibitem [{\citenamefont {Isoda}\ \emph {et~al.}(2011)\citenamefont {Isoda},
		\citenamefont {Nakano},\ and\ \citenamefont {Sakai}}]{IsodaNakano:XXZ}%
	\BibitemOpen
	\bibfield  {author} {\bibinfo {author} {\bibfnamefont {M.}~\bibnamefont
			{Isoda}}, \bibinfo {author} {\bibfnamefont {H.}~\bibnamefont {Nakano}},\ and\
		\bibinfo {author} {\bibfnamefont {T.}~\bibnamefont {Sakai}},\ }\bibfield
	{title} {\bibinfo {title} {Specific heat and magnetic susceptibility of
			{I}sing-like anisotropic {H}eisenberg model on kagome lattice},\ }\href
	{https://doi.org/10.1143/JPSJ.80.084704} {\bibfield  {journal} {\bibinfo
			{journal} {Journal of the Physical Society of Japan}\ }\textbf {\bibinfo
			{volume} {80}},\ \bibinfo {pages} {084704} (\bibinfo {year}
		{2011})}\BibitemShut {NoStop}%
	\bibitem [{\citenamefont {Prelov\ifmmode~\check{s}\else \v{s}\fi{}ek}\ and\
		\citenamefont {Kokalj}(2018)}]{PrelovsekKokalj:THAF}%
	\BibitemOpen
	\bibfield  {author} {\bibinfo {author} {\bibfnamefont {P.}~\bibnamefont
			{Prelov\ifmmode~\check{s}\else \v{s}\fi{}ek}}\ and\ \bibinfo {author}
		{\bibfnamefont {J.}~\bibnamefont {Kokalj}},\ }\bibfield  {title} {\bibinfo
		{title} {Finite-temperature properties of the extended {H}eisenberg model on
			a triangular lattice},\ }\href {https://doi.org/10.1103/PhysRevB.98.035107}
	{\bibfield  {journal} {\bibinfo  {journal} {Phys. Rev. B}\ }\textbf {\bibinfo
			{volume} {98}},\ \bibinfo {pages} {035107} (\bibinfo {year}
		{2018})}\BibitemShut {NoStop}%
	\bibitem [{\citenamefont {Schnack}\ \emph {et~al.}(2018)\citenamefont
		{Schnack}, \citenamefont {Schulenburg},\ and\ \citenamefont
		{Richter}}]{SchnackSchulenberg:KHAF}%
	\BibitemOpen
	\bibfield  {author} {\bibinfo {author} {\bibfnamefont {J.}~\bibnamefont
			{Schnack}}, \bibinfo {author} {\bibfnamefont {J.}~\bibnamefont
			{Schulenburg}},\ and\ \bibinfo {author} {\bibfnamefont {J.}~\bibnamefont
			{Richter}},\ }\bibfield  {title} {\bibinfo {title} {Magnetism of the {$N=42$}
			kagome lattice antiferromagnet},\ }\href
	{https://doi.org/10.1103/PhysRevB.98.094423} {\bibfield  {journal} {\bibinfo
			{journal} {Phys. Rev. B}\ }\textbf {\bibinfo {volume} {98}},\ \bibinfo
		{pages} {094423} (\bibinfo {year} {2018})}\BibitemShut {NoStop}%
	\bibitem [{\citenamefont {Chen}\ \emph {et~al.}(2019)\citenamefont {Chen},
		\citenamefont {Qu}, \citenamefont {Li}, \citenamefont {Chen}, \citenamefont
		{Gong}, \citenamefont {von Delft}, \citenamefont {Weichselbaum},\ and\
		\citenamefont {Li}}]{ChenQu:THAF}%
	\BibitemOpen
	\bibfield  {author} {\bibinfo {author} {\bibfnamefont {L.}~\bibnamefont
			{Chen}}, \bibinfo {author} {\bibfnamefont {D.-W.}\ \bibnamefont {Qu}},
		\bibinfo {author} {\bibfnamefont {H.}~\bibnamefont {Li}}, \bibinfo {author}
		{\bibfnamefont {B.-B.}\ \bibnamefont {Chen}}, \bibinfo {author}
		{\bibfnamefont {S.-S.}\ \bibnamefont {Gong}}, \bibinfo {author}
		{\bibfnamefont {J.}~\bibnamefont {von Delft}}, \bibinfo {author}
		{\bibfnamefont {A.}~\bibnamefont {Weichselbaum}},\ and\ \bibinfo {author}
		{\bibfnamefont {W.}~\bibnamefont {Li}},\ }\bibfield  {title} {\bibinfo
		{title} {Two-temperature scales in the triangular-lattice {H}eisenberg
			antiferromagnet},\ }\href {https://doi.org/10.1103/PhysRevB.99.140404}
	{\bibfield  {journal} {\bibinfo  {journal} {Phys. Rev. B}\ }\textbf {\bibinfo
			{volume} {99}},\ \bibinfo {pages} {140404} (\bibinfo {year}
		{2019})}\BibitemShut {NoStop}%
	\bibitem [{\citenamefont {Prelov\ifmmode~\check{s}\else \v{s}\fi{}ek}\ \emph
		{et~al.}(2020)\citenamefont {Prelov\ifmmode~\check{s}\else \v{s}\fi{}ek},
		\citenamefont {Morita}, \citenamefont {Tohyama},\ and\ \citenamefont
		{Herbrych}}]{Prelovsek:TriangularWilsonRatio}%
	\BibitemOpen
	\bibfield  {author} {\bibinfo {author} {\bibfnamefont {P.}~\bibnamefont
			{Prelov\ifmmode~\check{s}\else \v{s}\fi{}ek}}, \bibinfo {author}
		{\bibfnamefont {K.}~\bibnamefont {Morita}}, \bibinfo {author} {\bibfnamefont
			{T.}~\bibnamefont {Tohyama}},\ and\ \bibinfo {author} {\bibfnamefont
			{J.}~\bibnamefont {Herbrych}},\ }\bibfield  {title} {\bibinfo {title}
		{Vanishing wilson ratio as the hallmark of quantum spin-liquid models},\
	}\href {https://doi.org/10.1103/PhysRevResearch.2.023024} {\bibfield
		{journal} {\bibinfo  {journal} {Phys. Rev. Res.}\ }\textbf {\bibinfo {volume}
			{2}},\ \bibinfo {pages} {023024} (\bibinfo {year} {2020})}\BibitemShut
	{NoStop}%
	\bibitem [{\citenamefont {Seki}\ and\ \citenamefont
		{Yunoki}(2020)}]{SekiYunoki:RingExchange}%
	\BibitemOpen
	\bibfield  {author} {\bibinfo {author} {\bibfnamefont {K.}~\bibnamefont
			{Seki}}\ and\ \bibinfo {author} {\bibfnamefont {S.}~\bibnamefont {Yunoki}},\
	}\bibfield  {title} {\bibinfo {title} {Thermodynamic properties of an
			{$S=\frac{1}{2}$} ring-exchange model on the triangular lattice},\ }\href
	{https://doi.org/10.1103/PhysRevB.101.235115} {\bibfield  {journal} {\bibinfo
			{journal} {Phys. Rev. B}\ }\textbf {\bibinfo {volume} {101}},\ \bibinfo
		{pages} {235115} (\bibinfo {year} {2020})}\BibitemShut {NoStop}%
	\bibitem [{\citenamefont {Hutak}\ \emph {et~al.}(2024)\citenamefont {Hutak},
		\citenamefont {Krokhmalskii}, \citenamefont {Schnack}, \citenamefont
		{Richter},\ and\ \citenamefont {Derzhko}}]{HutakKrokhmalskii:HKHAF}%
	\BibitemOpen
	\bibfield  {author} {\bibinfo {author} {\bibfnamefont {T.}~\bibnamefont
			{Hutak}}, \bibinfo {author} {\bibfnamefont {T.}~\bibnamefont {Krokhmalskii}},
		\bibinfo {author} {\bibfnamefont {J.}~\bibnamefont {Schnack}}, \bibinfo
		{author} {\bibfnamefont {J.}~\bibnamefont {Richter}},\ and\ \bibinfo {author}
		{\bibfnamefont {O.}~\bibnamefont {Derzhko}},\ }\bibfield  {title} {\bibinfo
		{title} {Thermodynamics of the $s=\frac{1}{2}$ hyperkagome-lattice heisenberg
			antiferromagnet},\ }\href {https://doi.org/10.1103/PhysRevB.110.054428}
	{\bibfield  {journal} {\bibinfo  {journal} {Phys. Rev. B}\ }\textbf {\bibinfo
			{volume} {110}},\ \bibinfo {pages} {054428} (\bibinfo {year}
		{2024})}\BibitemShut {NoStop}%
	\bibitem [{\citenamefont {Ulaga}\ \emph {et~al.}(2024)\citenamefont {Ulaga},
		\citenamefont {Kokalj}, \citenamefont {Wietek}, \citenamefont {Zorko},\ and\
		\citenamefont {Prelov\ifmmode~\check{s}\else
			\v{s}\fi{}ek}}]{UlagaPrelovsek:TriangularKagomeTwoPeak}%
	\BibitemOpen
	\bibfield  {author} {\bibinfo {author} {\bibfnamefont {M.}~\bibnamefont
			{Ulaga}}, \bibinfo {author} {\bibfnamefont {J.}~\bibnamefont {Kokalj}},
		\bibinfo {author} {\bibfnamefont {A.}~\bibnamefont {Wietek}}, \bibinfo
		{author} {\bibfnamefont {A.}~\bibnamefont {Zorko}},\ and\ \bibinfo {author}
		{\bibfnamefont {P.}~\bibnamefont {Prelov\ifmmode~\check{s}\else
				\v{s}\fi{}ek}},\ }\bibfield  {title} {\bibinfo {title} {{Finite-temperature
				properties of the easy-axis Heisenberg model on frustrated lattices}},\
	}\href {https://doi.org/10.1103/PhysRevB.109.035110} {\bibfield  {journal}
		{\bibinfo  {journal} {Phys. Rev. B}\ }\textbf {\bibinfo {volume} {109}},\
		\bibinfo {pages} {035110} (\bibinfo {year} {2024})}\BibitemShut {NoStop}%
	\bibitem [{\citenamefont {{Ulaga}}\ \emph {et~al.}(2024)\citenamefont
		{{Ulaga}}, \citenamefont {{Kokalj}},\ and\ \citenamefont
		{{Prelov{\v{s}}ek}}}]{UlagaPrelovsek:latest}%
	\BibitemOpen
	\bibfield  {author} {\bibinfo {author} {\bibfnamefont {M.}~\bibnamefont
			{{Ulaga}}}, \bibinfo {author} {\bibfnamefont {J.}~\bibnamefont {{Kokalj}}},\
		and\ \bibinfo {author} {\bibfnamefont {P.}~\bibnamefont
			{{Prelov{\v{s}}ek}}},\ }\bibfield  {title} {\bibinfo {title} {{Easy-axis
				Heisenberg model on the triangular lattice: from supersolid to gapped
				solid}},\ }\href {https://doi.org/10.48550/arXiv.2408.05034} {\bibfield
		{journal} {\bibinfo  {journal} {arXiv e-prints}\ ,\ \bibinfo {eid}
			{arXiv:2408.05034}} (\bibinfo {year} {2024})},\ \Eprint
	{https://arxiv.org/abs/2408.05034} {arXiv:2408.05034 [cond-mat.str-el]}
	\BibitemShut {NoStop}%
	\bibitem [{\citenamefont {Hwang}\ \emph {et~al.}(2008)\citenamefont {Hwang},
		\citenamefont {Kim}, \citenamefont {Kang},\ and\ \citenamefont
		{Kim}}]{HwangKim:TIAF}%
	\BibitemOpen
	\bibfield  {author} {\bibinfo {author} {\bibfnamefont {C.-O.}\ \bibnamefont
			{Hwang}}, \bibinfo {author} {\bibfnamefont {S.-Y.}\ \bibnamefont {Kim}},
		\bibinfo {author} {\bibfnamefont {D.}~\bibnamefont {Kang}},\ and\ \bibinfo
		{author} {\bibfnamefont {J.~M.}\ \bibnamefont {Kim}},\ }\bibfield  {title}
	{\bibinfo {title} {{Thermodynamic properties of the triangular-lattice
				{I}sing antiferromagnet in a uniform magnetic field}},\ }\href
	{https://doi.org/https://doi.org/10.3938/jkps.52.203} {\bibfield  {journal}
		{\bibinfo  {journal} {J. Korean Phys. Soc.}\ }\textbf {\bibinfo {volume}
			{52}},\ \bibinfo {pages} {203} (\bibinfo {year} {2008})}\BibitemShut
	{NoStop}%
	\bibitem [{\citenamefont {Kim}(2015)}]{Kim:TIAF}%
	\BibitemOpen
	\bibfield  {author} {\bibinfo {author} {\bibfnamefont {S.-Y.}\ \bibnamefont
			{Kim}},\ }\bibfield  {title} {\bibinfo {title} {Ising antiferromagnet on a
			finite triangular lattice with free boundary conditions},\ }\href@noop {}
	{\bibfield  {journal} {\bibinfo  {journal} {J. Korean Phys. Soc.}\ }\textbf
		{\bibinfo {volume} {67}},\ \bibinfo {pages} {1517} (\bibinfo {year}
		{2015})}\BibitemShut {NoStop}%
	\bibitem [{\citenamefont {Kanô}\ and\ \citenamefont
		{Naya}(1953)}]{KanoNaya:KIAF}%
	\BibitemOpen
	\bibfield  {author} {\bibinfo {author} {\bibfnamefont {K.}~\bibnamefont
			{Kanô}}\ and\ \bibinfo {author} {\bibfnamefont {S.}~\bibnamefont {Naya}},\
	}\bibfield  {title} {\bibinfo {title} {{Antiferromagnetism. the kagomé
				{I}sing net}},\ }\href {https://doi.org/10.1143/ptp/10.2.158} {\bibfield
		{journal} {\bibinfo  {journal} {Progress of Theoretical Physics}\ }\textbf
		{\bibinfo {volume} {10}},\ \bibinfo {pages} {158} (\bibinfo {year}
		{1953})}\BibitemShut {NoStop}%
	\bibitem [{\citenamefont {Singh}\ and\ \citenamefont
		{Rigol}(2009)}]{SinghRigol:KIAF}%
	\BibitemOpen
	\bibfield  {author} {\bibinfo {author} {\bibfnamefont {R.~R.~P.}\
			\bibnamefont {Singh}}\ and\ \bibinfo {author} {\bibfnamefont
			{M.}~\bibnamefont {Rigol}},\ }\bibfield  {title} {\bibinfo {title}
		{Thermodynamic properties of kagome antiferromagnets with different
			perturbations},\ }\href {https://doi.org/10.1088/1742-6596/145/1/012003}
	{\bibfield  {journal} {\bibinfo  {journal} {Journal of Physics: Conference
				Series}\ }\textbf {\bibinfo {volume} {145}},\ \bibinfo {pages} {012003}
		(\bibinfo {year} {2009})}\BibitemShut {NoStop}%
	\bibitem [{\citenamefont {Lau}\ \emph {et~al.}(2006)\citenamefont {Lau},
		\citenamefont {Freitas}, \citenamefont {Ueland}, \citenamefont {Muegge},
		\citenamefont {Duncan}, \citenamefont {Schiffer},\ and\ \citenamefont
		{Cava}}]{LauFreitas:SpinIce}%
	\BibitemOpen
	\bibfield  {author} {\bibinfo {author} {\bibfnamefont {G.~C.}\ \bibnamefont
			{Lau}}, \bibinfo {author} {\bibfnamefont {R.~S.}\ \bibnamefont {Freitas}},
		\bibinfo {author} {\bibfnamefont {B.~G.}\ \bibnamefont {Ueland}}, \bibinfo
		{author} {\bibfnamefont {B.~D.}\ \bibnamefont {Muegge}}, \bibinfo {author}
		{\bibfnamefont {E.~L.}\ \bibnamefont {Duncan}}, \bibinfo {author}
		{\bibfnamefont {P.}~\bibnamefont {Schiffer}},\ and\ \bibinfo {author}
		{\bibfnamefont {R.~J.}\ \bibnamefont {Cava}},\ }\bibfield  {title} {\bibinfo
		{title} {Zero-point entropy in stuffed spin-ice},\ }\href
	{https://doi.org/10.1038/nphys270} {\bibfield  {journal} {\bibinfo  {journal}
			{Nature Physics}\ }\textbf {\bibinfo {volume} {2}},\ \bibinfo {pages}
		{249–253} (\bibinfo {year} {2006})}\BibitemShut {NoStop}%
	\bibitem [{\citenamefont {Mart\'{\i}nez}\ \emph {et~al.}(1992)\citenamefont
		{Mart\'{\i}nez}, \citenamefont {Sandiumenge}, \citenamefont {Rouco},
		\citenamefont {Labarta}, \citenamefont {Rodr\'{\i}guez-Carvajal},
		\citenamefont {Tovar}, \citenamefont {Causa}, \citenamefont {Gal\'{\i}},\
		and\ \citenamefont {Obradors}}]{Martinez:KagomeSCGO}%
	\BibitemOpen
	\bibfield  {author} {\bibinfo {author} {\bibfnamefont {B.}~\bibnamefont
			{Mart\'{\i}nez}}, \bibinfo {author} {\bibfnamefont {F.}~\bibnamefont
			{Sandiumenge}}, \bibinfo {author} {\bibfnamefont {A.}~\bibnamefont {Rouco}},
		\bibinfo {author} {\bibfnamefont {A.}~\bibnamefont {Labarta}}, \bibinfo
		{author} {\bibfnamefont {J.}~\bibnamefont {Rodr\'{\i}guez-Carvajal}},
		\bibinfo {author} {\bibfnamefont {M.}~\bibnamefont {Tovar}}, \bibinfo
		{author} {\bibfnamefont {M.}~\bibnamefont {Causa}}, \bibinfo {author}
		{\bibfnamefont {S.}~\bibnamefont {Gal\'{\i}}},\ and\ \bibinfo {author}
		{\bibfnamefont {X.}~\bibnamefont {Obradors}},\ }\bibfield  {title} {\bibinfo
		{title} {{Magnetic dilution in the strongly frustrated kagome antiferromagnet
				${\mathrm{SrGa}}_{12\mathrm{\ensuremath{-}}\mathit{x}}$${\mathrm{Cr}}_{\mathit{x}}$${\mathrm{O}}_{19}$}},\
	}\href {https://doi.org/10.1103/PhysRevB.46.10786} {\bibfield  {journal}
		{\bibinfo  {journal} {Phys. Rev. B}\ }\textbf {\bibinfo {volume} {46}},\
		\bibinfo {pages} {10786} (\bibinfo {year} {1992})}\BibitemShut {NoStop}%
	\bibitem [{\citenamefont {Anderson}\ \emph {et~al.}(1972)\citenamefont
		{Anderson}, \citenamefont {Halperin},\ and\ \citenamefont
		{Varma}}]{AndersonHalperinVarma:GlassHeatCapacity}%
	\BibitemOpen
	\bibfield  {author} {\bibinfo {author} {\bibfnamefont {P.~W.}\ \bibnamefont
			{Anderson}}, \bibinfo {author} {\bibfnamefont {B.~I.}\ \bibnamefont
			{Halperin}},\ and\ \bibinfo {author} {\bibfnamefont {C.~M.}\ \bibnamefont
			{Varma}},\ }\bibfield  {title} {\bibinfo {title} {Anomalous low-temperature
			thermal properties of glasses and spin glasses},\ }\href
	{https://doi.org/10.1080/14786437208229210} {\bibfield  {journal} {\bibinfo
			{journal} {The Philosophical Magazine: A Journal of Theoretical Experimental
				and Applied Physics}\ }\textbf {\bibinfo {volume} {25}},\ \bibinfo {pages}
		{1} (\bibinfo {year} {1972})},\ \Eprint
	{https://arxiv.org/abs/https://doi.org/10.1080/14786437208229210}
	{https://doi.org/10.1080/14786437208229210} \BibitemShut {NoStop}%
	\bibitem [{\citenamefont {Broholm}\ \emph
		{et~al.}(1990{\natexlab{b}})\citenamefont {Broholm}, \citenamefont {Aeppli},
		\citenamefont {Espinosa},\ and\ \citenamefont
		{Cooper}}]{Broholm:SCGOcorrelationLength}%
	\BibitemOpen
	\bibfield  {author} {\bibinfo {author} {\bibfnamefont {C.}~\bibnamefont
			{Broholm}}, \bibinfo {author} {\bibfnamefont {G.}~\bibnamefont {Aeppli}},
		\bibinfo {author} {\bibfnamefont {G.~P.}\ \bibnamefont {Espinosa}},\ and\
		\bibinfo {author} {\bibfnamefont {A.~S.}\ \bibnamefont {Cooper}},\ }\bibfield
	{title} {\bibinfo {title} {Antiferromagnetic fluctuations and short-range
			order in a kagom\'e lattice},\ }\href
	{https://doi.org/10.1103/PhysRevLett.65.3173} {\bibfield  {journal} {\bibinfo
			{journal} {Phys. Rev. Lett.}\ }\textbf {\bibinfo {volume} {65}},\ \bibinfo
		{pages} {3173} (\bibinfo {year} {1990}{\natexlab{b}})}\BibitemShut {NoStop}%
	\bibitem [{\citenamefont {Ramirez}\ \emph {et~al.}(1992)\citenamefont
		{Ramirez}, \citenamefont {Espinosa},\ and\ \citenamefont
		{Cooper}}]{Ramirez:T2SCGO}%
	\BibitemOpen
	\bibfield  {author} {\bibinfo {author} {\bibfnamefont {A.~P.}\ \bibnamefont
			{Ramirez}}, \bibinfo {author} {\bibfnamefont {G.~P.}\ \bibnamefont
			{Espinosa}},\ and\ \bibinfo {author} {\bibfnamefont {A.~S.}\ \bibnamefont
			{Cooper}},\ }\bibfield  {title} {\bibinfo {title} {Elementary excitations in
			a diluted antiferromagnetic kagom\'e lattice},\ }\href
	{https://doi.org/10.1103/PhysRevB.45.2505} {\bibfield  {journal} {\bibinfo
			{journal} {Phys. Rev. B}\ }\textbf {\bibinfo {volume} {45}},\ \bibinfo
		{pages} {2505} (\bibinfo {year} {1992})}\BibitemShut {NoStop}%
	\bibitem [{\citenamefont {Halperin}\ and\ \citenamefont
		{Saslow}(1977)}]{HalperinSaslowModes}%
	\BibitemOpen
	\bibfield  {author} {\bibinfo {author} {\bibfnamefont {B.~I.}\ \bibnamefont
			{Halperin}}\ and\ \bibinfo {author} {\bibfnamefont {W.~M.}\ \bibnamefont
			{Saslow}},\ }\bibfield  {title} {\bibinfo {title} {Hydrodynamic theory of
			spin waves in spin glasses and other systems with noncollinear spin
			orientations},\ }\href {https://doi.org/10.1103/PhysRevB.16.2154} {\bibfield
		{journal} {\bibinfo  {journal} {Phys. Rev. B}\ }\textbf {\bibinfo {volume}
			{16}},\ \bibinfo {pages} {2154} (\bibinfo {year} {1977})}\BibitemShut
	{NoStop}%
	\bibitem [{\citenamefont {Podolsky}\ and\ \citenamefont
		{Kim}(2009)}]{PodolskyKim:HalperinSaslowNiGaS}%
	\BibitemOpen
	\bibfield  {author} {\bibinfo {author} {\bibfnamefont {D.}~\bibnamefont
			{Podolsky}}\ and\ \bibinfo {author} {\bibfnamefont {Y.~B.}\ \bibnamefont
			{Kim}},\ }\bibfield  {title} {\bibinfo {title} {Halperin-saslow modes as the
			origin of the low-temperature anomaly in ${\text{niga}}_{2}{\text{s}}_{4}$},\
	}\href {https://doi.org/10.1103/PhysRevB.79.140402} {\bibfield  {journal}
		{\bibinfo  {journal} {Phys. Rev. B}\ }\textbf {\bibinfo {volume} {79}},\
		\bibinfo {pages} {140402} (\bibinfo {year} {2009})}\BibitemShut {NoStop}%
	\bibitem [{\citenamefont {Waldtmann}\ \emph {et~al.}(1998)\citenamefont
		{Waldtmann}, \citenamefont {Everts}, \citenamefont {Bernu}, \citenamefont
		{Lhuillier}, \citenamefont {Sindzingre}, \citenamefont {Lecheminant},\ and\
		\citenamefont {Pierre}}]{WaldtmannEverts:KHAF}%
	\BibitemOpen
	\bibfield  {author} {\bibinfo {author} {\bibfnamefont {C.}~\bibnamefont
			{Waldtmann}}, \bibinfo {author} {\bibfnamefont {H.-U.}\ \bibnamefont
			{Everts}}, \bibinfo {author} {\bibfnamefont {B.}~\bibnamefont {Bernu}},
		\bibinfo {author} {\bibfnamefont {C.}~\bibnamefont {Lhuillier}}, \bibinfo
		{author} {\bibfnamefont {P.}~\bibnamefont {Sindzingre}}, \bibinfo {author}
		{\bibfnamefont {P.}~\bibnamefont {Lecheminant}},\ and\ \bibinfo {author}
		{\bibfnamefont {L.}~\bibnamefont {Pierre}},\ }\bibfield  {title} {\bibinfo
		{title} {First excitations of the spin $1/2$ {H}eisenberg antiferromagnet on
			the kagomé lattice},\ }\href {https://doi.org/10.1007/s100510050274}
	{\bibfield  {journal} {\bibinfo  {journal} {The European Physical Journal B}\
		}\textbf {\bibinfo {volume} {2}},\ \bibinfo {pages} {501–507} (\bibinfo
		{year} {1998})}\BibitemShut {NoStop}%
	\bibitem [{\citenamefont {Ramirez}\ \emph {et~al.}(2000)\citenamefont
		{Ramirez}, \citenamefont {Hessen},\ and\ \citenamefont
		{Winklemann}}]{Ramirez:SCGOentropy}%
	\BibitemOpen
	\bibfield  {author} {\bibinfo {author} {\bibfnamefont {A.~P.}\ \bibnamefont
			{Ramirez}}, \bibinfo {author} {\bibfnamefont {B.}~\bibnamefont {Hessen}},\
		and\ \bibinfo {author} {\bibfnamefont {M.}~\bibnamefont {Winklemann}},\
	}\bibfield  {title} {\bibinfo {title} {Entropy balance and evidence for local
			spin singlets in a kagom\'e-like magnet},\ }\href
	{https://doi.org/10.1103/PhysRevLett.84.2957} {\bibfield  {journal} {\bibinfo
			{journal} {Phys. Rev. Lett.}\ }\textbf {\bibinfo {volume} {84}},\ \bibinfo
		{pages} {2957} (\bibinfo {year} {2000})}\BibitemShut {NoStop}%
	\bibitem [{\citenamefont {Limot}\ \emph {et~al.}(2002)\citenamefont {Limot},
		\citenamefont {Mendels}, \citenamefont {Collin}, \citenamefont {Mondelli},
		\citenamefont {Ouladdiaf}, \citenamefont {Mutka}, \citenamefont {Blanchard},\
		and\ \citenamefont {Mekata}}]{Limo:SCGOnmr}%
	\BibitemOpen
	\bibfield  {author} {\bibinfo {author} {\bibfnamefont {L.}~\bibnamefont
			{Limot}}, \bibinfo {author} {\bibfnamefont {P.}~\bibnamefont {Mendels}},
		\bibinfo {author} {\bibfnamefont {G.}~\bibnamefont {Collin}}, \bibinfo
		{author} {\bibfnamefont {C.}~\bibnamefont {Mondelli}}, \bibinfo {author}
		{\bibfnamefont {B.}~\bibnamefont {Ouladdiaf}}, \bibinfo {author}
		{\bibfnamefont {H.}~\bibnamefont {Mutka}}, \bibinfo {author} {\bibfnamefont
			{N.}~\bibnamefont {Blanchard}},\ and\ \bibinfo {author} {\bibfnamefont
			{M.}~\bibnamefont {Mekata}},\ }\bibfield  {title} {\bibinfo {title}
		{{Susceptibility and dilution effects of the kagom\'e bilayer geometrically
				frustrated network: A Ga NMR study of
				${\mathrm{SrCr}}_{9p}{\mathrm{Ga}}_{12\ensuremath{-}9p}{\mathrm{O}}_{19}$}},\
	}\href {https://doi.org/10.1103/PhysRevB.65.144447} {\bibfield  {journal}
		{\bibinfo  {journal} {Phys. Rev. B}\ }\textbf {\bibinfo {volume} {65}},\
		\bibinfo {pages} {144447} (\bibinfo {year} {2002})}\BibitemShut {NoStop}%
	\bibitem [{\citenamefont {Phelan}\ \emph {et~al.}(2023)\citenamefont {Phelan},
		\citenamefont {Ye}, \citenamefont {Zheng}, \citenamefont {Krivyakina},
		\citenamefont {Samarakoon}, \citenamefont {LaBarre}, \citenamefont {Neu},
		\citenamefont {Siegrist}, \citenamefont {Rosenkranz}, \citenamefont
		{Syzranov} \emph {et~al.}}]{Phelan:surfboards}%
	\BibitemOpen
	\bibfield  {author} {\bibinfo {author} {\bibfnamefont {D.}~\bibnamefont
			{Phelan}}, \bibinfo {author} {\bibfnamefont {F.}~\bibnamefont {Ye}}, \bibinfo
		{author} {\bibfnamefont {H.}~\bibnamefont {Zheng}}, \bibinfo {author}
		{\bibfnamefont {E.}~\bibnamefont {Krivyakina}}, \bibinfo {author}
		{\bibfnamefont {A.}~\bibnamefont {Samarakoon}}, \bibinfo {author}
		{\bibfnamefont {P.}~\bibnamefont {LaBarre}}, \bibinfo {author} {\bibfnamefont
			{J.}~\bibnamefont {Neu}}, \bibinfo {author} {\bibfnamefont {T.}~\bibnamefont
			{Siegrist}}, \bibinfo {author} {\bibfnamefont {S.}~\bibnamefont
			{Rosenkranz}}, \bibinfo {author} {\bibfnamefont {S.}~\bibnamefont
			{Syzranov}}, \emph {et~al.},\ }\bibfield  {title} {\bibinfo {title} {The
			geometrically frustrated spin glass {$(Fe_{1-p}Ga_p)_2TiO_5$}},\ }\href@noop
	{} {\bibfield  {journal} {\bibinfo  {journal} {Acta Crystallographica Section
				A: Foundations and Advances}\ }\textbf {\bibinfo {volume} {79}},\ \bibinfo
		{pages} {a247} (\bibinfo {year} {2023})}\BibitemShut {NoStop}%
\end{thebibliography}
\end{document}